%% file: zak2.tex
\documentstyle[preprint,eqsecnum,aps]{revtex}
\begin{document}
\newcommand{\beq}{\begin{equation}}
\newcommand{\eeq}{\end{equation}}
\newcommand{\bea}{\begin{eqnarray}}
\newcommand{\eea}{\end{eqnarray}}
\newcommand{\ek}{\not{\varepsilon}}
\newcommand{\eek}{\vec{\varepsilon}}
\newcommand{\gp}{ {\cal G\/}_{F} }

\draft
\preprint{BETHLEHEM-Phys-HEP980727}
\title{
Hot Nuclear Matter in the Quark Meson Coupling Model with Dilatons
}

\author{I. Zakout and H. R. Jaqaman}
\address{Department of Physics, Bethlehem University,
P.O. Box 9, Bethlehem, Palestine}
\maketitle
\begin{abstract}
We study hot nuclear matter in an explicit quark model based on a mean field 
description of nonoverlapping nucleon bags bound by the self-consistent 
exchange of scalar and vector mesons as well as the glueball field.
The glueball exchange as well as 
a realization of the broken scale invariance of quantum 
chromodynamics is achieved  through the introduction of 
a dilaton field.
The calculations also take into account the medium-dependence of the  
bag constant.
The effective potential with dilatons is applied to nuclear matter.
The nucleon properties at finite temperature as calculated here are found 
to be appreciably different from cold nuclear matter.
The introduction of the dilaton  potential improves the shape of 
the saturation curve at T=0 and is found to affect hot
nuclear matter significantly.
\end{abstract}
%
\narrowtext

\section{introduction}

The description of nuclear phenomena in relativistic mean-field theory 
has been successfully formulated using the hadronic degrees of freedom
\cite{Walecka,SerotA}. 
However, due to the observations which revealed  the medium 
modification of the internal structure of the
nucleon\cite{EMC}, 
it has become essential to explicitly incorporate the 
quark-gluon degrees of freedom while respecting the 
established model based on the hadronic degrees of 
freedom in nuclei. It was in this spirit that Guichon proposed 
the quark-meson coupling model which describes nuclear matter 
as a collection of non-overlapping MIT bags interacting through 
the self-consistent exchange of mesons in the mean
field approximation with the meson fields directly
coupled to the quarks\cite{Guichon,Saito,Jin96,Panda}.
It has been further suggested that including a medium-dependent 
bag constant may be essential for the success
of relativistic nuclear phenomenology\cite{Jin96}. This modification has
been recently applied in a study of the properties of  nuclear matter 
at finite temperature\cite{zak1} where it was found that the bag constant 
decreases appreciably above a critical temperature 
$T_{c}\approx 200 MeV$ indicating the onset of quark deconfinement.

On the other hand, it is important to incorporate the breaking of the 
classical scale invariance of Quantum Chromdynamics (QCD) resulting from 
the QCD trace anomaly\cite{dilaton}. 
It has been determined that it is possible to incorporate the broken global 
chiral and scale symmetries of QCD into an effective lagrangian through 
the introduction of a scalar, chiral isoscalar glueball-dilaton field
$\phi$ \cite{dilaton,Rodriguez,Mishustin,Heide1,Heide2,Ellisn,kalbermann}.
There have also been attempts to introduce the dilaton field 
in an appropriate way in order to obtain reasonable values for the
compression modulus of nuclear matter\cite{Heide1,Heide2,FurnstahlS}.
It has been noted that the scale invariant term which leads to 
an omega meson, after symmetry breaking, is strongly
favored to be of the form 
$\omega_{\mu}\omega^{\mu}\frac{\phi^{2}}{\phi^{2}_{0}}$
by the bulk properties of nuclei\cite{Heide2}.
The effective lagrangian
\begin{eqnarray}
{\cal L}={\cal L}_{0}-V_{G}, \label{effpot} 
\end{eqnarray}
consists of a chiral and scale invariant part
${\cal L}_{0}$ and an explicitly scale-breaking 
potential $V_{G}$.
The divergence of the scale current in QCD is given 
by the trace of the improved energy momentum tensor 
and so the scalar potential $V_{G}$ was chosen
to reproduce, via Noether's theorem, 
the effective trace anomaly\cite{dilaton}
\begin{eqnarray}
\theta^{\mu}_{\mu}=4V_{G}(\Phi)-
\sum_{i}\Phi_{i}\frac{\partial V_{G}}{\partial \Phi_{i}}
=4\epsilon_{\mbox{vac}}
\left(\frac{\phi}{\phi_{0}}\right)^{4},
\label{Noether} 
\end{eqnarray}
where $\Phi_{i}$ runs over the scalar fields
$\{\sigma,\vec{\pi},\phi\}$ and $\epsilon_{\mbox{vac}}$
is the vacuum energy\cite{Heide1,Heide2}. 
The proportionality $\theta^{\mu}_{\mu}\propto 
\left(\frac{\phi}{\phi_{0}}\right)^{4}$ is suggested 
by the form of the QCD trace anomaly.
In the mean field approximation, we have $<\vec{\pi}>=0$
for symmetric nuclear matter so that the isovector pionic field has 
no effect on the properties of nuclear matter in this approximation.

The outline of the present paper is as follows.
In Sect. II, we formalize the quark-meson coupling model with
dilatons at finite temperature. This is essentially an extension of 
our earlier work\cite{zak1} where the dilaton field was not included. 
Finally in Sect. III, we present our results and conclusions.

\section{The Quark Meson Coupling Model with Dilatons at Finite Temperature}

The quark field $\psi_{q}(\vec{r},t)$ inside the bag satisfies
\begin{eqnarray}
\left[
i\gamma^{\mu}\partial_{\mu} - 
\left(m^{0}_{q}\frac{\phi}{\phi_{0}}-g_{\sigma}^{q}\sigma\right)
-g^{q}_{\omega}\omega\gamma^{0}\right]
\psi_{q}(\vec{r},t)=0,
\label{bag} 
\end{eqnarray}
where $m^{0}_{q}$ is the current quark mass while 
$g^{q}_{\sigma}$ and $g^{q}_{\omega}$ 
are the quark couplings with the scalar $\sigma$ 
and vector $\omega$ mesons, respectively, 
and $\frac{\phi}{\phi_{0}}$ is the scale invariance breaking.

The single-particle quark and antiquark energies in units of 
$R^{-1}$, where $R$ is the bag radius, are given by\cite{Panda}
\begin{eqnarray}
\epsilon^{n\kappa}_{\pm}=\Omega^{n\kappa}\pm g^{q}_{\omega}\omega R
\label{eng} 
\end{eqnarray} 
where 
$\Omega^{n\kappa}=\sqrt{ x^{2}_{n\kappa}+R^{2}{m^{*}_{q}}^{2} }$
and $m^{*}_{q}=\frac{\phi}{\phi_{0}}m^{0}_{q}-g^{q}_{\sigma}\sigma$
is the effective quark mass. The quark momentum $x_{n\kappa}$
in the state characterized by specific values of 
$n$ and $\kappa$ is determined by the boundary condition at the bag surface 
\begin{eqnarray}
i\gamma\cdot n \psi^{n\kappa}_{q}|_{R}=
\psi^{n\kappa}_{q}|_{R},
\label{bound} 
\end{eqnarray}
which reduces for the ground state \cite{Saito,Jin96} to 
\begin{eqnarray}
j_{0}(x_{n\kappa})=\beta^{n\kappa}_{q}j_{1}(x_{n\kappa})
\label{sbound} 
\end{eqnarray}
where 
\begin{eqnarray}
\beta^{n\kappa}_{q}=\sqrt{\frac{\Omega^{n\kappa}-Rm^{*}_{q}}
{\Omega^{n\kappa}+Rm^{*}_{q}}}
\label{beta} 
\end{eqnarray}
and the coefficient $\kappa=-1$ for the $s$-state.

The total energy from the quarks and antiquarks reads
\begin{eqnarray}
E_{tot}=3\sum_{n\kappa}\frac{\Omega^{n\kappa}}{R}
\left[\frac{1}{e^{(\epsilon_{+}^{n\kappa}/R -\mu_{q})/T}+1}
+\frac{1}{e^{(\epsilon_{-}^{n\kappa}/R +\mu_{q})/T}+1}  
\right].
\label{Etot} 
\end{eqnarray}
The bag energy now becomes
\begin{eqnarray}
E_{bag}=E_{tot}-\frac{\phi}{\phi_{0}}\frac{Z}{R}
+\frac{4\pi}{3}R^{3}\frac{\phi}{\phi_{0}}B(\sigma).
\label{Ebag} 
\end{eqnarray}
where the $\phi$ dependence  is suggested by the asymptotic behavior
of the bag energy in the purely hadronic models\cite{kalbermann} 
\begin{eqnarray}
E_{bag}\propto \frac{\phi}{\phi_{0}} M_{N}. \label{Dscale}
\end{eqnarray}
The medium effects are taken into account by adopting the bag parameter
$B(\sigma)=B_{0}\exp\left(-\frac{4g_{\sigma}^{B}\sigma}%
{\frac{\phi}{\phi_{0}}M_{N}}\right)$
where $g_{\sigma}^{B}$ is an additional 
fitting parameter\cite{Jin96}.
The spurious center-of-mass motion in the bag is subtracted to obtain the
effective nucleon mass\cite{Fleck}
\begin{eqnarray}
M^{*}_{N}=\sqrt{E^{2}_{bag}-<p^{2}_{cm}>}. \label{Ecm}
\end{eqnarray}
The spurious center-of-mass average momentum squared is given by
\begin{eqnarray}
<p^{2}_{cm}>=\frac{<x^{2}>}{R^{2}}, \label{Pcm2}
\end{eqnarray}
where
\begin{eqnarray}
<x^{2}>=3\sum_{n\kappa} x^{2}_{n\kappa}
\left[\frac{1}{e^{(\epsilon^{n\kappa}_{+}/R-\mu_{q})/T}+1}
+ \frac{1}{e^{(\epsilon^{n\kappa}_{-}/R+\mu_{q})/T}+1}
\right] \label{Xsq}
\end{eqnarray}
is written  in terms of the sum of the quark and antiquark Fermi 
distribution functions since the center of mass motion
does not distinguish between a quark and an antiquark.
The quark chemical potential $\mu_{q}$, assuming that there are three
quarks in the nucleon bag, is determined through the constraint
\begin{eqnarray}
n_{q}=3=3\sum_{n\kappa}
\left[
\frac{1}{e^{(\epsilon^{n\kappa}_{+}/R-\mu_{q})/T}+1}
-
\frac{1}{e^{(\epsilon^{n\kappa}_{-}/R+\mu_{q})/T}+1}
\right]. \label{quarkc}
\end{eqnarray}
The temperature-dependent bag radius $R$ is obtained through 
the condition
\begin{eqnarray}
\frac{\partial M^{*}_{N}}{\partial R}=0. \label{Condi} 
\end{eqnarray}
The baryon-antibaryon thermal distribution functions
are given by
\begin{eqnarray}
f_{B}(\mu_{B})=\frac{1}{e^{(\epsilon^{*}(k)-\mu^{*}_{B})/T}+1}
\label{fB} 
\end{eqnarray}
and 
\begin{eqnarray}
\overline{f}_{B}(\mu_{B})=
\frac{1}{e^{(\epsilon^{*}(k)+\mu^{*}_{B})/T}+1}
\label{fBAR} 
\end{eqnarray}
with 
$\epsilon^{*}(\vec{k})=\sqrt{ \vec{k}^{2}+{M^{*}_{N}}^{2} }$
and the effective baryon chemical potential
$\mu^{*}_{B}=\mu_{B}-g_{\omega}\omega$.
The baryon chemical potential, $\mu_{B}$, 
is obtained from the nontrivial solution of
\begin{eqnarray}
\rho_{B}=\frac{\gamma}{(2\pi)^{3}}\int d^{3} k
\left(f_{B}(\mu_{B}) - \overline{f}_{B}(\mu_{B})\right).
\label{rhoB} 
\end{eqnarray}
However, for a given chemical potential $\mu_{B}$, 
the baryon density $\rho_{B}$ can be calculated easily. 
Subsequently, the vector mean field reads
\begin{eqnarray}
\omega=\frac{1}{m^{2}_{\omega}(\frac{\phi}{\phi_{0}})^{2}}
g_{\omega}\rho_{B}.
\label{selfomg} 
\end{eqnarray}
The total energy density at finite temperature and baryon
density $\rho_{B}$ reads
\begin{eqnarray}
\epsilon&=&\frac{\gamma}{(2\pi)^{3}}
\int d^{3} k \sqrt{ k^{2}+{M^{*}_{N}}^{2}}
\left[f_{B}(\mu_{B})+\overline{f}_{B}(\mu_{B})\right]
\nonumber \\
&+&\frac{1}{2}m^{2}_{\omega}\left(\frac{\phi}{\phi_{0}}\right)^{2}\omega^{2}
+\frac{1}{2}m^{2}_{\sigma}\left(\frac{\phi}{\phi_{0}}\right)^{2}\sigma^{2}
+U[\phi],
\label{Energy}
\end{eqnarray}
with the dilaton potential $U[\phi]$ given 
by $U[\phi]=V_{G}[\phi]-V_{G}[\phi_{0}]$
where\cite{Heide1,Heide2,Ellisn,kalbermann}
\begin{eqnarray}
V_{G}[\phi]=\epsilon_{\mbox{vac}}
\left(\frac{\phi}{\phi_{0}}\right)^{4}
\left[\log\left(\frac{\phi^{4}}{\phi^{4}_{0}}\right) - 1\right].
\label{diltp}
\end{eqnarray}
Here $\epsilon_{\mbox{vac}}$ is  the dilaton potential constant 
and corresponds to the vacuum energy.
The pressure of nuclear matter reads
\begin{eqnarray}
P=\frac{1}{3}\frac{\gamma}{(2\pi)^{3}}
\int d^{3}k \frac{\vec{k}^{2}}{\epsilon^{*}(k)}
\left[f_{B}(\mu_{B})+\overline{f}_{B}(\mu_{B})\right]+
\frac{1}{2}m^{2}_{\omega}\left(\frac{\phi}{\phi_{0}}\right)^{2}\omega^{2}
-\frac{1}{2}m^{2}_{\sigma}\left(\frac{\phi}{\phi_{0}}\right)^{2}\sigma^{2}
-U[\phi].
\label{Press} 
\end{eqnarray}
The scalar and dilaton fields are determined 
by the extremization of the pressure through the conditions
\cite{Saito,Jin96,Panda,zak1,kalbermann}
\begin{eqnarray}
\left.\frac{\partial P}{\partial \sigma}\right|_{\sigma}=
\left(\frac{\partial P}{\partial M^{*}_{N}}\right)_{\mu_{B},T}
\left(\frac{\partial M^{*}_{N}}{\partial \sigma}\right) 
+\left(\frac{\partial P}{\partial \sigma}\right)_{M^{*}_{N}}=0,
\label{CondII} 
\end{eqnarray}
and
\begin{eqnarray}
\left.\frac{\partial P}{\partial \chi}\right|_{\phi/\phi_{0}}=
\left(\frac{\partial P}{\partial M^{*}_{N}}\right)_{\mu_{B},T}
\left(\frac{\partial M^{*}_{N}}{\partial \chi}\right)   
+\left(\frac{\partial P}{\partial \chi}\right)_{M^{*}_{N}}=0,
\label{CondIII} 
\end{eqnarray}
for the $\sigma$ and $\chi=\frac{\phi}{\phi_{0}}$ fields, respectively. 
This extremization is carried out  while taking into consideration the 
full coupling of the scalar mean fields to the internal quark structure 
of the bag by means of the solution of the point-like Dirac equation with 
the required boundary condition of confinement at the surface of the bag 
as suggested by Refs.\cite{Saito,Jin96,zak1}.
The variation of the effective chemical potential $\mu^{*}_{B}$ is taken as
\begin{eqnarray}
\frac{ \partial \mu^{*}_{B} }{\partial M^{*}_{N} }=-g_{\omega}
\frac{ \partial \omega }{\partial M^{*}_{N} }
\label{ChemM} 
\end{eqnarray}
where
\begin{eqnarray}
\frac{\partial \omega}{\partial M^{*}_{N}}
=-
\frac{
\frac{g_{\omega}}{(\phi/\phi_{0})^{2}m^{2}_{\omega}}
\frac{\gamma}{(2\pi)^{3}}\frac{1}{T}
\int d^{3} k \frac{M^{*}_{N}}{\epsilon^{*}}
\left[f_{B}(1-f_{B})-\overline{f}_{B}(1-\overline{f}_{B})\right]
}{
1+ 
\frac{g^{2}_{\omega}}{(\phi/\phi_{0})^{2}m^{2}_{\omega}}
\frac{\gamma}{(2\pi)^{3}}\frac{1}{T}
\int d^{3} k
\left[f_{B}(1-f_{B})+\overline{f}_{B}(1-\overline{f}_{B})\right]
}.
\label{OmegM}
\end{eqnarray}

\section{Results and Conclusions}
We have used the quark meson coupling to study nuclear matter at zero and finite 
temperatures.
Glueball exchange 
and the breaking of scale invariance are taken into account by introducing 
the dilaton field.
We have adopted the fitting parameters that are used in the earlier
calculations, in the absence of the dilaton 
potential, which fit the saturation properties of 
nuclear matter\cite{Jin96,Panda}.
The bag parameters 
$B^{1/4}_{0}=188.1$ MeV and $Z_{0}=2.03$
are chosen to reproduce the free nucleon mass
$M_{N}$ at its experimental
value 939 MeV and bag radius $R_{0}=0.60$ fm.   
The current quark mass $m_{q}$ is taken equal to zero.
For $g_{\sigma}^{q}=1$, the values of the vector meson coupling
and the parameter $g_{\sigma}^{B}$ as fitted from the saturation
properties of nuclear  matter, are given
as $g^{2}_{\omega}/4\pi=5.24$
and ${g^{B}_{\sigma}}^{2}/4\pi$=3.69.

For specific values of the temperature and $\mu_{B}$, the thermodynamic
potential is given in terms of the effective nucleon mass which depends
on the bag radius R, the quark chemical potential $\mu_{q}$ and 
the mean fields $\sigma$ and $\omega$ as well as the dilaton field 
$\frac{\phi}{\phi_{0}}$.
For given values of the mean fields $\sigma$ and $\omega$ 
as well as the dilaton field $\frac{\phi}{\phi_{0}}$, 
the quark chemical potential $\mu_{q}$ 
and the bag radius $R$ are determined using 
the  conditions Eqns.(\ref{quarkc}) and (\ref{Condi}), respectively. 
We then determine the values of $\sigma$ and $\frac{\phi}{\phi_{0}}$ 
by minimizing the thermodynamic potential 
$\Omega$ through the conditions of Eqns.(\ref{CondII}) and (\ref{CondIII}), 
respectively, together with the self-consistency condition\cite{Panda}
for the $\omega$ mean field as given in Eq.(\ref{selfomg}).
For given values of the temperature and baryon chemical potential
$\mu_{B}$, we have calculated the different 
thermodynamic quantities.

The breaking of scale invariance is tested by studying the behavior 
of the dilaton scale $\chi=\left(\frac{\phi}{\phi_{0}}\right)$.
When the glueball field is frozen, the dilaton scale takes 
the value $\chi=1$\cite{kalbermann,FurnstahlS}.
We display the dilaton scale $\chi$ as a function of 
the baryon density $\rho_{B}$ 
for different values of temperature in Fig. 1. 
The dilaton potential constant is taken 
as $\epsilon_{\mbox{vac}}=(250 \mbox{MeV})^{4}$.
The dilaton field for cold nuclear matter ( T=0 )increases weakly
at small baryon density (for values $\rho_{B}\leq 0.12$ fm$^{-3}$). 
When $\rho_{B}$ reaches $0.12$ fm$^{-3}$, it starts to decrease.
Furthermore, it takes values $\chi<1$ for baryon densities
$\rho_{B}>0.22$ fm$^{-3}$.
The behavior is quite different for hot nuclear matter where 
the dilaton scale $\chi$ is monotonically  
increasing with  $\rho_{B}$ at all temperatures. It is interesting to note
that in contrast to the results of calculations involving 
only baryonic degrees of freedom\cite{kalbermann} where the dilaton 
scale at finite temperature takes values $\chi<1$  it here takes 
values $\chi>1$.
Moreover, the dilaton scale is seen to attain values $\chi>1$ at zero 
baryon density  for temperatures $T\geq  T_c \approx 200$ MeV. This  is
related to the  phase transition seen in the earlier calculations 
without the dilaton field\cite{zak1} when the system becomes a dilute 
gas of baryons in a sea of baryon-antibaryon pairs and quark 
deconfinement sets in.
Therefore at the phase transition, the scale invariance is
broken even at zero baryon density.

The effective nucleon mass $M^{*}_{N}$ decreases monotonically with 
the  baryon density $\rho_{B}$ as can be seen in Fig. 2. 
It also increases  with temperature for all values of $\rho_{B}$ up to 
the critical temperature and then suddenly starts to decrease rapidly 
with temperature for $T\ge 200$ MeV at low baryon density $\rho_{B}$.
This rapid fall of $M^{*}_{N}$ with increasing temperature is also related
to the above-mentioned phase transition and was observed in the earlier 
calculation\cite{zak1}.
The general behavior of the pressure with temperature and baryon density 
is also found to be similar to that obtained
in the earlier calculation in the absence of the dilaton field
\cite{zak1,Furnstahl}.
For example, it takes a nonzero value at $\rho_{B}=0$ 
for temperatures $T\ge 200$ MeV. 

The bag radius $R$ as a function of the baryon density 
$\rho_{B}$ is displayed in Fig. 3 for several values of temperature.
The bag radius $R$ increases monotonically with the increase in the 
baryon density and decreases when the temperature is increased.
However, when the temperature reaches 200 MeV, the bag radius $R$ 
suddenly starts to increase with temperature for low baryon densities.
For sufficiently high temperatures, for instance $T=240$ MeV, 
the bag radius $R$ tends to be approximately constant with very 
little variation with the baryon density $\rho_{B}$.

Fig. 4 shows the scalar field $\sigma$ as a function of the baryon 
density $\rho_{B}$.
The scalar field $\sigma$ increases with  $\rho_{B}$ and,
at first, tends to decrease with  temperature until
the temperature reaches 200 MeV when it suddenly starts to increase for 
low baryon densities.
Furthermore, the scalar field $\sigma$ takes  nonzero values at and above 
the critical temperature $T_{c}=200$ MeV.
This behavior was also observed in the earlier calculations with
a frozen glueball\cite{Panda,zak1} and is related to the phase transition 
and the onset of quark deconfinement.

To study further the effect of the dilaton field on the quark meson
coupling model,
we investigate the effect of the value of the dilaton potential constant, 
$\epsilon_{\mbox{vac}}$,  on the properties of  cold and hot
nuclear matter. 
In Fig. 5 we plot the dilaton scale $\chi$ as a function of  $\rho_{B}$
for several values of $\epsilon_{\mbox{vac}}$ .
The cold nuclear matter case is displayed in Fig. 5(a) where it is seen 
that the dilaton field increases with  $\rho_{B}$ 
for low baryon densities until it reaches its maximum value at 
$\rho_{B}=0.12$ fm$^{-3}$ and then starts to decrease. 
The curves of the dilaton scale $\chi$ for several values of 
$\epsilon_{\mbox{vac}}$ all intersect with the frozen glueball scale
($\chi=1$) at $\rho_{B}=0.24$ fm$^{-3}$ 
and then continue to decrease below $\chi=1$, 
for  baryon densities $\rho_{B}>0.24$ fm$^{-3}$.
It is seen that for the lowest value $\epsilon_{\mbox{vac}}=(200\mbox{MeV})^{4}$ 
the variation in $\chi$ with $\rho_{B}$ is very dramatic and tend 
to very small values for $\rho_{B}>0.35$ fm$^{-3}$ where the solution 
becomes unstable indicating that such low values of $\epsilon_{\mbox{vac}}$ 
are physically unacceptable.  For larger values of $\epsilon_{\mbox{vac}}$ 
the variation becomes less and less dramatic. 
For sufficiently large values, for instance, 
$\epsilon_{\mbox{vac}}=(800\mbox{MeV})^{4}$, it tends to act almost as 
a frozen glueball field and the dilaton scale  becomes equal to one.
The variation of $\chi$ with $\epsilon_{\mbox{vac}}$ for hot nuclear
matter at the temperature 
$T=$200 MeV is displayed in Fig. 5(b).
The dilaton scale $\chi$ increases monotonically with  $\rho_{B}$ with 
$\chi$ always $>1$ and is found to decrease with $\epsilon_{\mbox{vac}}$.
Its variation with density  becomes very steep for 
$\epsilon_{\mbox{vac}}=(200 \mbox{MeV})^{4}$ while it tends to be constant 
for $\epsilon_{\mbox{vac}}=(800\mbox{MeV})^{4}$. 
It is thus seen that the behavior of the dilaton scale $\chi$ is 
completely different for  cold and hot nuclear matter for low
values of the dilaton potential constant $\epsilon_{\mbox{vac}}$.
However, for  sufficiently large values of  $\epsilon_{\mbox{vac}}$,
the dilaton field changes very slightly and cannot be distinguished 
from the case without dilatons for both  cold and hot nuclear matter. 

The dependence of the effective nucleon mass $M_{N}^{*}$ 
on the value of $\epsilon_{\mbox{vac}}$ is displayed
in Figs. 6(a) and 6(b) for  cold and  hot nuclear matter, respectively.
In the case of  cold nuclear matter, the nucleon mass $M^{*}_{N}$
does not seem to be affected by the value of $\epsilon_{\mbox{vac}}$. 
However, for the hot nuclear matter case at a temperature $T=$200 MeV,
$M^{*}_{N}$ increases slightly with  $\epsilon_{\mbox{vac}}$.
In both cases,  $M^{*}_{N}$  decreases with $\rho_{B}$ as already observed
in Fig. 2 as well as in the calculations without the 
dilaton\cite{zak1,Furnstahl}.

The variation of the bag radius $R$ with the value of 
$\epsilon_{\mbox{vac}}$ is displayed in Fig. 7(a) for  cold nuclear matter. 
It is seen that $R$ increases monotonically with  the baryon density 
$\rho_{B}$ for all $\epsilon_{\mbox{vac}}$ values and it increases weakly
with $\epsilon_{\mbox{vac}}$ for low baryon densities 
$\rho_{B}< 0.24 \mbox{fm}^{-3}$.
However, when $\rho_{B}=0.24$ fm$^{-3}$ the bag radius takes the same value
for all values of $\epsilon_{\mbox{vac}}$. 
When $\rho_{B}$ exceeds $0.24$ fm$^{-3}$, the bag radius $R$ starts to
decrease with the increase in the value of $\epsilon_{\mbox{vac}}$.
The variation of the bag radius for hot nuclear matter at $T=200$ MeV 
is displayed in Fig. 7(b).
It is seen that the bag radius increases with both $\rho_{B}$ and 
$\epsilon_{\mbox{vac}}$ except for 
the smallest value $\epsilon_{\mbox{vac}}=(200 \mbox{MeV})^{4}$ 
where it is found that
the bag radius $R$ at first increases weakly with 
$\rho_{B}$ and then starts to decrease.
 
The scalar field $\sigma$ as a function of the baryon density 
$\rho_{B}$ is displayed in Fig. 8.
For  cold nuclear matter, Fig. 8(a), it is seen that
$\sigma$ increases with increasing $\rho_{B}$ 
and is almost independent of $\epsilon_{\mbox{vac}}$
for $\rho_{B}\le 0.24$ fm$^{-3}$.
For $\rho_{B}>0.24$ fm$^{-3}$, it decreases  
weakly with $\epsilon_{\mbox{vac}}$.
On the other hand for the hot nuclear matter case  at 
$T=$200 MeV, it is seen in Fig. 8(b) that 
the scalar field $\sigma$ increases with  
both $\rho_{B}$ and $\epsilon_{\mbox{vac}}$ .

We also examined the saturation properties of nuclear matter at T=0.
In Fig. 9, we display $E_{tot}-M_{N}$ as a function 
of the baryon density 
for  several values of $\epsilon_{\mbox{vac}}$.
It is seen that as $\epsilon_{\mbox{vac}}$ decreases, the equation
of state becomes stiffer and the compressibility, $K$, 
increases\cite{Heide1,Heide2}.
The compressibility was found to have the reasonable value $K$=300 MeV for
$\epsilon_{\mbox{vac}}=(800 \mbox{MeV})^{4}$. 
However, for  smaller values of 
$\epsilon_{\mbox{vac}}$, the compressibility increases.
For example, $K$ is increased by about a factor of two if
$\epsilon_{\mbox{vac}}$ is changed from $(800 \mbox{MeV})^{4}$ 
to $(200\mbox{MeV})^{4}$.

\acknowledgments
Financial support by  the Deutsche Forschungsgemeinschaft through the 
grant GR 243/51-1 is gratefully acknowledged.


\clearpage

\begin{figure}
\caption{
The dilaton scale $\chi=\frac{\phi}{\phi_{0}}$
for nuclear matter as a function of the baryon density
$\rho_{B}$ with a dilaton potential constant 
$\epsilon_{\mbox{vac}}=(250 \mbox{MeV})^{4}$
for various values of temperature.}
\label{fa1}
\end{figure}

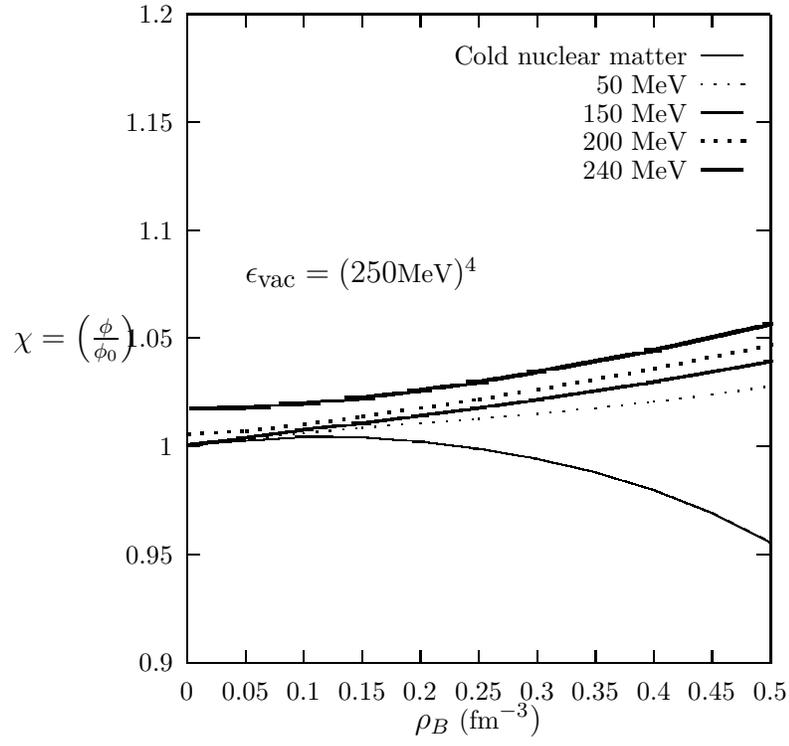
\begin{figure}
\caption{
The effective nucleon mass $M^{*}_{N}$ for nuclear matter as 
a function of the baryon density $\rho_{B}$
with a dilaton potential constant 
$\epsilon_{\mbox{vac}}=(250 \mbox{MeV})^{4}$
for various values of temperature.}
\label{fa2}
\end{figure}

\begin{figure}
\caption{
The bag radius $R$ for nuclear matter as a function of the baryon density $\rho_{B}$ with a
dilaton potential constant 
$\epsilon_{\mbox{vac}}=(250 \mbox{MeV})^{4}$
 for various values of temperature.}
\label{fa3}
\end{figure}

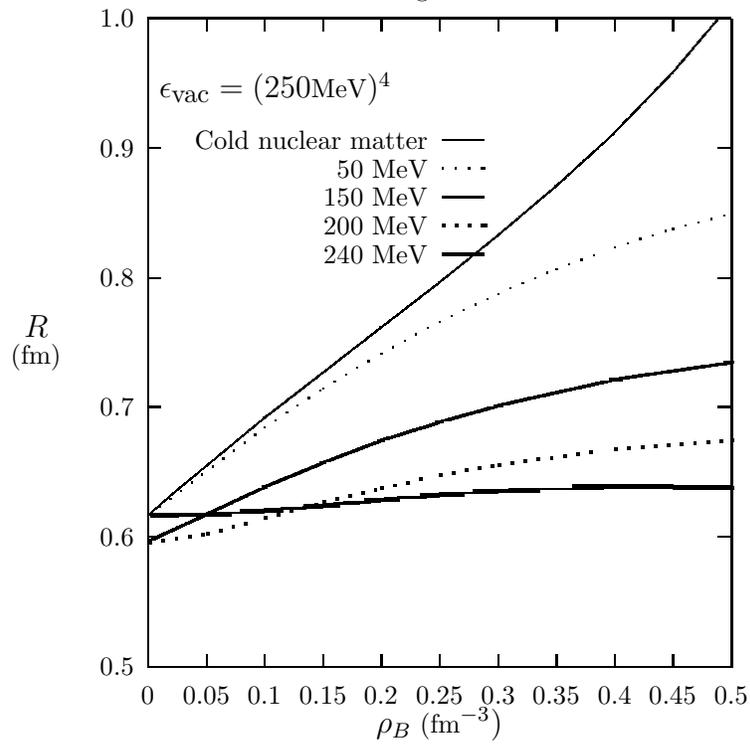
\begin{figure}
\caption{
The scalar mean field $\sigma$ for nuclear matter
as a function of the baryon density $\rho_{B}$
with a dilaton potential constant
$\epsilon_{\mbox{vac}}=(250 \mbox{MeV})^{4}$
at various temperatures.}
\label{fa4}
\end{figure}

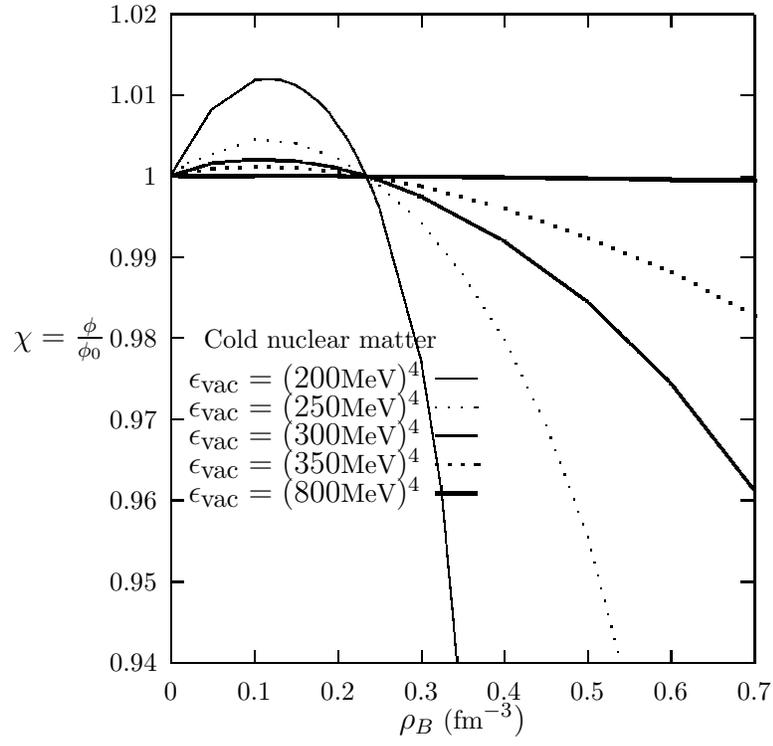
\begin{figure}
\caption{
The dilaton scale $\chi=\frac{\phi}{\phi_{0}}$ for a nuclear matter
as a function of the baryon density $\rho_{B}$ for several
values of the dilaton potential constant $\epsilon_{\mbox{vac}}$,
(a) for cold nuclear matter,
(b) for hot nuclear matter at  $T=200$ MeV.}
\label{fa5}                                                          
\end{figure}  

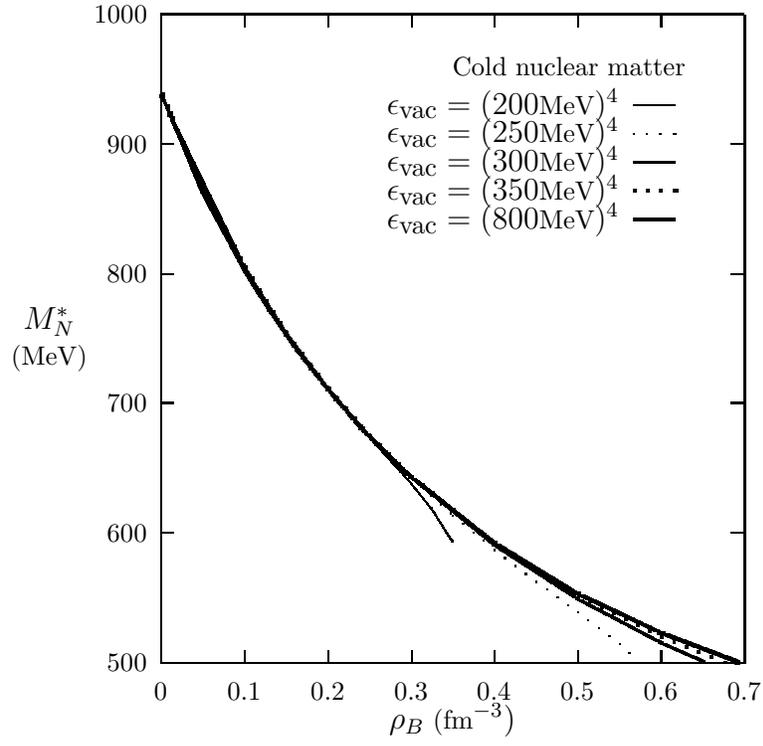
\begin{figure}
\caption{
The effective nucleon mass $M^{*}_{N}$ for nuclear matter
as a function of the baryon density $\rho_{B}$ for 
several values of $\epsilon_{\mbox{vac}}$,
(a) for cold nuclear matter,
(b) for hot nuclear matter at $T=200$ MeV.}
\label{fa6}
\end{figure}

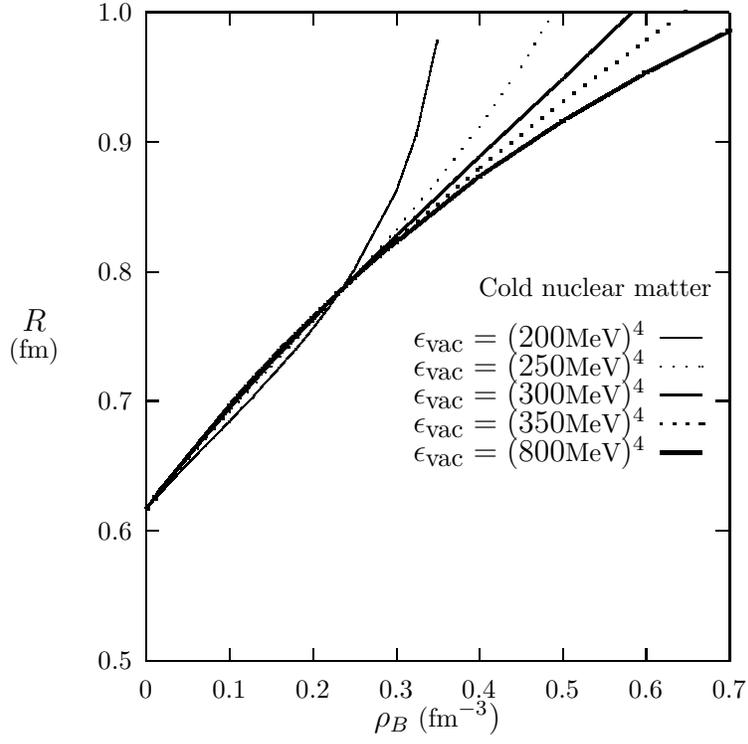
\begin{figure}
\caption{
The bag radius, $R$, as a function of the baryon density $\rho_{B}$ 
for several values
of  $\epsilon_{\mbox{vac}}$,
(a) for cold nuclear matter,  
(b) for hot nuclear matter at $T=200$ MeV.}
\label{fa7}
\end{figure}

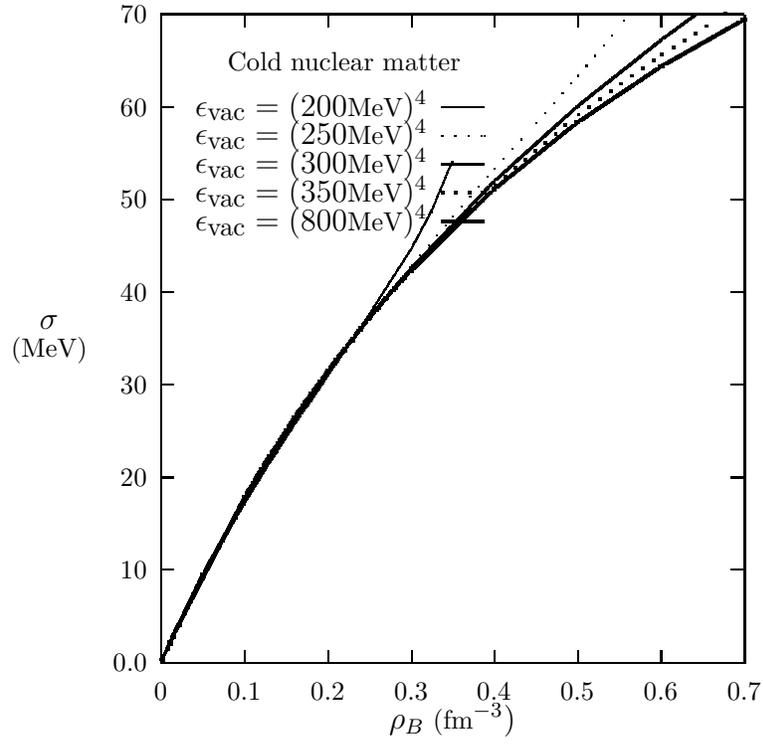
\begin{figure}
\caption{
The scalar mean field $\sigma$ as a function of baryon density $\rho_{B}$
for several values of  $\epsilon_{\mbox{vac}}$,
(a) for cold nuclear matter,
(b) for hot nuclear matter at $T=200$ MeV.}   
\label{fa8}
\end{figure}

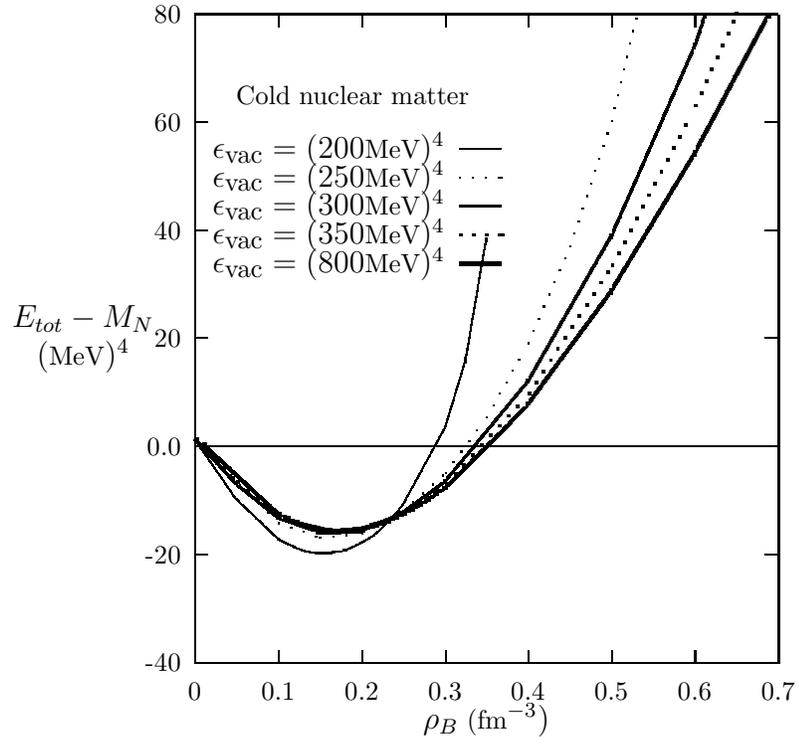
\begin{figure}
\caption{
The energy per nucleon for cold nuclear matter as a function
of baryon density $\rho_{B}$ with 
different values for $\epsilon_{\mbox{vac}}$.}
\label{f9}
\end{figure}

%
%

\clearpage
\begin{figure}[htbp]
\begin{center}
\input{fa1.tex}
\end{center}
\end{figure}
\begin{figure}[htbp]
\begin{center}
\input{fa2.tex}
\end{center}
\end{figure}  
\clearpage
\begin{figure}[htbp]
\begin{center}
\input{fa3.tex}
\end{center}
\end{figure}   
\begin{figure}[htbp]
\begin{center}
\input{fa4.tex}
\end{center}
\end{figure}
\clearpage
\begin{figure}[htbp]
\begin{center}
\input{fa5a.tex}
\end{center}
\end{figure}   
\begin{figure}[htbp]
\begin{center}
\input{fa5b.tex}
\end{center}
\end{figure}
\clearpage
\begin{figure}[htbp]
\begin{center} 
\input{fa6a.tex}
\end{center}
\end{figure}  
\begin{figure}[htbp]
\begin{center}
\input{fa6b.tex}
\end{center}
\end{figure}
\clearpage
\begin{figure}[htbp]
\begin{center}
\input{fa7a.tex}
\end{center}   
\end{figure}
\begin{figure}[htbp]
\begin{center}
\input{fa7b.tex}
\end{center}   
\end{figure}   
\clearpage  
\begin{figure}[htbp]
\begin{center}  
\input{fa8a.tex}
\end{center}
\end{figure}
\begin{figure}[htbp]
\begin{center}  
\input{fa8b.tex}
\end{center}
\end{figure}
\clearpage
\begin{figure}[htbp]
\begin{center}
\input{fa9.tex}
\end{center}
\end{figure}
\clearpage
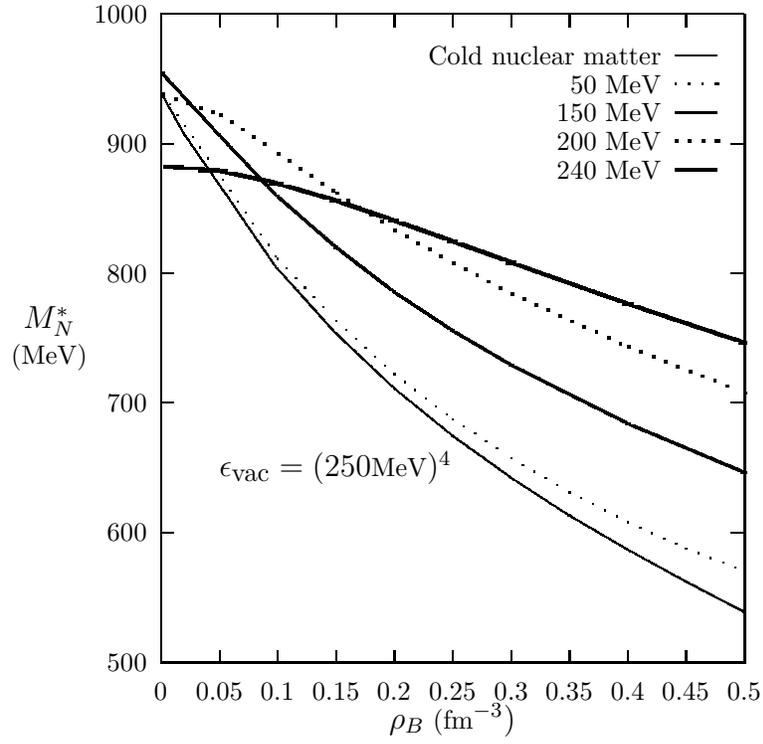
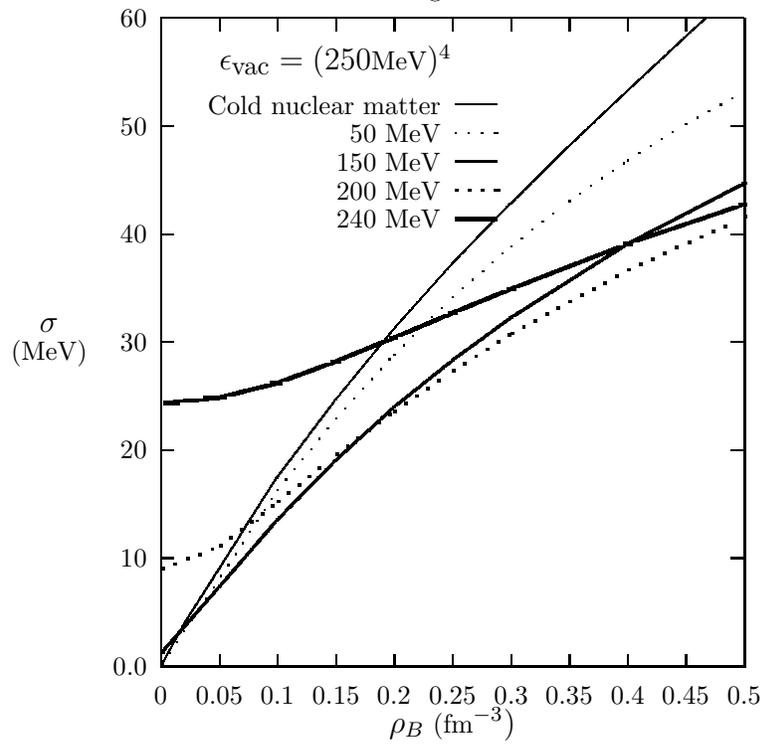
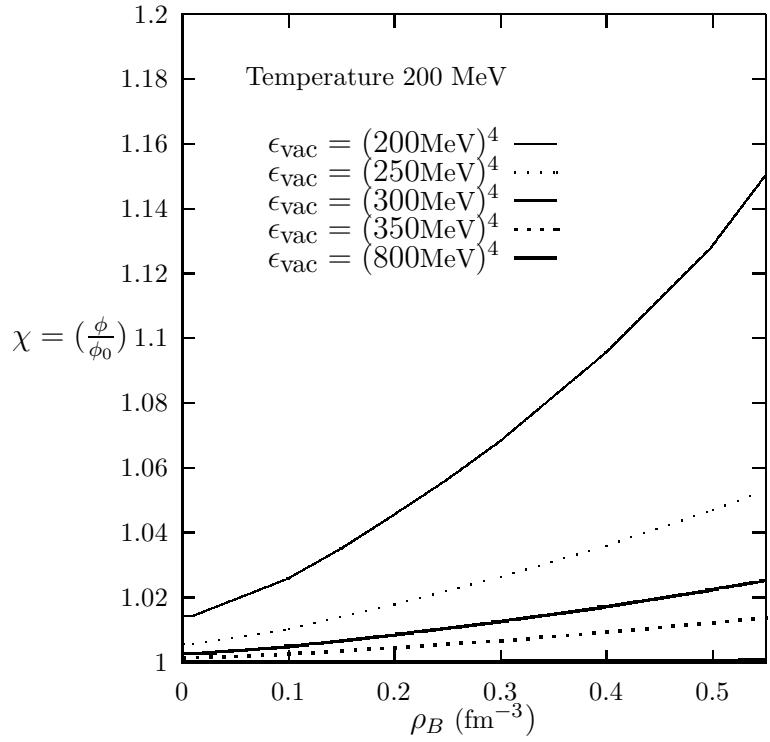
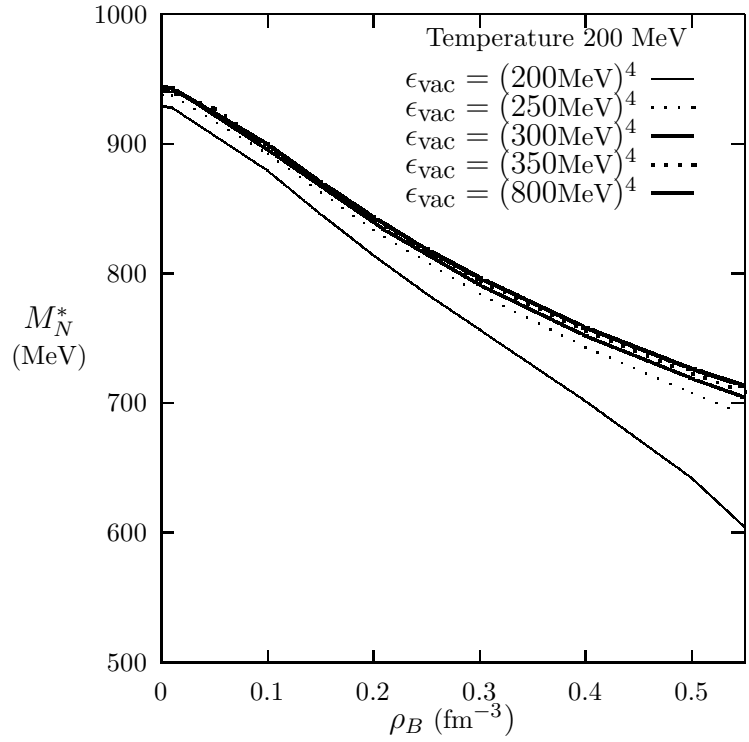
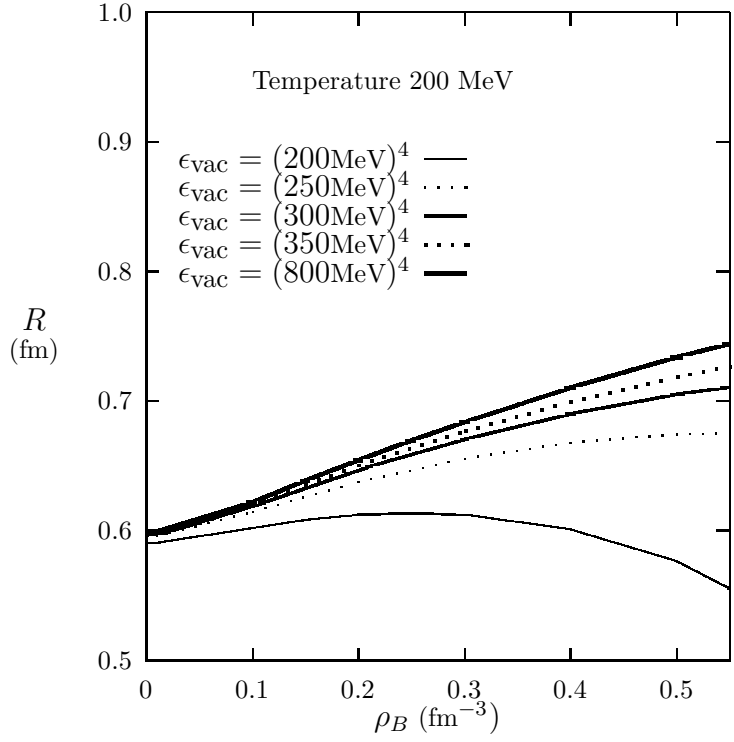
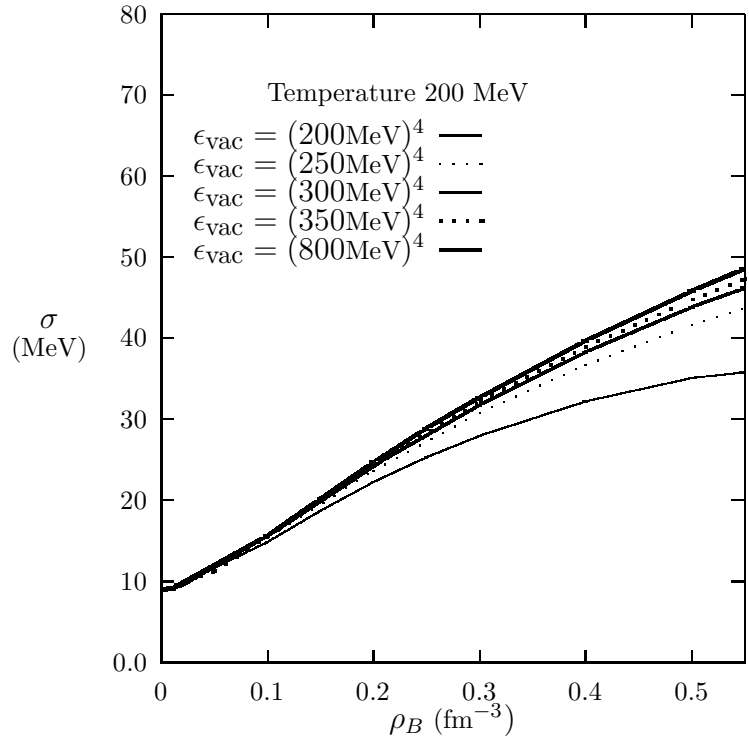
\end{document}

%% file: fa1.tex
\setlength{\unitlength}{0.240900pt}
\ifx\plotpoint\undefined\newsavebox{\plotpoint}\fi
\sbox{\plotpoint}{\rule[-0.200pt]{0.400pt}{0.400pt}}%
\begin{picture}(1200,1200)(0,0)
\font\gnuplot=cmr10 at 10pt
\gnuplot
\sbox{\plotpoint}{\rule[-0.200pt]{0.400pt}{0.400pt}}%
\put(220.0,113.0){\rule[-0.200pt]{0.400pt}{245.477pt}}
\put(220.0,113.0){\rule[-0.200pt]{4.818pt}{0.400pt}}
\put(198,113){\makebox(0,0)[r]{0.9}}
\put(1116.0,113.0){\rule[-0.200pt]{4.818pt}{0.400pt}}
\put(220.0,283.0){\rule[-0.200pt]{4.818pt}{0.400pt}}
\put(198,283){\makebox(0,0)[r]{0.95}}
\put(1116.0,283.0){\rule[-0.200pt]{4.818pt}{0.400pt}}
\put(220.0,453.0){\rule[-0.200pt]{4.818pt}{0.400pt}}
\put(198,453){\makebox(0,0)[r]{1}}
\put(1116.0,453.0){\rule[-0.200pt]{4.818pt}{0.400pt}}
\put(220.0,623.0){\rule[-0.200pt]{4.818pt}{0.400pt}}
\put(198,623){\makebox(0,0)[r]{1.05}}
\put(1116.0,623.0){\rule[-0.200pt]{4.818pt}{0.400pt}}
\put(220.0,792.0){\rule[-0.200pt]{4.818pt}{0.400pt}}
\put(198,792){\makebox(0,0)[r]{1.1}}
\put(1116.0,792.0){\rule[-0.200pt]{4.818pt}{0.400pt}}
\put(220.0,962.0){\rule[-0.200pt]{4.818pt}{0.400pt}}
\put(198,962){\makebox(0,0)[r]{1.15}}
\put(1116.0,962.0){\rule[-0.200pt]{4.818pt}{0.400pt}}
\put(220.0,1132.0){\rule[-0.200pt]{4.818pt}{0.400pt}}
\put(198,1132){\makebox(0,0)[r]{1.2}}
\put(1116.0,1132.0){\rule[-0.200pt]{4.818pt}{0.400pt}}
\put(220.0,113.0){\rule[-0.200pt]{0.400pt}{4.818pt}}
\put(220,68){\makebox(0,0){0}}
\put(220.0,1112.0){\rule[-0.200pt]{0.400pt}{4.818pt}}
\put(312.0,113.0){\rule[-0.200pt]{0.400pt}{4.818pt}}
\put(312,68){\makebox(0,0){0.05}}
\put(312.0,1112.0){\rule[-0.200pt]{0.400pt}{4.818pt}}
\put(403.0,113.0){\rule[-0.200pt]{0.400pt}{4.818pt}}
\put(403,68){\makebox(0,0){0.1}}
\put(403.0,1112.0){\rule[-0.200pt]{0.400pt}{4.818pt}}
\put(495.0,113.0){\rule[-0.200pt]{0.400pt}{4.818pt}}
\put(495,68){\makebox(0,0){0.15}}
\put(495.0,1112.0){\rule[-0.200pt]{0.400pt}{4.818pt}}
\put(586.0,113.0){\rule[-0.200pt]{0.400pt}{4.818pt}}
\put(586,68){\makebox(0,0){0.2}}
\put(586.0,1112.0){\rule[-0.200pt]{0.400pt}{4.818pt}}
\put(678.0,113.0){\rule[-0.200pt]{0.400pt}{4.818pt}}
\put(678,68){\makebox(0,0){0.25}}
\put(678.0,1112.0){\rule[-0.200pt]{0.400pt}{4.818pt}}
\put(770.0,113.0){\rule[-0.200pt]{0.400pt}{4.818pt}}
\put(770,68){\makebox(0,0){0.3}}
\put(770.0,1112.0){\rule[-0.200pt]{0.400pt}{4.818pt}}
\put(861.0,113.0){\rule[-0.200pt]{0.400pt}{4.818pt}}
\put(861,68){\makebox(0,0){0.35}}
\put(861.0,1112.0){\rule[-0.200pt]{0.400pt}{4.818pt}}
\put(953.0,113.0){\rule[-0.200pt]{0.400pt}{4.818pt}}
\put(953,68){\makebox(0,0){0.4}}
\put(953.0,1112.0){\rule[-0.200pt]{0.400pt}{4.818pt}}
\put(1044.0,113.0){\rule[-0.200pt]{0.400pt}{4.818pt}}
\put(1044,68){\makebox(0,0){0.45}}
\put(1044.0,1112.0){\rule[-0.200pt]{0.400pt}{4.818pt}}
\put(1136.0,113.0){\rule[-0.200pt]{0.400pt}{4.818pt}}
\put(1136,68){\makebox(0,0){0.5}}
\put(1136.0,1112.0){\rule[-0.200pt]{0.400pt}{4.818pt}}
\put(220.0,113.0){\rule[-0.200pt]{220.664pt}{0.400pt}}
\put(1136.0,113.0){\rule[-0.200pt]{0.400pt}{245.477pt}}
\put(220.0,1132.0){\rule[-0.200pt]{220.664pt}{0.400pt}}
\put(45,622){\makebox(0,0){$\chi=\left(\frac{\phi}{\phi_{0}}\right)$}}
\put(678,23){\makebox(0,0){$\rho_{B}$ (fm$^{-3})$}}
\put(678,1177){\makebox(0,0){Figure 1}}
\put(312,724){\makebox(0,0)[l]{$\epsilon_{\mbox{vac}}=(250 \mbox{MeV})^{4}$}}
\put(220.0,113.0){\rule[-0.200pt]{0.400pt}{245.477pt}}
\put(1006,1067){\makebox(0,0)[r]{Cold nuclear matter}}
\put(1028.0,1067.0){\rule[-0.200pt]{15.899pt}{0.400pt}}
\put(222,453){\usebox{\plotpoint}}
\multiput(222.00,453.59)(3.827,0.477){7}{\rule{2.900pt}{0.115pt}}
\multiput(222.00,452.17)(28.981,5.000){2}{\rule{1.450pt}{0.400pt}}
\multiput(257.00,458.58)(7.584,0.491){17}{\rule{5.940pt}{0.118pt}}
\multiput(257.00,457.17)(133.671,10.000){2}{\rule{2.970pt}{0.400pt}}
\put(403,466.67){\rule{22.163pt}{0.400pt}}
\multiput(403.00,467.17)(46.000,-1.000){2}{\rule{11.081pt}{0.400pt}}
\multiput(495.00,465.93)(6.900,-0.485){11}{\rule{5.300pt}{0.117pt}}
\multiput(495.00,466.17)(80.000,-7.000){2}{\rule{2.650pt}{0.400pt}}
\multiput(586.00,458.92)(4.317,-0.492){19}{\rule{3.445pt}{0.118pt}}
\multiput(586.00,459.17)(84.849,-11.000){2}{\rule{1.723pt}{0.400pt}}
\multiput(678.00,447.92)(2.930,-0.494){29}{\rule{2.400pt}{0.119pt}}
\multiput(678.00,448.17)(87.019,-16.000){2}{\rule{1.200pt}{0.400pt}}
\multiput(770.00,431.92)(2.194,-0.496){39}{\rule{1.833pt}{0.119pt}}
\multiput(770.00,432.17)(87.195,-21.000){2}{\rule{0.917pt}{0.400pt}}
\multiput(861.00,410.92)(1.656,-0.497){53}{\rule{1.414pt}{0.120pt}}
\multiput(861.00,411.17)(89.065,-28.000){2}{\rule{0.707pt}{0.400pt}}
\multiput(953.00,382.92)(1.270,-0.498){69}{\rule{1.111pt}{0.120pt}}
\multiput(953.00,383.17)(88.694,-36.000){2}{\rule{0.556pt}{0.400pt}}
\multiput(1044.00,346.92)(1.003,-0.498){89}{\rule{0.900pt}{0.120pt}}
\multiput(1044.00,347.17)(90.132,-46.000){2}{\rule{0.450pt}{0.400pt}}
\put(1136,302){\usebox{\plotpoint}}
\put(1006,1022){\makebox(0,0)[r]{ 50 MeV}}
\multiput(1028,1022)(20.756,0.000){4}{\usebox{\plotpoint}}
\put(1094,1022){\usebox{\plotpoint}}
\put(222,453){\usebox{\plotpoint}}
\multiput(222,453)(20.617,2.392){9}{\usebox{\plotpoint}}
\multiput(403,474)(20.677,1.798){5}{\usebox{\plotpoint}}
\multiput(495,482)(20.694,1.592){4}{\usebox{\plotpoint}}
\multiput(586,489)(20.696,1.575){5}{\usebox{\plotpoint}}
\multiput(678,496)(20.677,1.798){4}{\usebox{\plotpoint}}
\multiput(770,504)(20.655,2.043){4}{\usebox{\plotpoint}}
\multiput(861,513)(20.634,2.243){5}{\usebox{\plotpoint}}
\multiput(953,523)(20.606,2.491){4}{\usebox{\plotpoint}}
\multiput(1044,534)(20.551,2.904){5}{\usebox{\plotpoint}}
\put(1136,547){\usebox{\plotpoint}}
\sbox{\plotpoint}{\rule[-0.400pt]{0.800pt}{0.800pt}}%
\put(1006,977){\makebox(0,0)[r]{150 MeV}}
\put(1028.0,977.0){\rule[-0.400pt]{15.899pt}{0.800pt}}
\put(222,455){\usebox{\plotpoint}}
\multiput(222.00,456.40)(4.428,0.512){15}{\rule{6.745pt}{0.123pt}}
\multiput(222.00,453.34)(75.999,11.000){2}{\rule{3.373pt}{0.800pt}}
\multiput(312.00,467.41)(3.718,0.509){19}{\rule{5.800pt}{0.123pt}}
\multiput(312.00,464.34)(78.962,13.000){2}{\rule{2.900pt}{0.800pt}}
\multiput(403.00,480.40)(5.051,0.514){13}{\rule{7.560pt}{0.124pt}}
\multiput(403.00,477.34)(76.309,10.000){2}{\rule{3.780pt}{0.800pt}}
\multiput(495.00,490.41)(4.062,0.511){17}{\rule{6.267pt}{0.123pt}}
\multiput(495.00,487.34)(77.993,12.000){2}{\rule{3.133pt}{0.800pt}}
\multiput(586.00,502.41)(4.107,0.511){17}{\rule{6.333pt}{0.123pt}}
\multiput(586.00,499.34)(78.855,12.000){2}{\rule{3.167pt}{0.800pt}}
\multiput(678.00,514.41)(3.759,0.509){19}{\rule{5.862pt}{0.123pt}}
\multiput(678.00,511.34)(79.834,13.000){2}{\rule{2.931pt}{0.800pt}}
\multiput(770.00,527.41)(3.343,0.504){49}{\rule{5.429pt}{0.121pt}}
\multiput(770.00,524.34)(171.733,28.000){2}{\rule{2.714pt}{0.800pt}}
\multiput(953.00,555.41)(2.914,0.503){57}{\rule{4.775pt}{0.121pt}}
\multiput(953.00,552.34)(173.089,32.000){2}{\rule{2.388pt}{0.800pt}}
\put(1136,586){\usebox{\plotpoint}}
\sbox{\plotpoint}{\rule[-0.500pt]{1.000pt}{1.000pt}}%
\put(1006,932){\makebox(0,0)[r]{200 MeV}}
\multiput(1028,932)(20.756,0.000){4}{\usebox{\plotpoint}}
\put(1094,932){\usebox{\plotpoint}}
\put(222,472){\usebox{\plotpoint}}
\multiput(222,472)(20.724,1.151){5}{\usebox{\plotpoint}}
\multiput(312,477)(20.631,2.267){4}{\usebox{\plotpoint}}
\multiput(403,487)(20.551,2.904){5}{\usebox{\plotpoint}}
\multiput(495,500)(20.547,2.935){4}{\usebox{\plotpoint}}
\multiput(586,513)(20.519,3.122){5}{\usebox{\plotpoint}}
\multiput(678,527)(20.485,3.340){4}{\usebox{\plotpoint}}
\multiput(770,542)(20.426,3.683){9}{\usebox{\plotpoint}}
\multiput(953,575)(20.344,4.113){9}{\usebox{\plotpoint}}
\put(1136,612){\usebox{\plotpoint}}
\sbox{\plotpoint}{\rule[-0.600pt]{1.200pt}{1.200pt}}%
\put(1006,887){\makebox(0,0)[r]{240 MeV}}
\put(1028.0,887.0){\rule[-0.600pt]{15.899pt}{1.200pt}}
\put(224,512){\usebox{\plotpoint}}
\put(224,510.51){\rule{21.199pt}{1.200pt}}
\multiput(224.00,509.51)(44.000,2.000){2}{\rule{10.600pt}{1.200pt}}
\multiput(312.00,516.24)(14.205,0.509){2}{\rule{18.500pt}{0.123pt}}
\multiput(312.00,511.51)(52.602,6.000){2}{\rule{9.250pt}{1.200pt}}
\multiput(403.00,522.24)(5.691,0.502){8}{\rule{12.567pt}{0.121pt}}
\multiput(403.00,517.51)(65.917,9.000){2}{\rule{6.283pt}{1.200pt}}
\multiput(495.00,531.24)(3.998,0.501){14}{\rule{9.400pt}{0.121pt}}
\multiput(495.00,526.51)(71.490,12.000){2}{\rule{4.700pt}{1.200pt}}
\multiput(586.00,543.24)(3.700,0.501){16}{\rule{8.792pt}{0.121pt}}
\multiput(586.00,538.51)(73.751,13.000){2}{\rule{4.396pt}{1.200pt}}
\multiput(678.00,556.24)(2.958,0.501){22}{\rule{7.200pt}{0.121pt}}
\multiput(678.00,551.51)(77.056,16.000){2}{\rule{3.600pt}{1.200pt}}
\multiput(770.00,572.24)(2.718,0.500){58}{\rule{6.759pt}{0.121pt}}
\multiput(770.00,567.51)(168.972,34.000){2}{\rule{3.379pt}{1.200pt}}
\multiput(953.00,606.24)(2.247,0.500){72}{\rule{5.656pt}{0.121pt}}
\multiput(953.00,601.51)(171.260,41.000){2}{\rule{2.828pt}{1.200pt}}
\put(1136,645){\usebox{\plotpoint}}
\end{picture}

%% file: fa2.tex
\setlength{\unitlength}{0.240900pt}
\ifx\plotpoint\undefined\newsavebox{\plotpoint}\fi
\sbox{\plotpoint}{\rule[-0.200pt]{0.400pt}{0.400pt}}%
\begin{picture}(1200,1200)(0,0)
\font\gnuplot=cmr10 at 10pt
\gnuplot
\sbox{\plotpoint}{\rule[-0.200pt]{0.400pt}{0.400pt}}%
\put(220.0,113.0){\rule[-0.200pt]{0.400pt}{245.477pt}}
\put(220.0,113.0){\rule[-0.200pt]{4.818pt}{0.400pt}}
\put(198,113){\makebox(0,0)[r]{500}}
\put(1116.0,113.0){\rule[-0.200pt]{4.818pt}{0.400pt}}
\put(220.0,317.0){\rule[-0.200pt]{4.818pt}{0.400pt}}
\put(198,317){\makebox(0,0)[r]{600}}
\put(1116.0,317.0){\rule[-0.200pt]{4.818pt}{0.400pt}}
\put(220.0,521.0){\rule[-0.200pt]{4.818pt}{0.400pt}}
\put(198,521){\makebox(0,0)[r]{700}}
\put(1116.0,521.0){\rule[-0.200pt]{4.818pt}{0.400pt}}
\put(220.0,724.0){\rule[-0.200pt]{4.818pt}{0.400pt}}
\put(198,724){\makebox(0,0)[r]{800}}
\put(1116.0,724.0){\rule[-0.200pt]{4.818pt}{0.400pt}}
\put(220.0,928.0){\rule[-0.200pt]{4.818pt}{0.400pt}}
\put(198,928){\makebox(0,0)[r]{900}}
\put(1116.0,928.0){\rule[-0.200pt]{4.818pt}{0.400pt}}
\put(220.0,1132.0){\rule[-0.200pt]{4.818pt}{0.400pt}}
\put(198,1132){\makebox(0,0)[r]{1000}}
\put(1116.0,1132.0){\rule[-0.200pt]{4.818pt}{0.400pt}}
\put(220.0,113.0){\rule[-0.200pt]{0.400pt}{4.818pt}}
\put(220,68){\makebox(0,0){0}}
\put(220.0,1112.0){\rule[-0.200pt]{0.400pt}{4.818pt}}
\put(312.0,113.0){\rule[-0.200pt]{0.400pt}{4.818pt}}
\put(312,68){\makebox(0,0){0.05}}
\put(312.0,1112.0){\rule[-0.200pt]{0.400pt}{4.818pt}}
\put(403.0,113.0){\rule[-0.200pt]{0.400pt}{4.818pt}}
\put(403,68){\makebox(0,0){0.1}}
\put(403.0,1112.0){\rule[-0.200pt]{0.400pt}{4.818pt}}
\put(495.0,113.0){\rule[-0.200pt]{0.400pt}{4.818pt}}
\put(495,68){\makebox(0,0){0.15}}
\put(495.0,1112.0){\rule[-0.200pt]{0.400pt}{4.818pt}}
\put(586.0,113.0){\rule[-0.200pt]{0.400pt}{4.818pt}}
\put(586,68){\makebox(0,0){0.2}}
\put(586.0,1112.0){\rule[-0.200pt]{0.400pt}{4.818pt}}
\put(678.0,113.0){\rule[-0.200pt]{0.400pt}{4.818pt}}
\put(678,68){\makebox(0,0){0.25}}
\put(678.0,1112.0){\rule[-0.200pt]{0.400pt}{4.818pt}}
\put(770.0,113.0){\rule[-0.200pt]{0.400pt}{4.818pt}}
\put(770,68){\makebox(0,0){0.3}}
\put(770.0,1112.0){\rule[-0.200pt]{0.400pt}{4.818pt}}
\put(861.0,113.0){\rule[-0.200pt]{0.400pt}{4.818pt}}
\put(861,68){\makebox(0,0){0.35}}
\put(861.0,1112.0){\rule[-0.200pt]{0.400pt}{4.818pt}}
\put(953.0,113.0){\rule[-0.200pt]{0.400pt}{4.818pt}}
\put(953,68){\makebox(0,0){0.4}}
\put(953.0,1112.0){\rule[-0.200pt]{0.400pt}{4.818pt}}
\put(1044.0,113.0){\rule[-0.200pt]{0.400pt}{4.818pt}}
\put(1044,68){\makebox(0,0){0.45}}
\put(1044.0,1112.0){\rule[-0.200pt]{0.400pt}{4.818pt}}
\put(1136.0,113.0){\rule[-0.200pt]{0.400pt}{4.818pt}}
\put(1136,68){\makebox(0,0){0.5}}
\put(1136.0,1112.0){\rule[-0.200pt]{0.400pt}{4.818pt}}
\put(220.0,113.0){\rule[-0.200pt]{220.664pt}{0.400pt}}
\put(1136.0,113.0){\rule[-0.200pt]{0.400pt}{245.477pt}}
\put(220.0,1132.0){\rule[-0.200pt]{220.664pt}{0.400pt}}
\put(45,622){\makebox(0,0){\shortstack{$M^{*}_{N}$\\(MeV)}}}
\put(678,23){\makebox(0,0){$\rho_{B}$ (fm$^{-3})$}}
\put(678,1177){\makebox(0,0){Figure 2}}
\put(312,419){\makebox(0,0)[l]{$\epsilon_{\mbox{vac}}=(250 \mbox{MeV})^{4}$}}
\put(220.0,113.0){\rule[-0.200pt]{0.400pt}{245.477pt}}
\put(1006,1067){\makebox(0,0)[r]{Cold nuclear matter}}
\put(1028.0,1067.0){\rule[-0.200pt]{15.899pt}{0.400pt}}
\put(222,1006){\usebox{\plotpoint}}
\multiput(222.58,1002.60)(0.498,-0.903){67}{\rule{0.120pt}{0.820pt}}
\multiput(221.17,1004.30)(35.000,-61.298){2}{\rule{0.400pt}{0.410pt}}
\multiput(257.58,940.19)(0.499,-0.723){289}{\rule{0.120pt}{0.678pt}}
\multiput(256.17,941.59)(146.000,-209.593){2}{\rule{0.400pt}{0.339pt}}
\multiput(403.58,729.74)(0.499,-0.554){181}{\rule{0.120pt}{0.543pt}}
\multiput(402.17,730.87)(92.000,-100.872){2}{\rule{0.400pt}{0.272pt}}
\multiput(495.00,628.92)(0.529,-0.499){169}{\rule{0.523pt}{0.120pt}}
\multiput(495.00,629.17)(89.914,-86.000){2}{\rule{0.262pt}{0.400pt}}
\multiput(586.00,542.92)(0.613,-0.499){147}{\rule{0.591pt}{0.120pt}}
\multiput(586.00,543.17)(90.774,-75.000){2}{\rule{0.295pt}{0.400pt}}
\multiput(678.00,467.92)(0.697,-0.499){129}{\rule{0.658pt}{0.120pt}}
\multiput(678.00,468.17)(90.635,-66.000){2}{\rule{0.329pt}{0.400pt}}
\multiput(770.00,401.92)(0.772,-0.499){115}{\rule{0.717pt}{0.120pt}}
\multiput(770.00,402.17)(89.512,-59.000){2}{\rule{0.358pt}{0.400pt}}
\multiput(861.00,342.92)(0.853,-0.498){105}{\rule{0.781pt}{0.120pt}}
\multiput(861.00,343.17)(90.378,-54.000){2}{\rule{0.391pt}{0.400pt}}
\multiput(953.00,288.92)(0.912,-0.498){97}{\rule{0.828pt}{0.120pt}}
\multiput(953.00,289.17)(89.281,-50.000){2}{\rule{0.414pt}{0.400pt}}
\multiput(1044.00,238.92)(0.961,-0.498){93}{\rule{0.867pt}{0.120pt}}
\multiput(1044.00,239.17)(90.201,-48.000){2}{\rule{0.433pt}{0.400pt}}
\put(1136,192){\usebox{\plotpoint}}
\put(1006,1022){\makebox(0,0)[r]{ 50 MeV}}
\multiput(1028,1022)(20.756,0.000){4}{\usebox{\plotpoint}}
\put(1094,1022){\usebox{\plotpoint}}
\put(222,1007){\usebox{\plotpoint}}
\multiput(222,1007)(11.858,-17.034){16}{\usebox{\plotpoint}}
\multiput(403,747)(14.206,-15.132){6}{\usebox{\plotpoint}}
\multiput(495,649)(15.335,-13.987){6}{\usebox{\plotpoint}}
\multiput(586,566)(16.431,-12.681){6}{\usebox{\plotpoint}}
\multiput(678,495)(17.298,-11.470){5}{\usebox{\plotpoint}}
\multiput(770,434)(17.849,-10.592){5}{\usebox{\plotpoint}}
\multiput(861,380)(18.483,-9.443){5}{\usebox{\plotpoint}}
\multiput(953,333)(18.923,-8.526){5}{\usebox{\plotpoint}}
\multiput(1044,292)(19.328,-7.563){5}{\usebox{\plotpoint}}
\put(1136,256){\usebox{\plotpoint}}
\sbox{\plotpoint}{\rule[-0.400pt]{0.800pt}{0.800pt}}%
\put(1006,977){\makebox(0,0)[r]{150 MeV}}
\put(1028.0,977.0){\rule[-0.400pt]{15.899pt}{0.800pt}}
\put(222,1039){\usebox{\plotpoint}}
\multiput(223.41,1034.55)(0.501,-0.544){173}{\rule{0.121pt}{1.071pt}}
\multiput(220.34,1036.78)(90.000,-95.777){2}{\rule{0.800pt}{0.536pt}}
\multiput(313.41,936.67)(0.501,-0.527){175}{\rule{0.121pt}{1.044pt}}
\multiput(310.34,938.83)(91.000,-93.833){2}{\rule{0.800pt}{0.522pt}}
\multiput(403.00,843.09)(0.575,-0.501){153}{\rule{1.120pt}{0.121pt}}
\multiput(403.00,843.34)(89.675,-80.000){2}{\rule{0.560pt}{0.800pt}}
\multiput(495.00,763.09)(0.650,-0.501){133}{\rule{1.240pt}{0.121pt}}
\multiput(495.00,763.34)(88.426,-70.000){2}{\rule{0.620pt}{0.800pt}}
\multiput(586.00,693.09)(0.756,-0.502){115}{\rule{1.407pt}{0.121pt}}
\multiput(586.00,693.34)(89.081,-61.000){2}{\rule{0.703pt}{0.800pt}}
\multiput(678.00,632.09)(0.855,-0.502){101}{\rule{1.563pt}{0.121pt}}
\multiput(678.00,632.34)(88.756,-54.000){2}{\rule{0.781pt}{0.800pt}}
\multiput(770.00,578.09)(0.998,-0.501){177}{\rule{1.791pt}{0.121pt}}
\multiput(770.00,578.34)(179.282,-92.000){2}{\rule{0.896pt}{0.800pt}}
\multiput(953.00,486.09)(1.194,-0.501){147}{\rule{2.101pt}{0.121pt}}
\multiput(953.00,486.34)(178.639,-77.000){2}{\rule{1.051pt}{0.800pt}}
\put(1136,411){\usebox{\plotpoint}}
\sbox{\plotpoint}{\rule[-0.500pt]{1.000pt}{1.000pt}}%
\put(1006,932){\makebox(0,0)[r]{200 MeV}}
\multiput(1028,932)(20.756,0.000){4}{\usebox{\plotpoint}}
\put(1094,932){\usebox{\plotpoint}}
\put(222,1005){\usebox{\plotpoint}}
\multiput(222,1005)(19.556,-6.953){5}{\usebox{\plotpoint}}
\multiput(312,973)(17.240,-11.557){5}{\usebox{\plotpoint}}
\multiput(403,912)(17.212,-11.599){6}{\usebox{\plotpoint}}
\multiput(495,850)(17.503,-11.156){5}{\usebox{\plotpoint}}
\multiput(586,792)(18.069,-10.213){5}{\usebox{\plotpoint}}
\multiput(678,740)(18.402,-9.601){5}{\usebox{\plotpoint}}
\multiput(770,692)(18.863,-8.659){10}{\usebox{\plotpoint}}
\multiput(953,608)(19.314,-7.599){9}{\usebox{\plotpoint}}
\put(1136,536){\usebox{\plotpoint}}
\sbox{\plotpoint}{\rule[-0.600pt]{1.200pt}{1.200pt}}%
\put(1006,887){\makebox(0,0)[r]{240 MeV}}
\put(1028.0,887.0){\rule[-0.600pt]{15.899pt}{1.200pt}}
\put(224,892){\usebox{\plotpoint}}
\multiput(224.00,889.26)(8.044,-0.505){4}{\rule{15.386pt}{0.122pt}}
\multiput(224.00,889.51)(56.066,-7.000){2}{\rule{7.693pt}{1.200pt}}
\multiput(312.00,882.26)(2.313,-0.501){30}{\rule{5.760pt}{0.121pt}}
\multiput(312.00,882.51)(79.045,-20.000){2}{\rule{2.880pt}{1.200pt}}
\multiput(403.00,862.26)(1.716,-0.500){44}{\rule{4.389pt}{0.121pt}}
\multiput(403.00,862.51)(82.891,-27.000){2}{\rule{2.194pt}{1.200pt}}
\multiput(495.00,835.26)(1.473,-0.500){52}{\rule{3.823pt}{0.121pt}}
\multiput(495.00,835.51)(83.066,-31.000){2}{\rule{1.911pt}{1.200pt}}
\multiput(586.00,804.26)(1.398,-0.500){56}{\rule{3.645pt}{0.121pt}}
\multiput(586.00,804.51)(84.434,-33.000){2}{\rule{1.823pt}{1.200pt}}
\multiput(678.00,771.26)(1.398,-0.500){56}{\rule{3.645pt}{0.121pt}}
\multiput(678.00,771.51)(84.434,-33.000){2}{\rule{1.823pt}{1.200pt}}
\multiput(770.00,738.26)(1.409,-0.500){120}{\rule{3.678pt}{0.120pt}}
\multiput(770.00,738.51)(175.365,-65.000){2}{\rule{1.839pt}{1.200pt}}
\multiput(953.00,673.26)(1.503,-0.500){112}{\rule{3.900pt}{0.120pt}}
\multiput(953.00,673.51)(174.905,-61.000){2}{\rule{1.950pt}{1.200pt}}
\put(1136,615){\usebox{\plotpoint}}
\end{picture}

%% file: fa3.tex
\setlength{\unitlength}{0.240900pt}
\ifx\plotpoint\undefined\newsavebox{\plotpoint}\fi
\sbox{\plotpoint}{\rule[-0.200pt]{0.400pt}{0.400pt}}%
\begin{picture}(1200,1200)(0,0)
\font\gnuplot=cmr10 at 10pt
\gnuplot
\sbox{\plotpoint}{\rule[-0.200pt]{0.400pt}{0.400pt}}%
\put(220.0,113.0){\rule[-0.200pt]{0.400pt}{245.477pt}}
\put(220.0,113.0){\rule[-0.200pt]{4.818pt}{0.400pt}}
\put(198,113){\makebox(0,0)[r]{0.5}}
\put(1116.0,113.0){\rule[-0.200pt]{4.818pt}{0.400pt}}
\put(220.0,317.0){\rule[-0.200pt]{4.818pt}{0.400pt}}
\put(198,317){\makebox(0,0)[r]{0.6}}
\put(1116.0,317.0){\rule[-0.200pt]{4.818pt}{0.400pt}}
\put(220.0,521.0){\rule[-0.200pt]{4.818pt}{0.400pt}}
\put(198,521){\makebox(0,0)[r]{0.7}}
\put(1116.0,521.0){\rule[-0.200pt]{4.818pt}{0.400pt}}
\put(220.0,724.0){\rule[-0.200pt]{4.818pt}{0.400pt}}
\put(198,724){\makebox(0,0)[r]{0.8}}
\put(1116.0,724.0){\rule[-0.200pt]{4.818pt}{0.400pt}}
\put(220.0,928.0){\rule[-0.200pt]{4.818pt}{0.400pt}}
\put(198,928){\makebox(0,0)[r]{0.9}}
\put(1116.0,928.0){\rule[-0.200pt]{4.818pt}{0.400pt}}
\put(220.0,1132.0){\rule[-0.200pt]{4.818pt}{0.400pt}}
\put(198,1132){\makebox(0,0)[r]{1.0}}
\put(1116.0,1132.0){\rule[-0.200pt]{4.818pt}{0.400pt}}
\put(220.0,113.0){\rule[-0.200pt]{0.400pt}{4.818pt}}
\put(220,68){\makebox(0,0){0}}
\put(220.0,1112.0){\rule[-0.200pt]{0.400pt}{4.818pt}}
\put(312.0,113.0){\rule[-0.200pt]{0.400pt}{4.818pt}}
\put(312,68){\makebox(0,0){0.05}}
\put(312.0,1112.0){\rule[-0.200pt]{0.400pt}{4.818pt}}
\put(403.0,113.0){\rule[-0.200pt]{0.400pt}{4.818pt}}
\put(403,68){\makebox(0,0){0.1}}
\put(403.0,1112.0){\rule[-0.200pt]{0.400pt}{4.818pt}}
\put(495.0,113.0){\rule[-0.200pt]{0.400pt}{4.818pt}}
\put(495,68){\makebox(0,0){0.15}}
\put(495.0,1112.0){\rule[-0.200pt]{0.400pt}{4.818pt}}
\put(586.0,113.0){\rule[-0.200pt]{0.400pt}{4.818pt}}
\put(586,68){\makebox(0,0){0.2}}
\put(586.0,1112.0){\rule[-0.200pt]{0.400pt}{4.818pt}}
\put(678.0,113.0){\rule[-0.200pt]{0.400pt}{4.818pt}}
\put(678,68){\makebox(0,0){0.25}}
\put(678.0,1112.0){\rule[-0.200pt]{0.400pt}{4.818pt}}
\put(770.0,113.0){\rule[-0.200pt]{0.400pt}{4.818pt}}
\put(770,68){\makebox(0,0){0.3}}
\put(770.0,1112.0){\rule[-0.200pt]{0.400pt}{4.818pt}}
\put(861.0,113.0){\rule[-0.200pt]{0.400pt}{4.818pt}}
\put(861,68){\makebox(0,0){0.35}}
\put(861.0,1112.0){\rule[-0.200pt]{0.400pt}{4.818pt}}
\put(953.0,113.0){\rule[-0.200pt]{0.400pt}{4.818pt}}
\put(953,68){\makebox(0,0){0.4}}
\put(953.0,1112.0){\rule[-0.200pt]{0.400pt}{4.818pt}}
\put(1044.0,113.0){\rule[-0.200pt]{0.400pt}{4.818pt}}
\put(1044,68){\makebox(0,0){0.45}}
\put(1044.0,1112.0){\rule[-0.200pt]{0.400pt}{4.818pt}}
\put(1136.0,113.0){\rule[-0.200pt]{0.400pt}{4.818pt}}
\put(1136,68){\makebox(0,0){0.5}}
\put(1136.0,1112.0){\rule[-0.200pt]{0.400pt}{4.818pt}}
\put(220.0,113.0){\rule[-0.200pt]{220.664pt}{0.400pt}}
\put(1136.0,113.0){\rule[-0.200pt]{0.400pt}{245.477pt}}
\put(220.0,1132.0){\rule[-0.200pt]{220.664pt}{0.400pt}}
\put(45,622){\makebox(0,0){\shortstack{$R$\\(fm)}}}
\put(678,23){\makebox(0,0){$\rho_{B}$ (fm$^{-3})$}}
\put(678,1177){\makebox(0,0){Figure 3}}
\put(238,1021){\makebox(0,0)[l]{$\epsilon_{\mbox{vac}}=(250 \mbox{MeV})^{4}$}}
\put(220.0,113.0){\rule[-0.200pt]{0.400pt}{245.477pt}}
\put(660,941){\makebox(0,0)[r]{Cold nuclear matter}}
\put(682.0,941.0){\rule[-0.200pt]{15.899pt}{0.400pt}}
\put(222,352){\usebox{\plotpoint}}
\multiput(222.00,352.58)(0.564,0.497){59}{\rule{0.552pt}{0.120pt}}
\multiput(222.00,351.17)(33.855,31.000){2}{\rule{0.276pt}{0.400pt}}
\multiput(257.00,383.58)(0.603,0.499){239}{\rule{0.583pt}{0.120pt}}
\multiput(257.00,382.17)(144.791,121.000){2}{\rule{0.291pt}{0.400pt}}
\multiput(403.00,504.58)(0.648,0.499){139}{\rule{0.618pt}{0.120pt}}
\multiput(403.00,503.17)(90.717,71.000){2}{\rule{0.309pt}{0.400pt}}
\multiput(495.00,575.58)(0.641,0.499){139}{\rule{0.613pt}{0.120pt}}
\multiput(495.00,574.17)(89.728,71.000){2}{\rule{0.306pt}{0.400pt}}
\multiput(586.00,646.58)(0.648,0.499){139}{\rule{0.618pt}{0.120pt}}
\multiput(586.00,645.17)(90.717,71.000){2}{\rule{0.309pt}{0.400pt}}
\multiput(678.00,717.58)(0.622,0.499){145}{\rule{0.597pt}{0.120pt}}
\multiput(678.00,716.17)(90.760,74.000){2}{\rule{0.299pt}{0.400pt}}
\multiput(770.00,791.58)(0.591,0.499){151}{\rule{0.573pt}{0.120pt}}
\multiput(770.00,790.17)(89.811,77.000){2}{\rule{0.286pt}{0.400pt}}
\multiput(861.00,868.58)(0.547,0.499){165}{\rule{0.538pt}{0.120pt}}
\multiput(861.00,867.17)(90.883,84.000){2}{\rule{0.269pt}{0.400pt}}
\multiput(953.58,952.00)(0.499,0.516){179}{\rule{0.120pt}{0.513pt}}
\multiput(952.17,952.00)(91.000,92.935){2}{\rule{0.400pt}{0.257pt}}
\multiput(1044.58,1046.00)(0.499,0.589){143}{\rule{0.120pt}{0.571pt}}
\multiput(1043.17,1046.00)(73.000,84.814){2}{\rule{0.400pt}{0.286pt}}
\put(660,896){\makebox(0,0)[r]{ 50 MeV}}
\multiput(682,896)(20.756,0.000){4}{\usebox{\plotpoint}}
\put(748,896){\usebox{\plotpoint}}
\put(222,352){\usebox{\plotpoint}}
\multiput(222,352)(16.549,12.526){11}{\usebox{\plotpoint}}
\multiput(403,489)(17.298,11.470){6}{\usebox{\plotpoint}}
\multiput(495,550)(17.763,10.736){5}{\usebox{\plotpoint}}
\multiput(586,605)(18.236,9.911){5}{\usebox{\plotpoint}}
\multiput(678,655)(18.724,8.955){5}{\usebox{\plotpoint}}
\multiput(770,699)(19.077,8.176){5}{\usebox{\plotpoint}}
\multiput(861,738)(19.469,7.195){4}{\usebox{\plotpoint}}
\multiput(953,772)(19.776,6.302){5}{\usebox{\plotpoint}}
\multiput(1044,801)(20.136,5.034){5}{\usebox{\plotpoint}}
\put(1136,824){\usebox{\plotpoint}}
\sbox{\plotpoint}{\rule[-0.400pt]{0.800pt}{0.800pt}}%
\put(660,851){\makebox(0,0)[r]{150 MeV}}
\put(682.0,851.0){\rule[-0.400pt]{15.899pt}{0.800pt}}
\put(222,310){\usebox{\plotpoint}}
\multiput(222.00,311.41)(1.080,0.502){77}{\rule{1.914pt}{0.121pt}}
\multiput(222.00,308.34)(86.027,42.000){2}{\rule{0.957pt}{0.800pt}}
\multiput(312.00,353.41)(1.066,0.502){79}{\rule{1.893pt}{0.121pt}}
\multiput(312.00,350.34)(87.071,43.000){2}{\rule{0.947pt}{0.800pt}}
\multiput(403.00,396.41)(1.222,0.503){69}{\rule{2.137pt}{0.121pt}}
\multiput(403.00,393.34)(87.565,38.000){2}{\rule{1.068pt}{0.800pt}}
\multiput(495.00,434.41)(1.315,0.503){63}{\rule{2.280pt}{0.121pt}}
\multiput(495.00,431.34)(86.268,35.000){2}{\rule{1.140pt}{0.800pt}}
\multiput(586.00,469.41)(1.613,0.504){51}{\rule{2.738pt}{0.121pt}}
\multiput(586.00,466.34)(86.317,29.000){2}{\rule{1.369pt}{0.800pt}}
\multiput(678.00,498.41)(1.805,0.504){45}{\rule{3.031pt}{0.121pt}}
\multiput(678.00,495.34)(85.709,26.000){2}{\rule{1.515pt}{0.800pt}}
\multiput(770.00,524.41)(2.261,0.502){75}{\rule{3.771pt}{0.121pt}}
\multiput(770.00,521.34)(175.174,41.000){2}{\rule{1.885pt}{0.800pt}}
\multiput(953.00,565.41)(3.471,0.504){47}{\rule{5.622pt}{0.121pt}}
\multiput(953.00,562.34)(171.331,27.000){2}{\rule{2.811pt}{0.800pt}}
\put(1136,591){\usebox{\plotpoint}}
\sbox{\plotpoint}{\rule[-0.500pt]{1.000pt}{1.000pt}}%
\put(660,806){\makebox(0,0)[r]{200 MeV}}
\multiput(682,806)(20.756,0.000){4}{\usebox{\plotpoint}}
\put(748,806){\usebox{\plotpoint}}
\put(222,308){\usebox{\plotpoint}}
\multiput(222,308)(20.542,2.967){5}{\usebox{\plotpoint}}
\multiput(312,321)(20.014,5.498){4}{\usebox{\plotpoint}}
\multiput(403,346)(20.029,5.443){5}{\usebox{\plotpoint}}
\multiput(495,371)(20.174,4.877){5}{\usebox{\plotpoint}}
\multiput(586,393)(20.235,4.619){4}{\usebox{\plotpoint}}
\multiput(678,414)(20.449,3.556){5}{\usebox{\plotpoint}}
\multiput(770,430)(20.564,2.809){8}{\usebox{\plotpoint}}
\multiput(953,455)(20.703,1.471){9}{\usebox{\plotpoint}}
\put(1136,468){\usebox{\plotpoint}}
\sbox{\plotpoint}{\rule[-0.600pt]{1.200pt}{1.200pt}}%
\put(660,761){\makebox(0,0)[r]{240 MeV}}
\put(682.0,761.0){\rule[-0.600pt]{15.899pt}{1.200pt}}
\put(224,350){\usebox{\plotpoint}}
\put(224,348.51){\rule{21.199pt}{1.200pt}}
\multiput(224.00,347.51)(44.000,2.000){2}{\rule{10.600pt}{1.200pt}}
\multiput(312.00,354.24)(14.205,0.509){2}{\rule{18.500pt}{0.123pt}}
\multiput(312.00,349.51)(52.602,6.000){2}{\rule{9.250pt}{1.200pt}}
\multiput(403.00,360.24)(6.694,0.503){6}{\rule{14.100pt}{0.121pt}}
\multiput(403.00,355.51)(62.735,8.000){2}{\rule{7.050pt}{1.200pt}}
\multiput(495.00,368.24)(5.627,0.502){8}{\rule{12.433pt}{0.121pt}}
\multiput(495.00,363.51)(65.194,9.000){2}{\rule{6.217pt}{1.200pt}}
\multiput(586.00,377.24)(6.694,0.503){6}{\rule{14.100pt}{0.121pt}}
\multiput(586.00,372.51)(62.735,8.000){2}{\rule{7.050pt}{1.200pt}}
\multiput(678.00,385.24)(14.374,0.509){2}{\rule{18.700pt}{0.123pt}}
\multiput(678.00,380.51)(53.187,6.000){2}{\rule{9.350pt}{1.200pt}}
\multiput(770.00,391.24)(17.176,0.505){4}{\rule{31.671pt}{0.122pt}}
\multiput(770.00,386.51)(117.264,7.000){2}{\rule{15.836pt}{1.200pt}}
\put(953,392.51){\rule{44.085pt}{1.200pt}}
\multiput(953.00,393.51)(91.500,-2.000){2}{\rule{22.042pt}{1.200pt}}
\put(1136,394){\usebox{\plotpoint}}
\end{picture}

%% file: fa4.tex
\setlength{\unitlength}{0.240900pt}
\ifx\plotpoint\undefined\newsavebox{\plotpoint}\fi
\sbox{\plotpoint}{\rule[-0.200pt]{0.400pt}{0.400pt}}%
\begin{picture}(1200,1200)(0,0)
\font\gnuplot=cmr10 at 10pt
\gnuplot
\sbox{\plotpoint}{\rule[-0.200pt]{0.400pt}{0.400pt}}%
\put(220.0,113.0){\rule[-0.200pt]{220.664pt}{0.400pt}}
\put(220.0,113.0){\rule[-0.200pt]{0.400pt}{245.477pt}}
\put(220.0,113.0){\rule[-0.200pt]{4.818pt}{0.400pt}}
\put(198,113){\makebox(0,0)[r]{0.0}}
\put(1116.0,113.0){\rule[-0.200pt]{4.818pt}{0.400pt}}
\put(220.0,283.0){\rule[-0.200pt]{4.818pt}{0.400pt}}
\put(198,283){\makebox(0,0)[r]{10}}
\put(1116.0,283.0){\rule[-0.200pt]{4.818pt}{0.400pt}}
\put(220.0,453.0){\rule[-0.200pt]{4.818pt}{0.400pt}}
\put(198,453){\makebox(0,0)[r]{20}}
\put(1116.0,453.0){\rule[-0.200pt]{4.818pt}{0.400pt}}
\put(220.0,623.0){\rule[-0.200pt]{4.818pt}{0.400pt}}
\put(198,623){\makebox(0,0)[r]{30}}
\put(1116.0,623.0){\rule[-0.200pt]{4.818pt}{0.400pt}}
\put(220.0,792.0){\rule[-0.200pt]{4.818pt}{0.400pt}}
\put(198,792){\makebox(0,0)[r]{40}}
\put(1116.0,792.0){\rule[-0.200pt]{4.818pt}{0.400pt}}
\put(220.0,962.0){\rule[-0.200pt]{4.818pt}{0.400pt}}
\put(198,962){\makebox(0,0)[r]{50}}
\put(1116.0,962.0){\rule[-0.200pt]{4.818pt}{0.400pt}}
\put(220.0,1132.0){\rule[-0.200pt]{4.818pt}{0.400pt}}
\put(198,1132){\makebox(0,0)[r]{60}}
\put(1116.0,1132.0){\rule[-0.200pt]{4.818pt}{0.400pt}}
\put(220.0,113.0){\rule[-0.200pt]{0.400pt}{4.818pt}}
\put(220,68){\makebox(0,0){0}}
\put(220.0,1112.0){\rule[-0.200pt]{0.400pt}{4.818pt}}
\put(312.0,113.0){\rule[-0.200pt]{0.400pt}{4.818pt}}
\put(312,68){\makebox(0,0){0.05}}
\put(312.0,1112.0){\rule[-0.200pt]{0.400pt}{4.818pt}}
\put(403.0,113.0){\rule[-0.200pt]{0.400pt}{4.818pt}}
\put(403,68){\makebox(0,0){0.1}}
\put(403.0,1112.0){\rule[-0.200pt]{0.400pt}{4.818pt}}
\put(495.0,113.0){\rule[-0.200pt]{0.400pt}{4.818pt}}
\put(495,68){\makebox(0,0){0.15}}
\put(495.0,1112.0){\rule[-0.200pt]{0.400pt}{4.818pt}}
\put(586.0,113.0){\rule[-0.200pt]{0.400pt}{4.818pt}}
\put(586,68){\makebox(0,0){0.2}}
\put(586.0,1112.0){\rule[-0.200pt]{0.400pt}{4.818pt}}
\put(678.0,113.0){\rule[-0.200pt]{0.400pt}{4.818pt}}
\put(678,68){\makebox(0,0){0.25}}
\put(678.0,1112.0){\rule[-0.200pt]{0.400pt}{4.818pt}}
\put(770.0,113.0){\rule[-0.200pt]{0.400pt}{4.818pt}}
\put(770,68){\makebox(0,0){0.3}}
\put(770.0,1112.0){\rule[-0.200pt]{0.400pt}{4.818pt}}
\put(861.0,113.0){\rule[-0.200pt]{0.400pt}{4.818pt}}
\put(861,68){\makebox(0,0){0.35}}
\put(861.0,1112.0){\rule[-0.200pt]{0.400pt}{4.818pt}}
\put(953.0,113.0){\rule[-0.200pt]{0.400pt}{4.818pt}}
\put(953,68){\makebox(0,0){0.4}}
\put(953.0,1112.0){\rule[-0.200pt]{0.400pt}{4.818pt}}
\put(1044.0,113.0){\rule[-0.200pt]{0.400pt}{4.818pt}}
\put(1044,68){\makebox(0,0){0.45}}
\put(1044.0,1112.0){\rule[-0.200pt]{0.400pt}{4.818pt}}
\put(1136.0,113.0){\rule[-0.200pt]{0.400pt}{4.818pt}}
\put(1136,68){\makebox(0,0){0.5}}
\put(1136.0,1112.0){\rule[-0.200pt]{0.400pt}{4.818pt}}
\put(220.0,113.0){\rule[-0.200pt]{220.664pt}{0.400pt}}
\put(1136.0,113.0){\rule[-0.200pt]{0.400pt}{245.477pt}}
\put(220.0,1132.0){\rule[-0.200pt]{220.664pt}{0.400pt}}
\put(45,622){\makebox(0,0){\shortstack{$\sigma$ \\ (MeV)}}}
\put(678,23){\makebox(0,0){$\rho_{B}$ (fm$^{-3})$}}
\put(678,1177){\makebox(0,0){Figure 4}}
\put(312,1064){\makebox(0,0)[l]{$\epsilon_{\mbox{vac}}=(250 \mbox{MeV})^{4}$}}
\put(220.0,113.0){\rule[-0.200pt]{0.400pt}{245.477pt}}
\put(660,996){\makebox(0,0)[r]{Cold nuclear matter}}
\put(682.0,996.0){\rule[-0.200pt]{15.899pt}{0.400pt}}
\put(222,116){\usebox{\plotpoint}}
\multiput(222.58,116.00)(0.498,0.917){67}{\rule{0.120pt}{0.831pt}}
\multiput(221.17,116.00)(35.000,62.274){2}{\rule{0.400pt}{0.416pt}}
\multiput(257.58,180.00)(0.499,0.788){289}{\rule{0.120pt}{0.730pt}}
\multiput(256.17,180.00)(146.000,228.485){2}{\rule{0.400pt}{0.365pt}}
\multiput(403.58,410.00)(0.499,0.663){181}{\rule{0.120pt}{0.630pt}}
\multiput(402.17,410.00)(92.000,120.691){2}{\rule{0.400pt}{0.315pt}}
\multiput(495.58,532.00)(0.499,0.610){179}{\rule{0.120pt}{0.588pt}}
\multiput(494.17,532.00)(91.000,109.780){2}{\rule{0.400pt}{0.294pt}}
\multiput(586.58,643.00)(0.499,0.560){181}{\rule{0.120pt}{0.548pt}}
\multiput(585.17,643.00)(92.000,101.863){2}{\rule{0.400pt}{0.274pt}}
\multiput(678.58,746.00)(0.499,0.516){181}{\rule{0.120pt}{0.513pt}}
\multiput(677.17,746.00)(92.000,93.935){2}{\rule{0.400pt}{0.257pt}}
\multiput(770.00,841.58)(0.505,0.499){177}{\rule{0.504pt}{0.120pt}}
\multiput(770.00,840.17)(89.953,90.000){2}{\rule{0.252pt}{0.400pt}}
\multiput(861.00,931.58)(0.528,0.499){171}{\rule{0.523pt}{0.120pt}}
\multiput(861.00,930.17)(90.915,87.000){2}{\rule{0.261pt}{0.400pt}}
\multiput(953.00,1018.58)(0.535,0.499){167}{\rule{0.528pt}{0.120pt}}
\multiput(953.00,1017.17)(89.904,85.000){2}{\rule{0.264pt}{0.400pt}}
\multiput(1044.00,1103.58)(0.551,0.497){55}{\rule{0.541pt}{0.120pt}}
\multiput(1044.00,1102.17)(30.876,29.000){2}{\rule{0.271pt}{0.400pt}}
\put(660,951){\makebox(0,0)[r]{ 50 MeV}}
\multiput(682,951)(20.756,0.000){4}{\usebox{\plotpoint}}
\put(748,951){\usebox{\plotpoint}}
\put(222,116){\usebox{\plotpoint}}
\multiput(222,116)(11.440,17.318){16}{\usebox{\plotpoint}}
\multiput(403,390)(13.104,16.096){7}{\usebox{\plotpoint}}
\multiput(495,503)(13.969,15.351){7}{\usebox{\plotpoint}}
\multiput(586,603)(14.837,14.514){6}{\usebox{\plotpoint}}
\multiput(678,693)(15.662,13.619){6}{\usebox{\plotpoint}}
\multiput(770,773)(16.364,12.768){5}{\usebox{\plotpoint}}
\multiput(861,844)(17.038,11.853){6}{\usebox{\plotpoint}}
\multiput(953,908)(17.590,11.018){5}{\usebox{\plotpoint}}
\multiput(1044,965)(18.153,10.063){5}{\usebox{\plotpoint}}
\put(1136,1016){\usebox{\plotpoint}}
\sbox{\plotpoint}{\rule[-0.400pt]{0.800pt}{0.800pt}}%
\put(660,906){\makebox(0,0)[r]{150 MeV}}
\put(682.0,906.0){\rule[-0.400pt]{15.899pt}{0.800pt}}
\put(222,135){\usebox{\plotpoint}}
\multiput(223.41,135.00)(0.501,0.572){173}{\rule{0.121pt}{1.116pt}}
\multiput(220.34,135.00)(90.000,100.685){2}{\rule{0.800pt}{0.558pt}}
\multiput(313.41,238.00)(0.501,0.577){175}{\rule{0.121pt}{1.123pt}}
\multiput(310.34,238.00)(91.000,102.669){2}{\rule{0.800pt}{0.562pt}}
\multiput(404.41,343.00)(0.501,0.510){177}{\rule{0.121pt}{1.017pt}}
\multiput(401.34,343.00)(92.000,91.888){2}{\rule{0.800pt}{0.509pt}}
\multiput(495.00,438.41)(0.548,0.501){159}{\rule{1.077pt}{0.121pt}}
\multiput(495.00,435.34)(88.764,83.000){2}{\rule{0.539pt}{0.800pt}}
\multiput(586.00,521.41)(0.622,0.501){141}{\rule{1.195pt}{0.121pt}}
\multiput(586.00,518.34)(89.521,74.000){2}{\rule{0.597pt}{0.800pt}}
\multiput(678.00,595.41)(0.687,0.501){127}{\rule{1.299pt}{0.121pt}}
\multiput(678.00,592.34)(89.305,67.000){2}{\rule{0.649pt}{0.800pt}}
\multiput(770.00,662.41)(0.790,0.501){225}{\rule{1.462pt}{0.121pt}}
\multiput(770.00,659.34)(179.965,116.000){2}{\rule{0.731pt}{0.800pt}}
\multiput(953.00,778.41)(0.966,0.501){183}{\rule{1.741pt}{0.121pt}}
\multiput(953.00,775.34)(179.386,95.000){2}{\rule{0.871pt}{0.800pt}}
\put(1136,872){\usebox{\plotpoint}}
\sbox{\plotpoint}{\rule[-0.500pt]{1.000pt}{1.000pt}}%
\put(660,861){\makebox(0,0)[r]{200 MeV}}
\multiput(682,861)(20.756,0.000){4}{\usebox{\plotpoint}}
\put(748,861){\usebox{\plotpoint}}
\put(222,266){\usebox{\plotpoint}}
\multiput(222,266)(19.271,7.708){5}{\usebox{\plotpoint}}
\multiput(312,302)(16.451,12.655){6}{\usebox{\plotpoint}}
\multiput(403,372)(16.173,13.009){5}{\usebox{\plotpoint}}
\multiput(495,446)(16.626,12.424){6}{\usebox{\plotpoint}}
\multiput(586,514)(17.038,11.853){5}{\usebox{\plotpoint}}
\multiput(678,578)(17.558,11.069){6}{\usebox{\plotpoint}}
\multiput(770,636)(18.214,9.953){10}{\usebox{\plotpoint}}
\multiput(953,736)(18.902,8.573){9}{\usebox{\plotpoint}}
\put(1136,819){\usebox{\plotpoint}}
\sbox{\plotpoint}{\rule[-0.600pt]{1.200pt}{1.200pt}}%
\put(660,816){\makebox(0,0)[r]{240 MeV}}
\put(682.0,816.0){\rule[-0.600pt]{15.899pt}{1.200pt}}
\put(224,526){\usebox{\plotpoint}}
\multiput(224.00,528.24)(5.435,0.502){8}{\rule{12.033pt}{0.121pt}}
\multiput(224.00,523.51)(63.024,9.000){2}{\rule{6.017pt}{1.200pt}}
\multiput(312.00,537.24)(2.001,0.501){36}{\rule{5.048pt}{0.121pt}}
\multiput(312.00,532.51)(80.523,23.000){2}{\rule{2.524pt}{1.200pt}}
\multiput(403.00,560.24)(1.356,0.500){58}{\rule{3.547pt}{0.121pt}}
\multiput(403.00,555.51)(84.638,34.000){2}{\rule{1.774pt}{1.200pt}}
\multiput(495.00,594.24)(1.230,0.500){64}{\rule{3.251pt}{0.121pt}}
\multiput(495.00,589.51)(84.252,37.000){2}{\rule{1.626pt}{1.200pt}}
\multiput(586.00,631.24)(1.179,0.500){68}{\rule{3.131pt}{0.121pt}}
\multiput(586.00,626.51)(85.502,39.000){2}{\rule{1.565pt}{1.200pt}}
\multiput(678.00,670.24)(1.211,0.500){66}{\rule{3.205pt}{0.121pt}}
\multiput(678.00,665.51)(85.347,38.000){2}{\rule{1.603pt}{1.200pt}}
\multiput(770.00,708.24)(1.289,0.500){132}{\rule{3.393pt}{0.120pt}}
\multiput(770.00,703.51)(175.958,71.000){2}{\rule{1.696pt}{1.200pt}}
\multiput(953.00,779.24)(1.478,0.500){114}{\rule{3.842pt}{0.120pt}}
\multiput(953.00,774.51)(175.026,62.000){2}{\rule{1.921pt}{1.200pt}}
\put(1136,839){\usebox{\plotpoint}}
\end{picture}

%% file: fa5a.tex
\setlength{\unitlength}{0.240900pt}
\ifx\plotpoint\undefined\newsavebox{\plotpoint}\fi
\sbox{\plotpoint}{\rule[-0.200pt]{0.400pt}{0.400pt}}%
\begin{picture}(1200,1200)(0,0)
\font\gnuplot=cmr10 at 10pt
\gnuplot
\sbox{\plotpoint}{\rule[-0.200pt]{0.400pt}{0.400pt}}%
\put(220.0,113.0){\rule[-0.200pt]{0.400pt}{245.477pt}}
\put(220.0,113.0){\rule[-0.200pt]{4.818pt}{0.400pt}}
\put(198,113){\makebox(0,0)[r]{0.94}}
\put(1116.0,113.0){\rule[-0.200pt]{4.818pt}{0.400pt}}
\put(220.0,240.0){\rule[-0.200pt]{4.818pt}{0.400pt}}
\put(198,240){\makebox(0,0)[r]{0.95}}
\put(1116.0,240.0){\rule[-0.200pt]{4.818pt}{0.400pt}}
\put(220.0,368.0){\rule[-0.200pt]{4.818pt}{0.400pt}}
\put(198,368){\makebox(0,0)[r]{0.96}}
\put(1116.0,368.0){\rule[-0.200pt]{4.818pt}{0.400pt}}
\put(220.0,495.0){\rule[-0.200pt]{4.818pt}{0.400pt}}
\put(198,495){\makebox(0,0)[r]{0.97}}
\put(1116.0,495.0){\rule[-0.200pt]{4.818pt}{0.400pt}}
\put(220.0,623.0){\rule[-0.200pt]{4.818pt}{0.400pt}}
\put(198,623){\makebox(0,0)[r]{0.98}}
\put(1116.0,623.0){\rule[-0.200pt]{4.818pt}{0.400pt}}
\put(220.0,750.0){\rule[-0.200pt]{4.818pt}{0.400pt}}
\put(198,750){\makebox(0,0)[r]{0.99}}
\put(1116.0,750.0){\rule[-0.200pt]{4.818pt}{0.400pt}}
\put(220.0,877.0){\rule[-0.200pt]{4.818pt}{0.400pt}}
\put(198,877){\makebox(0,0)[r]{1}}
\put(1116.0,877.0){\rule[-0.200pt]{4.818pt}{0.400pt}}
\put(220.0,1005.0){\rule[-0.200pt]{4.818pt}{0.400pt}}
\put(198,1005){\makebox(0,0)[r]{1.01}}
\put(1116.0,1005.0){\rule[-0.200pt]{4.818pt}{0.400pt}}
\put(220.0,1132.0){\rule[-0.200pt]{4.818pt}{0.400pt}}
\put(198,1132){\makebox(0,0)[r]{1.02}}
\put(1116.0,1132.0){\rule[-0.200pt]{4.818pt}{0.400pt}}
\put(220.0,113.0){\rule[-0.200pt]{0.400pt}{4.818pt}}
\put(220,68){\makebox(0,0){0}}
\put(220.0,1112.0){\rule[-0.200pt]{0.400pt}{4.818pt}}
\put(351.0,113.0){\rule[-0.200pt]{0.400pt}{4.818pt}}
\put(351,68){\makebox(0,0){0.1}}
\put(351.0,1112.0){\rule[-0.200pt]{0.400pt}{4.818pt}}
\put(482.0,113.0){\rule[-0.200pt]{0.400pt}{4.818pt}}
\put(482,68){\makebox(0,0){0.2}}
\put(482.0,1112.0){\rule[-0.200pt]{0.400pt}{4.818pt}}
\put(613.0,113.0){\rule[-0.200pt]{0.400pt}{4.818pt}}
\put(613,68){\makebox(0,0){0.3}}
\put(613.0,1112.0){\rule[-0.200pt]{0.400pt}{4.818pt}}
\put(743.0,113.0){\rule[-0.200pt]{0.400pt}{4.818pt}}
\put(743,68){\makebox(0,0){0.4}}
\put(743.0,1112.0){\rule[-0.200pt]{0.400pt}{4.818pt}}
\put(874.0,113.0){\rule[-0.200pt]{0.400pt}{4.818pt}}
\put(874,68){\makebox(0,0){0.5}}
\put(874.0,1112.0){\rule[-0.200pt]{0.400pt}{4.818pt}}
\put(1005.0,113.0){\rule[-0.200pt]{0.400pt}{4.818pt}}
\put(1005,68){\makebox(0,0){0.6}}
\put(1005.0,1112.0){\rule[-0.200pt]{0.400pt}{4.818pt}}
\put(1136.0,113.0){\rule[-0.200pt]{0.400pt}{4.818pt}}
\put(1136,68){\makebox(0,0){0.7}}
\put(1136.0,1112.0){\rule[-0.200pt]{0.400pt}{4.818pt}}
\put(220.0,113.0){\rule[-0.200pt]{220.664pt}{0.400pt}}
\put(1136.0,113.0){\rule[-0.200pt]{0.400pt}{245.477pt}}
\put(220.0,1132.0){\rule[-0.200pt]{220.664pt}{0.400pt}}
\put(45,622){\makebox(0,0){$\chi=\frac{\phi}{\phi_{0}}$}}
\put(678,23){\makebox(0,0){$\rho_{B}$ (fm$^{-3})$}}
\put(678,1177){\makebox(0,0){Fig. 5(a)}}
\put(272,623){\makebox(0,0)[l]{Cold nuclear matter}}
\put(220.0,113.0){\rule[-0.200pt]{0.400pt}{245.477pt}}
\put(613,559){\makebox(0,0)[r]{$\epsilon_{\mbox{vac}}=(200\mbox{MeV})^{4}$}}
\put(635.0,559.0){\rule[-0.200pt]{15.899pt}{0.400pt}}
\put(221,880){\usebox{\plotpoint}}
\multiput(221.58,880.00)(0.499,0.806){125}{\rule{0.120pt}{0.744pt}}
\multiput(220.17,880.00)(64.000,101.456){2}{\rule{0.400pt}{0.372pt}}
\multiput(285.00,983.58)(0.734,0.498){87}{\rule{0.687pt}{0.120pt}}
\multiput(285.00,982.17)(64.575,45.000){2}{\rule{0.343pt}{0.400pt}}
\put(351,1028.17){\rule{2.700pt}{0.400pt}}
\multiput(351.00,1027.17)(7.396,2.000){2}{\rule{1.350pt}{0.400pt}}
\put(377,1028.67){\rule{3.132pt}{0.400pt}}
\multiput(377.00,1029.17)(6.500,-1.000){2}{\rule{1.566pt}{0.400pt}}
\multiput(390.00,1027.94)(1.797,-0.468){5}{\rule{1.400pt}{0.113pt}}
\multiput(390.00,1028.17)(10.094,-4.000){2}{\rule{0.700pt}{0.400pt}}
\multiput(403.00,1023.93)(1.378,-0.477){7}{\rule{1.140pt}{0.115pt}}
\multiput(403.00,1024.17)(10.634,-5.000){2}{\rule{0.570pt}{0.400pt}}
\multiput(416.00,1018.93)(0.824,-0.488){13}{\rule{0.750pt}{0.117pt}}
\multiput(416.00,1019.17)(11.443,-8.000){2}{\rule{0.375pt}{0.400pt}}
\multiput(429.00,1010.92)(0.652,-0.491){17}{\rule{0.620pt}{0.118pt}}
\multiput(429.00,1011.17)(11.713,-10.000){2}{\rule{0.310pt}{0.400pt}}
\multiput(442.00,1000.92)(0.637,-0.492){19}{\rule{0.609pt}{0.118pt}}
\multiput(442.00,1001.17)(12.736,-11.000){2}{\rule{0.305pt}{0.400pt}}
\multiput(456.58,988.80)(0.493,-0.536){23}{\rule{0.119pt}{0.531pt}}
\multiput(455.17,989.90)(13.000,-12.898){2}{\rule{0.400pt}{0.265pt}}
\multiput(469.58,974.03)(0.493,-0.774){23}{\rule{0.119pt}{0.715pt}}
\multiput(468.17,975.52)(13.000,-18.515){2}{\rule{0.400pt}{0.358pt}}
\multiput(482.58,954.41)(0.493,-0.655){23}{\rule{0.119pt}{0.623pt}}
\multiput(481.17,955.71)(13.000,-15.707){2}{\rule{0.400pt}{0.312pt}}
\multiput(495.58,936.65)(0.493,-0.893){23}{\rule{0.119pt}{0.808pt}}
\multiput(494.17,938.32)(13.000,-21.324){2}{\rule{0.400pt}{0.404pt}}
\multiput(508.59,913.26)(0.482,-1.033){9}{\rule{0.116pt}{0.900pt}}
\multiput(507.17,915.13)(6.000,-10.132){2}{\rule{0.400pt}{0.450pt}}
\multiput(514.58,901.10)(0.496,-1.058){37}{\rule{0.119pt}{0.940pt}}
\multiput(513.17,903.05)(20.000,-40.049){2}{\rule{0.400pt}{0.470pt}}
\multiput(534.58,858.37)(0.493,-1.290){23}{\rule{0.119pt}{1.115pt}}
\multiput(533.17,860.68)(13.000,-30.685){2}{\rule{0.400pt}{0.558pt}}
\multiput(547.58,823.52)(0.499,-1.832){129}{\rule{0.120pt}{1.561pt}}
\multiput(546.17,826.76)(66.000,-237.761){2}{\rule{0.400pt}{0.780pt}}
\multiput(613.58,578.05)(0.497,-3.201){61}{\rule{0.120pt}{2.638pt}}
\multiput(612.17,583.53)(32.000,-197.526){2}{\rule{0.400pt}{1.319pt}}
\multiput(645.58,367.45)(0.497,-5.532){47}{\rule{0.120pt}{4.468pt}}
\multiput(644.17,376.73)(25.000,-263.726){2}{\rule{0.400pt}{2.234pt}}
\put(364.0,1030.0){\rule[-0.200pt]{3.132pt}{0.400pt}}
\put(613,514){\makebox(0,0)[r]{$\epsilon_{\mbox{vac}}=(250 \mbox{MeV})^{4}$}}
\multiput(635,514)(20.756,0.000){4}{\usebox{\plotpoint}}
\put(701,514){\usebox{\plotpoint}}
\put(221,879){\usebox{\plotpoint}}
\multiput(221,879)(16.525,12.559){2}{\usebox{\plotpoint}}
\multiput(246,898)(19.576,6.898){5}{\usebox{\plotpoint}}
\multiput(351,935)(20.668,-1.908){4}{\usebox{\plotpoint}}
\multiput(416,929)(19.506,-7.093){3}{\usebox{\plotpoint}}
\multiput(482,905)(17.433,-11.264){4}{\usebox{\plotpoint}}
\multiput(547,863)(15.358,-13.962){4}{\usebox{\plotpoint}}
\multiput(613,803)(13.088,-16.109){5}{\usebox{\plotpoint}}
\multiput(678,723)(11.077,-17.553){6}{\usebox{\plotpoint}}
\multiput(743,620)(9.171,-18.620){7}{\usebox{\plotpoint}}
\multiput(809,486)(7.191,-19.470){9}{\usebox{\plotpoint}}
\multiput(874,310)(5.297,-20.068){10}{\usebox{\plotpoint}}
\put(926,113){\usebox{\plotpoint}}
\sbox{\plotpoint}{\rule[-0.400pt]{0.800pt}{0.800pt}}%
\put(613,469){\makebox(0,0)[r]{$\epsilon_{\mbox{vac}}=(300 \mbox{MeV})^{4}$}}
\put(635.0,469.0){\rule[-0.400pt]{15.899pt}{0.800pt}}
\put(221,878){\usebox{\plotpoint}}
\multiput(221.00,879.41)(1.642,0.505){33}{\rule{2.760pt}{0.122pt}}
\multiput(221.00,876.34)(58.271,20.000){2}{\rule{1.380pt}{0.800pt}}
\multiput(285.00,899.39)(7.160,0.536){5}{\rule{9.000pt}{0.129pt}}
\multiput(285.00,896.34)(47.320,6.000){2}{\rule{4.500pt}{0.800pt}}
\put(351,900.84){\rule{15.658pt}{0.800pt}}
\multiput(351.00,902.34)(32.500,-3.000){2}{\rule{7.829pt}{0.800pt}}
\multiput(416.00,899.08)(3.232,-0.512){15}{\rule{5.000pt}{0.123pt}}
\multiput(416.00,899.34)(55.622,-11.000){2}{\rule{2.500pt}{0.800pt}}
\multiput(482.00,888.09)(1.760,-0.506){31}{\rule{2.937pt}{0.122pt}}
\multiput(482.00,888.34)(58.904,-19.000){2}{\rule{1.468pt}{0.800pt}}
\multiput(547.00,869.09)(1.290,-0.504){45}{\rule{2.231pt}{0.121pt}}
\multiput(547.00,869.34)(61.370,-26.000){2}{\rule{1.115pt}{0.800pt}}
\multiput(613.00,843.09)(0.932,-0.501){133}{\rule{1.686pt}{0.121pt}}
\multiput(613.00,843.34)(126.501,-70.000){2}{\rule{0.843pt}{0.800pt}}
\multiput(743.00,773.09)(0.690,-0.501){183}{\rule{1.303pt}{0.121pt}}
\multiput(743.00,773.34)(128.295,-95.000){2}{\rule{0.652pt}{0.800pt}}
\multiput(874.00,678.09)(0.511,-0.501){249}{\rule{1.019pt}{0.121pt}}
\multiput(874.00,678.34)(128.886,-128.000){2}{\rule{0.509pt}{0.800pt}}
\multiput(1006.41,546.91)(0.501,-0.641){255}{\rule{0.121pt}{1.226pt}}
\multiput(1003.34,549.46)(131.000,-165.455){2}{\rule{0.800pt}{0.613pt}}
\sbox{\plotpoint}{\rule[-0.500pt]{1.000pt}{1.000pt}}%
\put(613,424){\makebox(0,0)[r]{$\epsilon_{\mbox{vac}}=(350 \mbox{MeV})^{4}$}}
\multiput(635,424)(20.756,0.000){4}{\usebox{\plotpoint}}
\put(701,424){\usebox{\plotpoint}}
\put(221,878){\usebox{\plotpoint}}
\multiput(221,878)(20.456,3.516){4}{\usebox{\plotpoint}}
\multiput(285,889)(20.734,0.942){3}{\usebox{\plotpoint}}
\multiput(351,892)(20.746,-0.638){3}{\usebox{\plotpoint}}
\multiput(416,890)(20.670,-1.879){3}{\usebox{\plotpoint}}
\multiput(482,884)(20.514,-3.156){3}{\usebox{\plotpoint}}
\multiput(547,874)(20.364,-4.011){4}{\usebox{\plotpoint}}
\multiput(613,861)(20.042,-5.396){6}{\usebox{\plotpoint}}
\multiput(743,826)(19.583,-6.877){7}{\usebox{\plotpoint}}
\multiput(874,780)(19.189,-7.910){7}{\usebox{\plotpoint}}
\multiput(1005,726)(18.422,-9.562){7}{\usebox{\plotpoint}}
\put(1136,658){\usebox{\plotpoint}}
\sbox{\plotpoint}{\rule[-0.600pt]{1.200pt}{1.200pt}}%
\put(613,379){\makebox(0,0)[r]{$\epsilon_{\mbox{vac}}=(800 \mbox{MeV})^{4}$}}
\put(635.0,379.0){\rule[-0.600pt]{15.899pt}{1.200pt}}
\put(233,877){\usebox{\plotpoint}}
\put(233,875.01){\rule{28.426pt}{1.200pt}}
\multiput(233.00,874.51)(59.000,1.000){2}{\rule{14.213pt}{1.200pt}}
\put(482,875.01){\rule{3.132pt}{1.200pt}}
\multiput(482.00,875.51)(6.500,-1.000){2}{\rule{1.566pt}{1.200pt}}
\put(351.0,878.0){\rule[-0.600pt]{31.558pt}{1.200pt}}
\put(613,873.51){\rule{31.317pt}{1.200pt}}
\multiput(613.00,874.51)(65.000,-2.000){2}{\rule{15.658pt}{1.200pt}}
\put(743,872.01){\rule{31.558pt}{1.200pt}}
\multiput(743.00,872.51)(65.500,-1.000){2}{\rule{15.779pt}{1.200pt}}
\put(874,870.51){\rule{31.558pt}{1.200pt}}
\multiput(874.00,871.51)(65.500,-2.000){2}{\rule{15.779pt}{1.200pt}}
\put(1005,868.51){\rule{31.558pt}{1.200pt}}
\multiput(1005.00,869.51)(65.500,-2.000){2}{\rule{15.779pt}{1.200pt}}
\put(495.0,877.0){\rule[-0.600pt]{28.426pt}{1.200pt}}
\put(1136,870){\usebox{\plotpoint}}
\end{picture}

%% file: fa5b.tex
\setlength{\unitlength}{0.240900pt}
\ifx\plotpoint\undefined\newsavebox{\plotpoint}\fi
\sbox{\plotpoint}{\rule[-0.200pt]{0.400pt}{0.400pt}}%
\begin{picture}(1200,1200)(0,0)
\font\gnuplot=cmr10 at 10pt
\gnuplot
\sbox{\plotpoint}{\rule[-0.200pt]{0.400pt}{0.400pt}}%
\put(220.0,113.0){\rule[-0.200pt]{0.400pt}{245.477pt}}
\put(220.0,113.0){\rule[-0.200pt]{4.818pt}{0.400pt}}
\put(198,113){\makebox(0,0)[r]{1}}
\put(1116.0,113.0){\rule[-0.200pt]{4.818pt}{0.400pt}}
\put(220.0,215.0){\rule[-0.200pt]{4.818pt}{0.400pt}}
\put(198,215){\makebox(0,0)[r]{1.02}}
\put(1116.0,215.0){\rule[-0.200pt]{4.818pt}{0.400pt}}
\put(220.0,317.0){\rule[-0.200pt]{4.818pt}{0.400pt}}
\put(198,317){\makebox(0,0)[r]{1.04}}
\put(1116.0,317.0){\rule[-0.200pt]{4.818pt}{0.400pt}}
\put(220.0,419.0){\rule[-0.200pt]{4.818pt}{0.400pt}}
\put(198,419){\makebox(0,0)[r]{1.06}}
\put(1116.0,419.0){\rule[-0.200pt]{4.818pt}{0.400pt}}
\put(220.0,521.0){\rule[-0.200pt]{4.818pt}{0.400pt}}
\put(198,521){\makebox(0,0)[r]{1.08}}
\put(1116.0,521.0){\rule[-0.200pt]{4.818pt}{0.400pt}}
\put(220.0,623.0){\rule[-0.200pt]{4.818pt}{0.400pt}}
\put(198,623){\makebox(0,0)[r]{1.1}}
\put(1116.0,623.0){\rule[-0.200pt]{4.818pt}{0.400pt}}
\put(220.0,724.0){\rule[-0.200pt]{4.818pt}{0.400pt}}
\put(198,724){\makebox(0,0)[r]{1.12}}
\put(1116.0,724.0){\rule[-0.200pt]{4.818pt}{0.400pt}}
\put(220.0,826.0){\rule[-0.200pt]{4.818pt}{0.400pt}}
\put(198,826){\makebox(0,0)[r]{1.14}}
\put(1116.0,826.0){\rule[-0.200pt]{4.818pt}{0.400pt}}
\put(220.0,928.0){\rule[-0.200pt]{4.818pt}{0.400pt}}
\put(198,928){\makebox(0,0)[r]{1.16}}
\put(1116.0,928.0){\rule[-0.200pt]{4.818pt}{0.400pt}}
\put(220.0,1030.0){\rule[-0.200pt]{4.818pt}{0.400pt}}
\put(198,1030){\makebox(0,0)[r]{1.18}}
\put(1116.0,1030.0){\rule[-0.200pt]{4.818pt}{0.400pt}}
\put(220.0,1132.0){\rule[-0.200pt]{4.818pt}{0.400pt}}
\put(198,1132){\makebox(0,0)[r]{1.2}}
\put(1116.0,1132.0){\rule[-0.200pt]{4.818pt}{0.400pt}}
\put(220.0,113.0){\rule[-0.200pt]{0.400pt}{4.818pt}}
\put(220,68){\makebox(0,0){0}}
\put(220.0,1112.0){\rule[-0.200pt]{0.400pt}{4.818pt}}
\put(387.0,113.0){\rule[-0.200pt]{0.400pt}{4.818pt}}
\put(387,68){\makebox(0,0){0.1}}
\put(387.0,1112.0){\rule[-0.200pt]{0.400pt}{4.818pt}}
\put(553.0,113.0){\rule[-0.200pt]{0.400pt}{4.818pt}}
\put(553,68){\makebox(0,0){0.2}}
\put(553.0,1112.0){\rule[-0.200pt]{0.400pt}{4.818pt}}
\put(720.0,113.0){\rule[-0.200pt]{0.400pt}{4.818pt}}
\put(720,68){\makebox(0,0){0.3}}
\put(720.0,1112.0){\rule[-0.200pt]{0.400pt}{4.818pt}}
\put(886.0,113.0){\rule[-0.200pt]{0.400pt}{4.818pt}}
\put(886,68){\makebox(0,0){0.4}}
\put(886.0,1112.0){\rule[-0.200pt]{0.400pt}{4.818pt}}
\put(1053.0,113.0){\rule[-0.200pt]{0.400pt}{4.818pt}}
\put(1053,68){\makebox(0,0){0.5}}
\put(1053.0,1112.0){\rule[-0.200pt]{0.400pt}{4.818pt}}
\put(220.0,113.0){\rule[-0.200pt]{220.664pt}{0.400pt}}
\put(1136.0,113.0){\rule[-0.200pt]{0.400pt}{245.477pt}}
\put(220.0,1132.0){\rule[-0.200pt]{220.664pt}{0.400pt}}
\put(45,622){\makebox(0,0){$\chi=(\frac{\phi}{\phi_{0}})$}}
\put(678,23){\makebox(0,0){$\rho_{B}$ (fm$^{-3})$}}
\put(678,1177){\makebox(0,0){Figure 5(b)}}
\put(320,1030){\makebox(0,0)[l]{Temperature 200 MeV}}
\put(220.0,113.0){\rule[-0.200pt]{0.400pt}{245.477pt}}
\put(720,928){\makebox(0,0)[r]{$\epsilon_{\mbox{vac}}=(200 \mbox{MeV})^{4}$}}
\put(742.0,928.0){\rule[-0.200pt]{15.899pt}{0.400pt}}
\put(222,186){\usebox{\plotpoint}}
\multiput(237.00,186.58)(1.275,0.499){115}{\rule{1.117pt}{0.120pt}}
\multiput(237.00,185.17)(147.682,59.000){2}{\rule{0.558pt}{0.400pt}}
\multiput(387.00,245.58)(0.885,0.498){91}{\rule{0.806pt}{0.120pt}}
\multiput(387.00,244.17)(81.326,47.000){2}{\rule{0.403pt}{0.400pt}}
\multiput(470.00,292.58)(0.784,0.498){103}{\rule{0.726pt}{0.120pt}}
\multiput(470.00,291.17)(81.492,53.000){2}{\rule{0.363pt}{0.400pt}}
\multiput(553.00,345.58)(0.755,0.499){107}{\rule{0.704pt}{0.120pt}}
\multiput(553.00,344.17)(81.540,55.000){2}{\rule{0.352pt}{0.400pt}}
\multiput(636.00,400.58)(0.689,0.499){119}{\rule{0.651pt}{0.120pt}}
\multiput(636.00,399.17)(82.649,61.000){2}{\rule{0.325pt}{0.400pt}}
\multiput(720.00,461.58)(0.597,0.499){275}{\rule{0.578pt}{0.120pt}}
\multiput(720.00,460.17)(164.801,139.000){2}{\rule{0.289pt}{0.400pt}}
\multiput(886.00,600.58)(0.500,0.500){331}{\rule{0.500pt}{0.120pt}}
\multiput(886.00,599.17)(165.962,167.000){2}{\rule{0.250pt}{0.400pt}}
\multiput(1053.58,767.00)(0.499,0.663){163}{\rule{0.120pt}{0.630pt}}
\multiput(1052.17,767.00)(83.000,108.692){2}{\rule{0.400pt}{0.315pt}}
\put(222.0,186.0){\rule[-0.200pt]{3.613pt}{0.400pt}}
\put(720,883){\makebox(0,0)[r]{$\epsilon_{\mbox{vac}}=(250 \mbox{MeV})^{4}$}}
\multiput(742,883)(20.756,0.000){4}{\usebox{\plotpoint}}
\put(808,883){\usebox{\plotpoint}}
\put(222,142){\usebox{\plotpoint}}
\put(222.00,142.00){\usebox{\plotpoint}}
\multiput(237,142)(20.516,3.146){8}{\usebox{\plotpoint}}
\multiput(387,165)(20.205,4.747){8}{\usebox{\plotpoint}}
\multiput(553,204)(20.100,5.175){8}{\usebox{\plotpoint}}
\multiput(720,247)(19.906,5.876){8}{\usebox{\plotpoint}}
\multiput(886,296)(19.679,6.599){9}{\usebox{\plotpoint}}
\multiput(1053,352)(19.444,7.262){4}{\usebox{\plotpoint}}
\put(1136,383){\usebox{\plotpoint}}
\sbox{\plotpoint}{\rule[-0.400pt]{0.800pt}{0.800pt}}%
\put(720,838){\makebox(0,0)[r]{$\epsilon_{\mbox{vac}}=(300 \mbox{MeV})^{4}$}}
\put(742.0,838.0){\rule[-0.400pt]{15.899pt}{0.800pt}}
\put(222,127){\usebox{\plotpoint}}
\multiput(253.00,128.41)(6.729,0.511){17}{\rule{10.200pt}{0.123pt}}
\multiput(253.00,125.34)(128.829,12.000){2}{\rule{5.100pt}{0.800pt}}
\multiput(403.00,140.41)(4.566,0.506){31}{\rule{7.232pt}{0.122pt}}
\multiput(403.00,137.34)(151.990,19.000){2}{\rule{3.616pt}{0.800pt}}
\multiput(570.00,159.41)(4.098,0.506){31}{\rule{6.516pt}{0.122pt}}
\multiput(570.00,156.34)(136.476,19.000){2}{\rule{3.258pt}{0.800pt}}
\multiput(720.00,178.41)(3.716,0.505){39}{\rule{5.974pt}{0.122pt}}
\multiput(720.00,175.34)(153.601,23.000){2}{\rule{2.987pt}{0.800pt}}
\multiput(886.00,201.41)(3.166,0.504){47}{\rule{5.148pt}{0.121pt}}
\multiput(886.00,198.34)(156.315,27.000){2}{\rule{2.574pt}{0.800pt}}
\multiput(1053.00,228.41)(2.900,0.508){23}{\rule{4.627pt}{0.122pt}}
\multiput(1053.00,225.34)(73.397,15.000){2}{\rule{2.313pt}{0.800pt}}
\put(222.0,127.0){\rule[-0.400pt]{7.468pt}{0.800pt}}
\sbox{\plotpoint}{\rule[-0.500pt]{1.000pt}{1.000pt}}%
\put(720,793){\makebox(0,0)[r]{$\epsilon_{\mbox{vac}}=(350 \mbox{MeV})^{4}$}}
\multiput(742,793)(20.756,0.000){4}{\usebox{\plotpoint}}
\put(808,793){\usebox{\plotpoint}}
\put(222,120){\usebox{\plotpoint}}
\multiput(222,120)(20.749,0.512){4}{\usebox{\plotpoint}}
\multiput(303,122)(20.730,1.036){5}{\usebox{\plotpoint}}
\multiput(403,127)(20.719,1.237){3}{\usebox{\plotpoint}}
\multiput(470,131)(20.718,1.248){4}{\usebox{\plotpoint}}
\multiput(553,136)(20.711,1.364){9}{\usebox{\plotpoint}}
\multiput(720,147)(20.692,1.620){8}{\usebox{\plotpoint}}
\multiput(886,160)(20.672,1.857){8}{\usebox{\plotpoint}}
\multiput(1053,175)(20.660,1.991){4}{\usebox{\plotpoint}}
\put(1136,183){\usebox{\plotpoint}}
\sbox{\plotpoint}{\rule[-0.600pt]{1.200pt}{1.200pt}}%
\put(720,748){\makebox(0,0)[r]{$\epsilon_{\mbox{vac}}=(800 \mbox{MeV})^{4}$}}
\put(742.0,748.0){\rule[-0.600pt]{15.899pt}{1.200pt}}
\put(222,113){\usebox{\plotpoint}}
\put(237,111.01){\rule{36.135pt}{1.200pt}}
\multiput(237.00,110.51)(75.000,1.000){2}{\rule{18.067pt}{1.200pt}}
\put(222.0,113.0){\rule[-0.600pt]{3.613pt}{1.200pt}}
\put(720,112.01){\rule{39.989pt}{1.200pt}}
\multiput(720.00,111.51)(83.000,1.000){2}{\rule{19.995pt}{1.200pt}}
\put(387.0,114.0){\rule[-0.600pt]{80.220pt}{1.200pt}}
\put(1053,113.01){\rule{19.995pt}{1.200pt}}
\multiput(1053.00,112.51)(41.500,1.000){2}{\rule{9.997pt}{1.200pt}}
\put(886.0,115.0){\rule[-0.600pt]{40.230pt}{1.200pt}}
\end{picture}

%% file: fa6a.tex
\setlength{\unitlength}{0.240900pt}
\ifx\plotpoint\undefined\newsavebox{\plotpoint}\fi
\sbox{\plotpoint}{\rule[-0.200pt]{0.400pt}{0.400pt}}%
\begin{picture}(1200,1200)(0,0)
\font\gnuplot=cmr10 at 10pt
\gnuplot
\sbox{\plotpoint}{\rule[-0.200pt]{0.400pt}{0.400pt}}%
\put(220.0,113.0){\rule[-0.200pt]{0.400pt}{245.477pt}}
\put(220.0,113.0){\rule[-0.200pt]{4.818pt}{0.400pt}}
\put(198,113){\makebox(0,0)[r]{500}}
\put(1116.0,113.0){\rule[-0.200pt]{4.818pt}{0.400pt}}
\put(220.0,317.0){\rule[-0.200pt]{4.818pt}{0.400pt}}
\put(198,317){\makebox(0,0)[r]{600}}
\put(1116.0,317.0){\rule[-0.200pt]{4.818pt}{0.400pt}}
\put(220.0,521.0){\rule[-0.200pt]{4.818pt}{0.400pt}}
\put(198,521){\makebox(0,0)[r]{700}}
\put(1116.0,521.0){\rule[-0.200pt]{4.818pt}{0.400pt}}
\put(220.0,724.0){\rule[-0.200pt]{4.818pt}{0.400pt}}
\put(198,724){\makebox(0,0)[r]{800}}
\put(1116.0,724.0){\rule[-0.200pt]{4.818pt}{0.400pt}}
\put(220.0,928.0){\rule[-0.200pt]{4.818pt}{0.400pt}}
\put(198,928){\makebox(0,0)[r]{900}}
\put(1116.0,928.0){\rule[-0.200pt]{4.818pt}{0.400pt}}
\put(220.0,1132.0){\rule[-0.200pt]{4.818pt}{0.400pt}}
\put(198,1132){\makebox(0,0)[r]{1000}}
\put(1116.0,1132.0){\rule[-0.200pt]{4.818pt}{0.400pt}}
\put(220.0,113.0){\rule[-0.200pt]{0.400pt}{4.818pt}}
\put(220,68){\makebox(0,0){0}}
\put(220.0,1112.0){\rule[-0.200pt]{0.400pt}{4.818pt}}
\put(351.0,113.0){\rule[-0.200pt]{0.400pt}{4.818pt}}
\put(351,68){\makebox(0,0){0.1}}
\put(351.0,1112.0){\rule[-0.200pt]{0.400pt}{4.818pt}}
\put(482.0,113.0){\rule[-0.200pt]{0.400pt}{4.818pt}}
\put(482,68){\makebox(0,0){0.2}}
\put(482.0,1112.0){\rule[-0.200pt]{0.400pt}{4.818pt}}
\put(613.0,113.0){\rule[-0.200pt]{0.400pt}{4.818pt}}
\put(613,68){\makebox(0,0){0.3}}
\put(613.0,1112.0){\rule[-0.200pt]{0.400pt}{4.818pt}}
\put(743.0,113.0){\rule[-0.200pt]{0.400pt}{4.818pt}}
\put(743,68){\makebox(0,0){0.4}}
\put(743.0,1112.0){\rule[-0.200pt]{0.400pt}{4.818pt}}
\put(874.0,113.0){\rule[-0.200pt]{0.400pt}{4.818pt}}
\put(874,68){\makebox(0,0){0.5}}
\put(874.0,1112.0){\rule[-0.200pt]{0.400pt}{4.818pt}}
\put(1005.0,113.0){\rule[-0.200pt]{0.400pt}{4.818pt}}
\put(1005,68){\makebox(0,0){0.6}}
\put(1005.0,1112.0){\rule[-0.200pt]{0.400pt}{4.818pt}}
\put(1136.0,113.0){\rule[-0.200pt]{0.400pt}{4.818pt}}
\put(1136,68){\makebox(0,0){0.7}}
\put(1136.0,1112.0){\rule[-0.200pt]{0.400pt}{4.818pt}}
\put(220.0,113.0){\rule[-0.200pt]{220.664pt}{0.400pt}}
\put(1136.0,113.0){\rule[-0.200pt]{0.400pt}{245.477pt}}
\put(220.0,1132.0){\rule[-0.200pt]{220.664pt}{0.400pt}}
\put(45,622){\makebox(0,0){\shortstack{$M^{*}_{N}$\\(MeV)}}}
\put(678,23){\makebox(0,0){$\rho_{B}$ (fm$^{-3})$}}
\put(678,1177){\makebox(0,0){Fig. 6(a)}}
\put(678,1050){\makebox(0,0)[l]{Cold nuclear matter}}
\put(220.0,113.0){\rule[-0.200pt]{0.400pt}{245.477pt}}
\put(940,989){\makebox(0,0)[r]{$\epsilon_{\mbox{vac}}=(200 \mbox{MeV})^{4}$}}
\put(962.0,989.0){\rule[-0.200pt]{15.899pt}{0.400pt}}
\put(221,1006){\usebox{\plotpoint}}
\multiput(221.58,1001.54)(0.499,-1.222){125}{\rule{0.120pt}{1.075pt}}
\multiput(220.17,1003.77)(64.000,-153.769){2}{\rule{0.400pt}{0.538pt}}
\multiput(285.58,846.52)(0.499,-0.926){129}{\rule{0.120pt}{0.839pt}}
\multiput(284.17,848.26)(66.000,-120.258){2}{\rule{0.400pt}{0.420pt}}
\multiput(351.58,724.77)(0.493,-0.853){23}{\rule{0.119pt}{0.777pt}}
\multiput(350.17,726.39)(13.000,-20.387){2}{\rule{0.400pt}{0.388pt}}
\multiput(364.58,703.03)(0.493,-0.774){23}{\rule{0.119pt}{0.715pt}}
\multiput(363.17,704.52)(13.000,-18.515){2}{\rule{0.400pt}{0.358pt}}
\multiput(377.58,683.03)(0.493,-0.774){23}{\rule{0.119pt}{0.715pt}}
\multiput(376.17,684.52)(13.000,-18.515){2}{\rule{0.400pt}{0.358pt}}
\multiput(390.58,663.16)(0.493,-0.734){23}{\rule{0.119pt}{0.685pt}}
\multiput(389.17,664.58)(13.000,-17.579){2}{\rule{0.400pt}{0.342pt}}
\multiput(403.58,644.16)(0.493,-0.734){23}{\rule{0.119pt}{0.685pt}}
\multiput(402.17,645.58)(13.000,-17.579){2}{\rule{0.400pt}{0.342pt}}
\multiput(416.58,625.41)(0.493,-0.655){23}{\rule{0.119pt}{0.623pt}}
\multiput(415.17,626.71)(13.000,-15.707){2}{\rule{0.400pt}{0.312pt}}
\multiput(429.58,608.29)(0.493,-0.695){23}{\rule{0.119pt}{0.654pt}}
\multiput(428.17,609.64)(13.000,-16.643){2}{\rule{0.400pt}{0.327pt}}
\multiput(442.58,590.57)(0.494,-0.607){25}{\rule{0.119pt}{0.586pt}}
\multiput(441.17,591.78)(14.000,-15.784){2}{\rule{0.400pt}{0.293pt}}
\multiput(456.58,573.54)(0.493,-0.616){23}{\rule{0.119pt}{0.592pt}}
\multiput(455.17,574.77)(13.000,-14.771){2}{\rule{0.400pt}{0.296pt}}
\multiput(469.58,557.54)(0.493,-0.616){23}{\rule{0.119pt}{0.592pt}}
\multiput(468.17,558.77)(13.000,-14.771){2}{\rule{0.400pt}{0.296pt}}
\multiput(482.58,541.54)(0.493,-0.616){23}{\rule{0.119pt}{0.592pt}}
\multiput(481.17,542.77)(13.000,-14.771){2}{\rule{0.400pt}{0.296pt}}
\multiput(495.58,525.67)(0.493,-0.576){23}{\rule{0.119pt}{0.562pt}}
\multiput(494.17,526.83)(13.000,-13.834){2}{\rule{0.400pt}{0.281pt}}
\multiput(508.59,510.37)(0.482,-0.671){9}{\rule{0.116pt}{0.633pt}}
\multiput(507.17,511.69)(6.000,-6.685){2}{\rule{0.400pt}{0.317pt}}
\multiput(514.58,502.76)(0.496,-0.549){37}{\rule{0.119pt}{0.540pt}}
\multiput(513.17,503.88)(20.000,-20.879){2}{\rule{0.400pt}{0.270pt}}
\multiput(534.58,480.67)(0.493,-0.576){23}{\rule{0.119pt}{0.562pt}}
\multiput(533.17,481.83)(13.000,-13.834){2}{\rule{0.400pt}{0.281pt}}
\multiput(547.58,465.75)(0.499,-0.553){129}{\rule{0.120pt}{0.542pt}}
\multiput(546.17,466.87)(66.000,-71.874){2}{\rule{0.400pt}{0.271pt}}
\multiput(613.58,392.51)(0.497,-0.625){61}{\rule{0.120pt}{0.600pt}}
\multiput(612.17,393.75)(32.000,-38.755){2}{\rule{0.400pt}{0.300pt}}
\multiput(645.58,351.97)(0.497,-0.790){63}{\rule{0.120pt}{0.730pt}}
\multiput(644.17,353.48)(33.000,-50.484){2}{\rule{0.400pt}{0.365pt}}
\put(940,944){\makebox(0,0)[r]{$\epsilon_{\mbox{vac}}=(250 \mbox{MeV})^{4}$}}
\multiput(962,944)(20.756,0.000){4}{\usebox{\plotpoint}}
\put(1028,944){\usebox{\plotpoint}}
\put(221,1006){\usebox{\plotpoint}}
\multiput(221,1006)(7.656,-19.292){4}{\usebox{\plotpoint}}
\multiput(246,943)(9.247,-18.582){11}{\usebox{\plotpoint}}
\multiput(351,732)(11.154,-17.504){6}{\usebox{\plotpoint}}
\multiput(416,630)(12.636,-16.466){5}{\usebox{\plotpoint}}
\multiput(482,544)(13.593,-15.685){5}{\usebox{\plotpoint}}
\multiput(547,469)(14.676,-14.676){4}{\usebox{\plotpoint}}
\multiput(613,403)(15.369,-13.950){5}{\usebox{\plotpoint}}
\multiput(678,344)(15.965,-13.263){4}{\usebox{\plotpoint}}
\multiput(743,290)(16.544,-12.533){4}{\usebox{\plotpoint}}
\multiput(809,240)(16.696,-12.330){4}{\usebox{\plotpoint}}
\multiput(874,192)(16.544,-12.533){4}{\usebox{\plotpoint}}
\multiput(940,142)(15.157,-14.179){2}{\usebox{\plotpoint}}
\put(971,113){\usebox{\plotpoint}}
\sbox{\plotpoint}{\rule[-0.400pt]{0.800pt}{0.800pt}}%
\put(940,899){\makebox(0,0)[r]{$\epsilon_{\mbox{vac}}=(300 \mbox{MeV})^{4}$}}
\put(962.0,899.0){\rule[-0.400pt]{15.899pt}{0.800pt}}
\put(221,1006){\usebox{\plotpoint}}
\multiput(222.41,997.44)(0.501,-1.170){121}{\rule{0.121pt}{2.063pt}}
\multiput(219.34,1001.72)(64.000,-144.719){2}{\rule{0.800pt}{1.031pt}}
\multiput(286.41,849.93)(0.501,-0.943){125}{\rule{0.121pt}{1.703pt}}
\multiput(283.34,853.47)(66.000,-120.465){2}{\rule{0.800pt}{0.852pt}}
\multiput(352.41,726.96)(0.501,-0.787){123}{\rule{0.121pt}{1.455pt}}
\multiput(349.34,729.98)(65.000,-98.979){2}{\rule{0.800pt}{0.728pt}}
\multiput(417.41,625.79)(0.501,-0.660){125}{\rule{0.121pt}{1.255pt}}
\multiput(414.34,628.40)(66.000,-84.396){2}{\rule{0.800pt}{0.627pt}}
\multiput(483.41,539.34)(0.501,-0.577){123}{\rule{0.121pt}{1.123pt}}
\multiput(480.34,541.67)(65.000,-72.669){2}{\rule{0.800pt}{0.562pt}}
\multiput(547.00,467.09)(0.507,-0.501){123}{\rule{1.012pt}{0.121pt}}
\multiput(547.00,467.34)(63.899,-65.000){2}{\rule{0.506pt}{0.800pt}}
\multiput(613.00,402.09)(0.613,-0.501){205}{\rule{1.181pt}{0.121pt}}
\multiput(613.00,402.34)(127.549,-106.000){2}{\rule{0.591pt}{0.800pt}}
\multiput(743.00,296.09)(0.772,-0.501){163}{\rule{1.433pt}{0.121pt}}
\multiput(743.00,296.34)(128.026,-85.000){2}{\rule{0.716pt}{0.800pt}}
\multiput(874.00,211.09)(0.953,-0.501){131}{\rule{1.719pt}{0.121pt}}
\multiput(874.00,211.34)(127.432,-69.000){2}{\rule{0.859pt}{0.800pt}}
\multiput(1005.00,142.09)(1.142,-0.503){55}{\rule{2.006pt}{0.121pt}}
\multiput(1005.00,142.34)(65.836,-31.000){2}{\rule{1.003pt}{0.800pt}}
\sbox{\plotpoint}{\rule[-0.500pt]{1.000pt}{1.000pt}}%
\put(940,854){\makebox(0,0)[r]{$\epsilon_{\mbox{vac}}=(350 \mbox{MeV})^{4}$}}
\multiput(962,854)(20.756,0.000){4}{\usebox{\plotpoint}}
\put(1028,854){\usebox{\plotpoint}}
\put(221,1006){\usebox{\plotpoint}}
\multiput(221,1006)(8.238,-19.051){8}{\usebox{\plotpoint}}
\multiput(285,858)(9.752,-18.322){7}{\usebox{\plotpoint}}
\multiput(351,734)(11.077,-17.553){6}{\usebox{\plotpoint}}
\multiput(416,631)(12.544,-16.536){5}{\usebox{\plotpoint}}
\multiput(482,544)(13.593,-15.685){5}{\usebox{\plotpoint}}
\multiput(547,469)(14.900,-14.449){4}{\usebox{\plotpoint}}
\multiput(613,405)(16.147,-13.041){8}{\usebox{\plotpoint}}
\multiput(743,300)(17.593,-11.012){8}{\usebox{\plotpoint}}
\multiput(874,218)(18.593,-9.225){7}{\usebox{\plotpoint}}
\multiput(1005,153)(19.271,-7.708){5}{\usebox{\plotpoint}}
\put(1105,113){\usebox{\plotpoint}}
\sbox{\plotpoint}{\rule[-0.600pt]{1.200pt}{1.200pt}}%
\put(940,809){\makebox(0,0)[r]{$\epsilon_{\mbox{vac}}=(800 \mbox{MeV})^{4}$}}
\put(962.0,809.0){\rule[-0.600pt]{15.899pt}{1.200pt}}
\put(233,977){\usebox{\plotpoint}}
\multiput(235.24,965.50)(0.500,-1.029){226}{\rule{0.120pt}{2.771pt}}
\multiput(230.51,971.25)(118.000,-237.248){2}{\rule{1.200pt}{1.386pt}}
\multiput(353.24,724.32)(0.501,-0.822){16}{\rule{0.121pt}{2.331pt}}
\multiput(348.51,729.16)(13.000,-17.162){2}{\rule{1.200pt}{1.165pt}}
\multiput(366.24,702.71)(0.501,-0.781){16}{\rule{0.121pt}{2.238pt}}
\multiput(361.51,707.35)(13.000,-16.354){2}{\rule{1.200pt}{1.119pt}}
\multiput(379.24,681.71)(0.501,-0.781){16}{\rule{0.121pt}{2.238pt}}
\multiput(374.51,686.35)(13.000,-16.354){2}{\rule{1.200pt}{1.119pt}}
\multiput(392.24,661.09)(0.501,-0.739){16}{\rule{0.121pt}{2.146pt}}
\multiput(387.51,665.55)(13.000,-15.546){2}{\rule{1.200pt}{1.073pt}}
\multiput(405.24,641.47)(0.501,-0.698){16}{\rule{0.121pt}{2.054pt}}
\multiput(400.51,645.74)(13.000,-14.737){2}{\rule{1.200pt}{1.027pt}}
\multiput(418.24,622.47)(0.501,-0.698){16}{\rule{0.121pt}{2.054pt}}
\multiput(413.51,626.74)(13.000,-14.737){2}{\rule{1.200pt}{1.027pt}}
\multiput(431.24,603.86)(0.501,-0.657){16}{\rule{0.121pt}{1.962pt}}
\multiput(426.51,607.93)(13.000,-13.929){2}{\rule{1.200pt}{0.981pt}}
\multiput(444.24,586.71)(0.501,-0.571){18}{\rule{0.121pt}{1.757pt}}
\multiput(439.51,590.35)(14.000,-13.353){2}{\rule{1.200pt}{0.879pt}}
\multiput(458.24,569.24)(0.501,-0.616){16}{\rule{0.121pt}{1.869pt}}
\multiput(453.51,573.12)(13.000,-13.120){2}{\rule{1.200pt}{0.935pt}}
\multiput(471.24,552.62)(0.501,-0.575){16}{\rule{0.121pt}{1.777pt}}
\multiput(466.51,556.31)(13.000,-12.312){2}{\rule{1.200pt}{0.888pt}}
\multiput(484.24,536.62)(0.501,-0.575){16}{\rule{0.121pt}{1.777pt}}
\multiput(479.51,540.31)(13.000,-12.312){2}{\rule{1.200pt}{0.888pt}}
\multiput(497.24,521.01)(0.501,-0.534){16}{\rule{0.121pt}{1.685pt}}
\multiput(492.51,524.50)(13.000,-11.503){2}{\rule{1.200pt}{0.842pt}}
\multiput(510.24,506.01)(0.501,-0.534){16}{\rule{0.121pt}{1.685pt}}
\multiput(505.51,509.50)(13.000,-11.503){2}{\rule{1.200pt}{0.842pt}}
\multiput(523.24,491.01)(0.501,-0.534){16}{\rule{0.121pt}{1.685pt}}
\multiput(518.51,494.50)(13.000,-11.503){2}{\rule{1.200pt}{0.842pt}}
\multiput(536.24,476.39)(0.501,-0.493){16}{\rule{0.121pt}{1.592pt}}
\multiput(531.51,479.70)(13.000,-10.695){2}{\rule{1.200pt}{0.796pt}}
\multiput(547.00,466.26)(0.452,-0.501){16}{\rule{1.500pt}{0.121pt}}
\multiput(547.00,466.51)(9.887,-13.000){2}{\rule{0.750pt}{1.200pt}}
\multiput(560.00,453.26)(0.452,-0.501){16}{\rule{1.500pt}{0.121pt}}
\multiput(560.00,453.51)(9.887,-13.000){2}{\rule{0.750pt}{1.200pt}}
\multiput(573.00,440.26)(0.452,-0.501){16}{\rule{1.500pt}{0.121pt}}
\multiput(573.00,440.51)(9.887,-13.000){2}{\rule{0.750pt}{1.200pt}}
\multiput(586.00,427.26)(0.452,-0.501){16}{\rule{1.500pt}{0.121pt}}
\multiput(586.00,427.51)(9.887,-13.000){2}{\rule{0.750pt}{1.200pt}}
\multiput(599.00,414.26)(0.534,-0.501){14}{\rule{1.700pt}{0.121pt}}
\multiput(599.00,414.51)(10.472,-12.000){2}{\rule{0.850pt}{1.200pt}}
\multiput(613.00,402.26)(0.628,-0.500){196}{\rule{1.815pt}{0.120pt}}
\multiput(613.00,402.51)(126.234,-103.000){2}{\rule{0.907pt}{1.200pt}}
\multiput(743.00,299.26)(0.816,-0.500){150}{\rule{2.265pt}{0.120pt}}
\multiput(743.00,299.51)(126.299,-80.000){2}{\rule{1.133pt}{1.200pt}}
\multiput(874.00,219.26)(1.055,-0.500){114}{\rule{2.835pt}{0.120pt}}
\multiput(874.00,219.51)(125.115,-62.000){2}{\rule{1.418pt}{1.200pt}}
\multiput(1005.00,157.26)(1.321,-0.500){84}{\rule{3.466pt}{0.121pt}}
\multiput(1005.00,157.51)(116.806,-47.000){2}{\rule{1.733pt}{1.200pt}}
\end{picture}

%% file: fa6b.tex
\setlength{\unitlength}{0.240900pt}
\ifx\plotpoint\undefined\newsavebox{\plotpoint}\fi
\sbox{\plotpoint}{\rule[-0.200pt]{0.400pt}{0.400pt}}%
\begin{picture}(1200,1200)(0,0)
\font\gnuplot=cmr10 at 10pt
\gnuplot
\sbox{\plotpoint}{\rule[-0.200pt]{0.400pt}{0.400pt}}%
\put(220.0,113.0){\rule[-0.200pt]{0.400pt}{245.477pt}}
\put(220.0,113.0){\rule[-0.200pt]{4.818pt}{0.400pt}}
\put(198,113){\makebox(0,0)[r]{500}}
\put(1116.0,113.0){\rule[-0.200pt]{4.818pt}{0.400pt}}
\put(220.0,317.0){\rule[-0.200pt]{4.818pt}{0.400pt}}
\put(198,317){\makebox(0,0)[r]{600}}
\put(1116.0,317.0){\rule[-0.200pt]{4.818pt}{0.400pt}}
\put(220.0,521.0){\rule[-0.200pt]{4.818pt}{0.400pt}}
\put(198,521){\makebox(0,0)[r]{700}}
\put(1116.0,521.0){\rule[-0.200pt]{4.818pt}{0.400pt}}
\put(220.0,724.0){\rule[-0.200pt]{4.818pt}{0.400pt}}
\put(198,724){\makebox(0,0)[r]{800}}
\put(1116.0,724.0){\rule[-0.200pt]{4.818pt}{0.400pt}}
\put(220.0,928.0){\rule[-0.200pt]{4.818pt}{0.400pt}}
\put(198,928){\makebox(0,0)[r]{900}}
\put(1116.0,928.0){\rule[-0.200pt]{4.818pt}{0.400pt}}
\put(220.0,1132.0){\rule[-0.200pt]{4.818pt}{0.400pt}}
\put(198,1132){\makebox(0,0)[r]{1000}}
\put(1116.0,1132.0){\rule[-0.200pt]{4.818pt}{0.400pt}}
\put(220.0,113.0){\rule[-0.200pt]{0.400pt}{4.818pt}}
\put(220,68){\makebox(0,0){0}}
\put(220.0,1112.0){\rule[-0.200pt]{0.400pt}{4.818pt}}
\put(387.0,113.0){\rule[-0.200pt]{0.400pt}{4.818pt}}
\put(387,68){\makebox(0,0){0.1}}
\put(387.0,1112.0){\rule[-0.200pt]{0.400pt}{4.818pt}}
\put(553.0,113.0){\rule[-0.200pt]{0.400pt}{4.818pt}}
\put(553,68){\makebox(0,0){0.2}}
\put(553.0,1112.0){\rule[-0.200pt]{0.400pt}{4.818pt}}
\put(720.0,113.0){\rule[-0.200pt]{0.400pt}{4.818pt}}
\put(720,68){\makebox(0,0){0.3}}
\put(720.0,1112.0){\rule[-0.200pt]{0.400pt}{4.818pt}}
\put(886.0,113.0){\rule[-0.200pt]{0.400pt}{4.818pt}}
\put(886,68){\makebox(0,0){0.4}}
\put(886.0,1112.0){\rule[-0.200pt]{0.400pt}{4.818pt}}
\put(1053.0,113.0){\rule[-0.200pt]{0.400pt}{4.818pt}}
\put(1053,68){\makebox(0,0){0.5}}
\put(1053.0,1112.0){\rule[-0.200pt]{0.400pt}{4.818pt}}
\put(220.0,113.0){\rule[-0.200pt]{220.664pt}{0.400pt}}
\put(1136.0,113.0){\rule[-0.200pt]{0.400pt}{245.477pt}}
\put(220.0,1132.0){\rule[-0.200pt]{220.664pt}{0.400pt}}
\put(45,622){\makebox(0,0){\shortstack{$M^{*}_{N}$\\ (MeV)}}}
\put(678,23){\makebox(0,0){$\rho_{B}$ (fm$^{-3})$}}
\put(678,1177){\makebox(0,0){Fig. 6(b)}}
\put(636,1091){\makebox(0,0)[l]{Temperature 200 MeV}}
\put(220.0,113.0){\rule[-0.200pt]{0.400pt}{245.477pt}}
\put(969,1030){\makebox(0,0)[r]{$\epsilon_{\mbox{vac}}=(200 \mbox{MeV})^{4}$}}
\put(991.0,1030.0){\rule[-0.200pt]{15.899pt}{0.400pt}}
\put(222,987){\usebox{\plotpoint}}
\put(222,985.17){\rule{3.100pt}{0.400pt}}
\multiput(222.00,986.17)(8.566,-2.000){2}{\rule{1.550pt}{0.400pt}}
\multiput(237.00,983.92)(0.758,-0.499){195}{\rule{0.706pt}{0.120pt}}
\multiput(237.00,984.17)(148.535,-99.000){2}{\rule{0.353pt}{0.400pt}}
\multiput(387.00,884.92)(0.610,-0.499){133}{\rule{0.588pt}{0.120pt}}
\multiput(387.00,885.17)(81.779,-68.000){2}{\rule{0.294pt}{0.400pt}}
\multiput(470.00,816.92)(0.639,-0.499){127}{\rule{0.611pt}{0.120pt}}
\multiput(470.00,817.17)(81.732,-65.000){2}{\rule{0.305pt}{0.400pt}}
\multiput(553.00,751.92)(0.692,-0.499){117}{\rule{0.653pt}{0.120pt}}
\multiput(553.00,752.17)(81.644,-60.000){2}{\rule{0.327pt}{0.400pt}}
\multiput(636.00,691.92)(0.738,-0.499){111}{\rule{0.689pt}{0.120pt}}
\multiput(636.00,692.17)(82.569,-57.000){2}{\rule{0.345pt}{0.400pt}}
\multiput(720.00,634.92)(0.741,-0.499){221}{\rule{0.693pt}{0.120pt}}
\multiput(720.00,635.17)(164.562,-112.000){2}{\rule{0.346pt}{0.400pt}}
\multiput(886.00,522.92)(0.690,-0.499){239}{\rule{0.652pt}{0.120pt}}
\multiput(886.00,523.17)(165.647,-121.000){2}{\rule{0.326pt}{0.400pt}}
\multiput(1053.00,401.92)(0.539,-0.499){151}{\rule{0.531pt}{0.120pt}}
\multiput(1053.00,402.17)(81.898,-77.000){2}{\rule{0.266pt}{0.400pt}}
\put(969,985){\makebox(0,0)[r]{$\epsilon_{\mbox{vac}}=(250 \mbox{MeV})^{4}$}}
\multiput(991,985)(20.756,0.000){4}{\usebox{\plotpoint}}
\put(1057,985){\usebox{\plotpoint}}
\put(222,1005){\usebox{\plotpoint}}
\put(222.00,1005.00){\usebox{\plotpoint}}
\multiput(237,1004)(17.693,-10.852){9}{\usebox{\plotpoint}}
\multiput(387,912)(16.821,-12.160){10}{\usebox{\plotpoint}}
\multiput(553,792)(17.807,-10.663){9}{\usebox{\plotpoint}}
\multiput(720,692)(18.519,-9.371){9}{\usebox{\plotpoint}}
\multiput(886,608)(19.060,-8.217){9}{\usebox{\plotpoint}}
\multiput(1053,536)(19.287,-7.668){4}{\usebox{\plotpoint}}
\put(1136,503){\usebox{\plotpoint}}
\sbox{\plotpoint}{\rule[-0.400pt]{0.800pt}{0.800pt}}%
\put(969,940){\makebox(0,0)[r]{$\epsilon_{\mbox{vac}}=(300 \mbox{MeV})^{4}$}}
\put(991.0,940.0){\rule[-0.400pt]{15.899pt}{0.800pt}}
\put(222,1011){\usebox{\plotpoint}}
\multiput(222.00,1009.06)(4.790,-0.560){3}{\rule{5.160pt}{0.135pt}}
\multiput(222.00,1009.34)(20.290,-5.000){2}{\rule{2.580pt}{0.800pt}}
\multiput(253.00,1004.09)(0.766,-0.501){189}{\rule{1.424pt}{0.121pt}}
\multiput(253.00,1004.34)(147.043,-98.000){2}{\rule{0.712pt}{0.800pt}}
\multiput(403.00,906.09)(0.727,-0.501){223}{\rule{1.362pt}{0.121pt}}
\multiput(403.00,906.34)(164.174,-115.000){2}{\rule{0.681pt}{0.800pt}}
\multiput(570.00,791.09)(0.864,-0.501){167}{\rule{1.579pt}{0.121pt}}
\multiput(570.00,791.34)(146.722,-87.000){2}{\rule{0.790pt}{0.800pt}}
\multiput(720.00,704.09)(1.041,-0.501){153}{\rule{1.860pt}{0.121pt}}
\multiput(720.00,704.34)(162.139,-80.000){2}{\rule{0.930pt}{0.800pt}}
\multiput(886.00,624.09)(1.253,-0.501){127}{\rule{2.194pt}{0.121pt}}
\multiput(886.00,624.34)(162.446,-67.000){2}{\rule{1.097pt}{0.800pt}}
\multiput(1053.00,557.09)(1.404,-0.503){53}{\rule{2.413pt}{0.121pt}}
\multiput(1053.00,557.34)(77.991,-30.000){2}{\rule{1.207pt}{0.800pt}}
\sbox{\plotpoint}{\rule[-0.500pt]{1.000pt}{1.000pt}}%
\put(969,895){\makebox(0,0)[r]{$\epsilon_{\mbox{vac}}=(350 \mbox{MeV})^{4}$}}
\multiput(991,895)(20.756,0.000){4}{\usebox{\plotpoint}}
\put(1057,895){\usebox{\plotpoint}}
\put(222,1014){\usebox{\plotpoint}}
\multiput(222,1014)(19.384,-7.419){5}{\usebox{\plotpoint}}
\multiput(303,983)(16.844,-12.128){6}{\usebox{\plotpoint}}
\multiput(403,911)(16.992,-11.920){4}{\usebox{\plotpoint}}
\multiput(470,864)(17.206,-11.609){4}{\usebox{\plotpoint}}
\multiput(553,808)(17.994,-10.344){10}{\usebox{\plotpoint}}
\multiput(720,712)(18.741,-8.919){9}{\usebox{\plotpoint}}
\multiput(886,633)(19.263,-7.728){8}{\usebox{\plotpoint}}
\multiput(1053,566)(19.667,-6.635){4}{\usebox{\plotpoint}}
\put(1136,538){\usebox{\plotpoint}}
\sbox{\plotpoint}{\rule[-0.600pt]{1.200pt}{1.200pt}}%
\put(969,850){\makebox(0,0)[r]{$\epsilon_{\mbox{vac}}=(800 \mbox{MeV})^{4}$}}
\put(991.0,850.0){\rule[-0.600pt]{15.899pt}{1.200pt}}
\put(222,1017){\usebox{\plotpoint}}
\put(222,1013.51){\rule{3.614pt}{1.200pt}}
\multiput(222.00,1014.51)(7.500,-2.000){2}{\rule{1.807pt}{1.200pt}}
\multiput(237.00,1012.26)(0.850,-0.500){166}{\rule{2.345pt}{0.120pt}}
\multiput(237.00,1012.51)(145.132,-88.000){2}{\rule{1.173pt}{1.200pt}}
\multiput(387.00,924.26)(0.699,-0.500){108}{\rule{1.988pt}{0.120pt}}
\multiput(387.00,924.51)(78.874,-59.000){2}{\rule{0.994pt}{1.200pt}}
\multiput(470.00,865.26)(0.750,-0.500){100}{\rule{2.111pt}{0.120pt}}
\multiput(470.00,865.51)(78.619,-55.000){2}{\rule{1.055pt}{1.200pt}}
\multiput(553.00,810.26)(0.809,-0.500){92}{\rule{2.253pt}{0.120pt}}
\multiput(553.00,810.51)(78.324,-51.000){2}{\rule{1.126pt}{1.200pt}}
\multiput(636.00,759.26)(0.930,-0.500){80}{\rule{2.540pt}{0.121pt}}
\multiput(636.00,759.51)(78.728,-45.000){2}{\rule{1.270pt}{1.200pt}}
\multiput(720.00,714.26)(1.063,-0.500){146}{\rule{2.854pt}{0.120pt}}
\multiput(720.00,714.51)(160.077,-78.000){2}{\rule{1.427pt}{1.200pt}}
\multiput(886.00,636.26)(1.285,-0.500){120}{\rule{3.383pt}{0.120pt}}
\multiput(886.00,636.51)(159.978,-65.000){2}{\rule{1.692pt}{1.200pt}}
\multiput(1053.00,571.26)(1.545,-0.500){44}{\rule{3.989pt}{0.121pt}}
\multiput(1053.00,571.51)(74.721,-27.000){2}{\rule{1.994pt}{1.200pt}}
\end{picture}

%% file: fa7a.tex
\setlength{\unitlength}{0.240900pt}
\ifx\plotpoint\undefined\newsavebox{\plotpoint}\fi
\sbox{\plotpoint}{\rule[-0.200pt]{0.400pt}{0.400pt}}%
\begin{picture}(1200,1200)(0,0)
\font\gnuplot=cmr10 at 10pt
\gnuplot
\sbox{\plotpoint}{\rule[-0.200pt]{0.400pt}{0.400pt}}%
\put(220.0,113.0){\rule[-0.200pt]{0.400pt}{245.477pt}}
\put(220.0,113.0){\rule[-0.200pt]{4.818pt}{0.400pt}}
\put(198,113){\makebox(0,0)[r]{0.5}}
\put(1116.0,113.0){\rule[-0.200pt]{4.818pt}{0.400pt}}
\put(220.0,317.0){\rule[-0.200pt]{4.818pt}{0.400pt}}
\put(198,317){\makebox(0,0)[r]{0.6}}
\put(1116.0,317.0){\rule[-0.200pt]{4.818pt}{0.400pt}}
\put(220.0,521.0){\rule[-0.200pt]{4.818pt}{0.400pt}}
\put(198,521){\makebox(0,0)[r]{0.7}}
\put(1116.0,521.0){\rule[-0.200pt]{4.818pt}{0.400pt}}
\put(220.0,724.0){\rule[-0.200pt]{4.818pt}{0.400pt}}
\put(198,724){\makebox(0,0)[r]{0.8}}
\put(1116.0,724.0){\rule[-0.200pt]{4.818pt}{0.400pt}}
\put(220.0,928.0){\rule[-0.200pt]{4.818pt}{0.400pt}}
\put(198,928){\makebox(0,0)[r]{0.9}}
\put(1116.0,928.0){\rule[-0.200pt]{4.818pt}{0.400pt}}
\put(220.0,1132.0){\rule[-0.200pt]{4.818pt}{0.400pt}}
\put(198,1132){\makebox(0,0)[r]{1.0}}
\put(1116.0,1132.0){\rule[-0.200pt]{4.818pt}{0.400pt}}
\put(220.0,113.0){\rule[-0.200pt]{0.400pt}{4.818pt}}
\put(220,68){\makebox(0,0){0}}
\put(220.0,1112.0){\rule[-0.200pt]{0.400pt}{4.818pt}}
\put(351.0,113.0){\rule[-0.200pt]{0.400pt}{4.818pt}}
\put(351,68){\makebox(0,0){0.1}}
\put(351.0,1112.0){\rule[-0.200pt]{0.400pt}{4.818pt}}
\put(482.0,113.0){\rule[-0.200pt]{0.400pt}{4.818pt}}
\put(482,68){\makebox(0,0){0.2}}
\put(482.0,1112.0){\rule[-0.200pt]{0.400pt}{4.818pt}}
\put(613.0,113.0){\rule[-0.200pt]{0.400pt}{4.818pt}}
\put(613,68){\makebox(0,0){0.3}}
\put(613.0,1112.0){\rule[-0.200pt]{0.400pt}{4.818pt}}
\put(743.0,113.0){\rule[-0.200pt]{0.400pt}{4.818pt}}
\put(743,68){\makebox(0,0){0.4}}
\put(743.0,1112.0){\rule[-0.200pt]{0.400pt}{4.818pt}}
\put(874.0,113.0){\rule[-0.200pt]{0.400pt}{4.818pt}}
\put(874,68){\makebox(0,0){0.5}}
\put(874.0,1112.0){\rule[-0.200pt]{0.400pt}{4.818pt}}
\put(1005.0,113.0){\rule[-0.200pt]{0.400pt}{4.818pt}}
\put(1005,68){\makebox(0,0){0.6}}
\put(1005.0,1112.0){\rule[-0.200pt]{0.400pt}{4.818pt}}
\put(1136.0,113.0){\rule[-0.200pt]{0.400pt}{4.818pt}}
\put(1136,68){\makebox(0,0){0.7}}
\put(1136.0,1112.0){\rule[-0.200pt]{0.400pt}{4.818pt}}
\put(220.0,113.0){\rule[-0.200pt]{220.664pt}{0.400pt}}
\put(1136.0,113.0){\rule[-0.200pt]{0.400pt}{245.477pt}}
\put(220.0,1132.0){\rule[-0.200pt]{220.664pt}{0.400pt}}
\put(45,622){\makebox(0,0){\shortstack{$R$\\ (fm)}}}
\put(678,23){\makebox(0,0){$\rho_{B}$ (fm$^{-3})$}}
\put(678,1177){\makebox(0,0){Fig. 7(a)}}
\put(743,700){\makebox(0,0)[l]{Cold nuclear matter}}
\put(220.0,113.0){\rule[-0.200pt]{0.400pt}{245.477pt}}
\put(1005,620){\makebox(0,0)[r]{$\epsilon_{\mbox{vac}}=(200 \mbox{MeV})^{4}$}}
\put(1027.0,620.0){\rule[-0.200pt]{15.899pt}{0.400pt}}
\put(221,352){\usebox{\plotpoint}}
\multiput(221.58,352.00)(0.499,0.539){125}{\rule{0.120pt}{0.531pt}}
\multiput(220.17,352.00)(64.000,67.897){2}{\rule{0.400pt}{0.266pt}}
\multiput(285.58,421.00)(0.499,0.507){129}{\rule{0.120pt}{0.506pt}}
\multiput(284.17,421.00)(66.000,65.950){2}{\rule{0.400pt}{0.253pt}}
\multiput(351.58,488.00)(0.493,0.536){23}{\rule{0.119pt}{0.531pt}}
\multiput(350.17,488.00)(13.000,12.898){2}{\rule{0.400pt}{0.265pt}}
\multiput(364.58,502.00)(0.493,0.536){23}{\rule{0.119pt}{0.531pt}}
\multiput(363.17,502.00)(13.000,12.898){2}{\rule{0.400pt}{0.265pt}}
\multiput(377.00,516.58)(0.497,0.493){23}{\rule{0.500pt}{0.119pt}}
\multiput(377.00,515.17)(11.962,13.000){2}{\rule{0.250pt}{0.400pt}}
\multiput(390.58,529.00)(0.493,0.576){23}{\rule{0.119pt}{0.562pt}}
\multiput(389.17,529.00)(13.000,13.834){2}{\rule{0.400pt}{0.281pt}}
\multiput(403.58,544.00)(0.493,0.536){23}{\rule{0.119pt}{0.531pt}}
\multiput(402.17,544.00)(13.000,12.898){2}{\rule{0.400pt}{0.265pt}}
\multiput(416.58,558.00)(0.493,0.536){23}{\rule{0.119pt}{0.531pt}}
\multiput(415.17,558.00)(13.000,12.898){2}{\rule{0.400pt}{0.265pt}}
\multiput(429.58,572.00)(0.493,0.576){23}{\rule{0.119pt}{0.562pt}}
\multiput(428.17,572.00)(13.000,13.834){2}{\rule{0.400pt}{0.281pt}}
\multiput(442.58,587.00)(0.494,0.534){25}{\rule{0.119pt}{0.529pt}}
\multiput(441.17,587.00)(14.000,13.903){2}{\rule{0.400pt}{0.264pt}}
\multiput(456.58,602.00)(0.493,0.616){23}{\rule{0.119pt}{0.592pt}}
\multiput(455.17,602.00)(13.000,14.771){2}{\rule{0.400pt}{0.296pt}}
\multiput(469.58,618.00)(0.493,0.616){23}{\rule{0.119pt}{0.592pt}}
\multiput(468.17,618.00)(13.000,14.771){2}{\rule{0.400pt}{0.296pt}}
\multiput(482.58,634.00)(0.493,0.655){23}{\rule{0.119pt}{0.623pt}}
\multiput(481.17,634.00)(13.000,15.707){2}{\rule{0.400pt}{0.312pt}}
\multiput(495.58,651.00)(0.493,0.655){23}{\rule{0.119pt}{0.623pt}}
\multiput(494.17,651.00)(13.000,15.707){2}{\rule{0.400pt}{0.312pt}}
\multiput(508.59,668.00)(0.482,0.762){9}{\rule{0.116pt}{0.700pt}}
\multiput(507.17,668.00)(6.000,7.547){2}{\rule{0.400pt}{0.350pt}}
\multiput(514.58,677.00)(0.496,0.702){37}{\rule{0.119pt}{0.660pt}}
\multiput(513.17,677.00)(20.000,26.630){2}{\rule{0.400pt}{0.330pt}}
\multiput(534.58,705.00)(0.493,0.814){23}{\rule{0.119pt}{0.746pt}}
\multiput(533.17,705.00)(13.000,19.451){2}{\rule{0.400pt}{0.373pt}}
\multiput(547.58,726.00)(0.499,0.934){129}{\rule{0.120pt}{0.845pt}}
\multiput(546.17,726.00)(66.000,121.245){2}{\rule{0.400pt}{0.423pt}}
\multiput(613.58,849.00)(0.497,1.415){61}{\rule{0.120pt}{1.225pt}}
\multiput(612.17,849.00)(32.000,87.457){2}{\rule{0.400pt}{0.613pt}}
\multiput(645.58,939.00)(0.497,2.276){63}{\rule{0.120pt}{1.906pt}}
\multiput(644.17,939.00)(33.000,145.044){2}{\rule{0.400pt}{0.953pt}}
\put(1005,575){\makebox(0,0)[r]{$\epsilon_{\mbox{vac}}=(250 \mbox{MeV})^{4}$}}
\multiput(1027,575)(20.756,0.000){4}{\usebox{\plotpoint}}
\put(1093,575){\usebox{\plotpoint}}
\put(221,352){\usebox{\plotpoint}}
\multiput(221,352)(13.029,16.156){2}{\usebox{\plotpoint}}
\multiput(246,383)(13.603,15.676){8}{\usebox{\plotpoint}}
\multiput(351,504)(14.015,15.309){5}{\usebox{\plotpoint}}
\multiput(416,575)(14.131,15.202){4}{\usebox{\plotpoint}}
\multiput(482,646)(14.015,15.309){5}{\usebox{\plotpoint}}
\multiput(547,717)(13.815,15.490){5}{\usebox{\plotpoint}}
\multiput(613,791)(13.388,15.860){5}{\usebox{\plotpoint}}
\multiput(678,868)(12.702,16.415){5}{\usebox{\plotpoint}}
\multiput(743,952)(11.927,16.987){5}{\usebox{\plotpoint}}
\multiput(809,1046)(10.587,17.852){5}{\usebox{\plotpoint}}
\put(860,1132){\usebox{\plotpoint}}
\sbox{\plotpoint}{\rule[-0.400pt]{0.800pt}{0.800pt}}%
\put(1005,530){\makebox(0,0)[r]{$\epsilon_{\mbox{vac}}=(300 \mbox{MeV})^{4}$}}
\put(1027.0,530.0){\rule[-0.400pt]{15.899pt}{0.800pt}}
\put(221,352){\usebox{\plotpoint}}
\multiput(222.41,352.00)(0.501,0.625){121}{\rule{0.121pt}{1.200pt}}
\multiput(219.34,352.00)(64.000,77.509){2}{\rule{0.800pt}{0.600pt}}
\multiput(286.41,432.00)(0.501,0.583){125}{\rule{0.121pt}{1.133pt}}
\multiput(283.34,432.00)(66.000,74.648){2}{\rule{0.800pt}{0.567pt}}
\multiput(352.41,509.00)(0.501,0.553){123}{\rule{0.121pt}{1.086pt}}
\multiput(349.34,509.00)(65.000,69.746){2}{\rule{0.800pt}{0.543pt}}
\multiput(417.41,581.00)(0.501,0.514){125}{\rule{0.121pt}{1.024pt}}
\multiput(414.34,581.00)(66.000,65.874){2}{\rule{0.800pt}{0.512pt}}
\multiput(483.41,649.00)(0.501,0.514){123}{\rule{0.121pt}{1.025pt}}
\multiput(480.34,649.00)(65.000,64.873){2}{\rule{0.800pt}{0.512pt}}
\multiput(547.00,717.41)(0.515,0.501){121}{\rule{1.025pt}{0.121pt}}
\multiput(547.00,714.34)(63.873,64.000){2}{\rule{0.513pt}{0.800pt}}
\multiput(613.00,781.41)(0.524,0.501){241}{\rule{1.039pt}{0.121pt}}
\multiput(613.00,778.34)(127.844,124.000){2}{\rule{0.519pt}{0.800pt}}
\multiput(743.00,905.41)(0.536,0.501){237}{\rule{1.059pt}{0.121pt}}
\multiput(743.00,902.34)(128.802,122.000){2}{\rule{0.530pt}{0.800pt}}
\multiput(874.00,1027.41)(0.518,0.501){205}{\rule{1.030pt}{0.121pt}}
\multiput(874.00,1024.34)(107.862,106.000){2}{\rule{0.515pt}{0.800pt}}
\sbox{\plotpoint}{\rule[-0.500pt]{1.000pt}{1.000pt}}%
\put(1005,485){\makebox(0,0)[r]{$\epsilon_{\mbox{vac}}=(350 \mbox{MeV})^{4}$}}
\multiput(1027,485)(20.756,0.000){4}{\usebox{\plotpoint}}
\put(1093,485){\usebox{\plotpoint}}
\put(221,352){\usebox{\plotpoint}}
\multiput(221,352)(12.770,16.362){6}{\usebox{\plotpoint}}
\multiput(285,434)(13.508,15.759){4}{\usebox{\plotpoint}}
\multiput(351,511)(13.908,15.406){5}{\usebox{\plotpoint}}
\multiput(416,583)(14.456,14.894){5}{\usebox{\plotpoint}}
\multiput(482,651)(14.790,14.562){4}{\usebox{\plotpoint}}
\multiput(547,715)(15.242,14.088){4}{\usebox{\plotpoint}}
\multiput(613,776)(15.725,13.547){9}{\usebox{\plotpoint}}
\multiput(743,888)(16.256,12.905){8}{\usebox{\plotpoint}}
\multiput(874,992)(16.680,12.351){8}{\usebox{\plotpoint}}
\multiput(1005,1089)(16.870,12.090){3}{\usebox{\plotpoint}}
\put(1065,1132){\usebox{\plotpoint}}
\sbox{\plotpoint}{\rule[-0.600pt]{1.200pt}{1.200pt}}%
\put(1005,440){\makebox(0,0)[r]{$\epsilon_{\mbox{vac}}=(800 \mbox{MeV})^{4}$}}
\put(1027.0,440.0){\rule[-0.600pt]{15.899pt}{1.200pt}}
\put(233,368){\usebox{\plotpoint}}
\multiput(235.24,368.00)(0.500,0.612){226}{\rule{0.120pt}{1.775pt}}
\multiput(230.51,368.00)(118.000,141.317){2}{\rule{1.200pt}{0.887pt}}
\multiput(353.24,513.00)(0.501,0.534){16}{\rule{0.121pt}{1.685pt}}
\multiput(348.51,513.00)(13.000,11.503){2}{\rule{1.200pt}{0.842pt}}
\multiput(366.24,528.00)(0.501,0.534){16}{\rule{0.121pt}{1.685pt}}
\multiput(361.51,528.00)(13.000,11.503){2}{\rule{1.200pt}{0.842pt}}
\multiput(379.24,543.00)(0.501,0.493){16}{\rule{0.121pt}{1.592pt}}
\multiput(374.51,543.00)(13.000,10.695){2}{\rule{1.200pt}{0.796pt}}
\multiput(392.24,557.00)(0.501,0.493){16}{\rule{0.121pt}{1.592pt}}
\multiput(387.51,557.00)(13.000,10.695){2}{\rule{1.200pt}{0.796pt}}
\multiput(405.24,571.00)(0.501,0.493){16}{\rule{0.121pt}{1.592pt}}
\multiput(400.51,571.00)(13.000,10.695){2}{\rule{1.200pt}{0.796pt}}
\multiput(418.24,585.00)(0.501,0.493){16}{\rule{0.121pt}{1.592pt}}
\multiput(413.51,585.00)(13.000,10.695){2}{\rule{1.200pt}{0.796pt}}
\multiput(431.24,599.00)(0.501,0.493){16}{\rule{0.121pt}{1.592pt}}
\multiput(426.51,599.00)(13.000,10.695){2}{\rule{1.200pt}{0.796pt}}
\multiput(442.00,615.24)(0.493,0.501){16}{\rule{1.592pt}{0.121pt}}
\multiput(442.00,610.51)(10.695,13.000){2}{\rule{0.796pt}{1.200pt}}
\multiput(456.00,628.24)(0.452,0.501){16}{\rule{1.500pt}{0.121pt}}
\multiput(456.00,623.51)(9.887,13.000){2}{\rule{0.750pt}{1.200pt}}
\multiput(469.00,641.24)(0.452,0.501){16}{\rule{1.500pt}{0.121pt}}
\multiput(469.00,636.51)(9.887,13.000){2}{\rule{0.750pt}{1.200pt}}
\multiput(482.00,654.24)(0.452,0.501){16}{\rule{1.500pt}{0.121pt}}
\multiput(482.00,649.51)(9.887,13.000){2}{\rule{0.750pt}{1.200pt}}
\multiput(495.00,667.24)(0.489,0.501){14}{\rule{1.600pt}{0.121pt}}
\multiput(495.00,662.51)(9.679,12.000){2}{\rule{0.800pt}{1.200pt}}
\multiput(508.00,679.24)(0.452,0.501){16}{\rule{1.500pt}{0.121pt}}
\multiput(508.00,674.51)(9.887,13.000){2}{\rule{0.750pt}{1.200pt}}
\multiput(521.00,692.24)(0.489,0.501){14}{\rule{1.600pt}{0.121pt}}
\multiput(521.00,687.51)(9.679,12.000){2}{\rule{0.800pt}{1.200pt}}
\multiput(534.00,704.24)(0.489,0.501){14}{\rule{1.600pt}{0.121pt}}
\multiput(534.00,699.51)(9.679,12.000){2}{\rule{0.800pt}{1.200pt}}
\multiput(547.00,716.24)(0.489,0.501){14}{\rule{1.600pt}{0.121pt}}
\multiput(547.00,711.51)(9.679,12.000){2}{\rule{0.800pt}{1.200pt}}
\multiput(560.00,728.24)(0.533,0.502){12}{\rule{1.718pt}{0.121pt}}
\multiput(560.00,723.51)(9.434,11.000){2}{\rule{0.859pt}{1.200pt}}
\multiput(573.00,739.24)(0.489,0.501){14}{\rule{1.600pt}{0.121pt}}
\multiput(573.00,734.51)(9.679,12.000){2}{\rule{0.800pt}{1.200pt}}
\multiput(586.00,751.24)(0.533,0.502){12}{\rule{1.718pt}{0.121pt}}
\multiput(586.00,746.51)(9.434,11.000){2}{\rule{0.859pt}{1.200pt}}
\multiput(599.00,762.24)(0.583,0.502){12}{\rule{1.827pt}{0.121pt}}
\multiput(599.00,757.51)(10.207,11.000){2}{\rule{0.914pt}{1.200pt}}
\multiput(613.00,773.24)(0.628,0.500){196}{\rule{1.815pt}{0.120pt}}
\multiput(613.00,768.51)(126.234,103.000){2}{\rule{0.907pt}{1.200pt}}
\multiput(743.00,876.24)(0.750,0.500){164}{\rule{2.107pt}{0.120pt}}
\multiput(743.00,871.51)(126.627,87.000){2}{\rule{1.053pt}{1.200pt}}
\multiput(874.00,963.24)(0.859,0.500){142}{\rule{2.368pt}{0.120pt}}
\multiput(874.00,958.51)(126.084,76.000){2}{\rule{1.184pt}{1.200pt}}
\multiput(1005.00,1039.24)(0.991,0.500){122}{\rule{2.682pt}{0.120pt}}
\multiput(1005.00,1034.51)(125.434,66.000){2}{\rule{1.341pt}{1.200pt}}
\put(1136,1103){\usebox{\plotpoint}}
\end{picture}

%% file: fa7b.tex
\setlength{\unitlength}{0.240900pt}
\ifx\plotpoint\undefined\newsavebox{\plotpoint}\fi
\sbox{\plotpoint}{\rule[-0.200pt]{0.400pt}{0.400pt}}%
\begin{picture}(1200,1200)(0,0)
\font\gnuplot=cmr10 at 10pt
\gnuplot
\sbox{\plotpoint}{\rule[-0.200pt]{0.400pt}{0.400pt}}%
\put(220.0,113.0){\rule[-0.200pt]{0.400pt}{245.477pt}}
\put(220.0,113.0){\rule[-0.200pt]{4.818pt}{0.400pt}}
\put(198,113){\makebox(0,0)[r]{0.5}}
\put(1116.0,113.0){\rule[-0.200pt]{4.818pt}{0.400pt}}
\put(220.0,317.0){\rule[-0.200pt]{4.818pt}{0.400pt}}
\put(198,317){\makebox(0,0)[r]{0.6}}
\put(1116.0,317.0){\rule[-0.200pt]{4.818pt}{0.400pt}}
\put(220.0,521.0){\rule[-0.200pt]{4.818pt}{0.400pt}}
\put(198,521){\makebox(0,0)[r]{0.7}}
\put(1116.0,521.0){\rule[-0.200pt]{4.818pt}{0.400pt}}
\put(220.0,724.0){\rule[-0.200pt]{4.818pt}{0.400pt}}
\put(198,724){\makebox(0,0)[r]{0.8}}
\put(1116.0,724.0){\rule[-0.200pt]{4.818pt}{0.400pt}}
\put(220.0,928.0){\rule[-0.200pt]{4.818pt}{0.400pt}}
\put(198,928){\makebox(0,0)[r]{0.9}}
\put(1116.0,928.0){\rule[-0.200pt]{4.818pt}{0.400pt}}
\put(220.0,1132.0){\rule[-0.200pt]{4.818pt}{0.400pt}}
\put(198,1132){\makebox(0,0)[r]{1.0}}
\put(1116.0,1132.0){\rule[-0.200pt]{4.818pt}{0.400pt}}
\put(220.0,113.0){\rule[-0.200pt]{0.400pt}{4.818pt}}
\put(220,68){\makebox(0,0){0}}
\put(220.0,1112.0){\rule[-0.200pt]{0.400pt}{4.818pt}}
\put(387.0,113.0){\rule[-0.200pt]{0.400pt}{4.818pt}}
\put(387,68){\makebox(0,0){0.1}}
\put(387.0,1112.0){\rule[-0.200pt]{0.400pt}{4.818pt}}
\put(553.0,113.0){\rule[-0.200pt]{0.400pt}{4.818pt}}
\put(553,68){\makebox(0,0){0.2}}
\put(553.0,1112.0){\rule[-0.200pt]{0.400pt}{4.818pt}}
\put(720.0,113.0){\rule[-0.200pt]{0.400pt}{4.818pt}}
\put(720,68){\makebox(0,0){0.3}}
\put(720.0,1112.0){\rule[-0.200pt]{0.400pt}{4.818pt}}
\put(886.0,113.0){\rule[-0.200pt]{0.400pt}{4.818pt}}
\put(886,68){\makebox(0,0){0.4}}
\put(886.0,1112.0){\rule[-0.200pt]{0.400pt}{4.818pt}}
\put(1053.0,113.0){\rule[-0.200pt]{0.400pt}{4.818pt}}
\put(1053,68){\makebox(0,0){0.5}}
\put(1053.0,1112.0){\rule[-0.200pt]{0.400pt}{4.818pt}}
\put(220.0,113.0){\rule[-0.200pt]{220.664pt}{0.400pt}}
\put(1136.0,113.0){\rule[-0.200pt]{0.400pt}{245.477pt}}
\put(220.0,1132.0){\rule[-0.200pt]{220.664pt}{0.400pt}}
\put(45,622){\makebox(0,0){\shortstack{$R$\\ (fm)}}}
\put(678,23){\makebox(0,0){$\rho_{B}$ (fm$^{-3})$}}
\put(678,1177){\makebox(0,0){Fig. 7(b)}}
\put(387,1021){\makebox(0,0)[l]{Temperature 200 MeV}}
\put(220.0,113.0){\rule[-0.200pt]{0.400pt}{245.477pt}}
\put(636,901){\makebox(0,0)[r]{$\epsilon_{\mbox{vac}}=(200\mbox{MeV})^{4}$}}
\put(658.0,901.0){\rule[-0.200pt]{15.899pt}{0.400pt}}
\put(222,297){\usebox{\plotpoint}}
\put(222,296.67){\rule{3.614pt}{0.400pt}}
\multiput(222.00,296.17)(7.500,1.000){2}{\rule{1.807pt}{0.400pt}}
\multiput(237.00,298.58)(3.304,0.496){43}{\rule{2.709pt}{0.120pt}}
\multiput(237.00,297.17)(144.378,23.000){2}{\rule{1.354pt}{0.400pt}}
\multiput(387.00,321.58)(3.272,0.493){23}{\rule{2.654pt}{0.119pt}}
\multiput(387.00,320.17)(77.492,13.000){2}{\rule{1.327pt}{0.400pt}}
\multiput(470.00,334.59)(5.446,0.488){13}{\rule{4.250pt}{0.117pt}}
\multiput(470.00,333.17)(74.179,8.000){2}{\rule{2.125pt}{0.400pt}}
\put(553,342.17){\rule{16.700pt}{0.400pt}}
\multiput(553.00,341.17)(48.338,2.000){2}{\rule{8.350pt}{0.400pt}}
\put(636,342.17){\rule{16.900pt}{0.400pt}}
\multiput(636.00,343.17)(48.923,-2.000){2}{\rule{8.450pt}{0.400pt}}
\multiput(720.00,340.92)(3.657,-0.496){43}{\rule{2.987pt}{0.120pt}}
\multiput(720.00,341.17)(159.800,-23.000){2}{\rule{1.493pt}{0.400pt}}
\multiput(886.00,317.92)(1.677,-0.498){97}{\rule{1.436pt}{0.120pt}}
\multiput(886.00,318.17)(164.020,-50.000){2}{\rule{0.718pt}{0.400pt}}
\multiput(1053.00,267.92)(0.968,-0.498){83}{\rule{0.872pt}{0.120pt}}
\multiput(1053.00,268.17)(81.190,-43.000){2}{\rule{0.436pt}{0.400pt}}
\put(636,856){\makebox(0,0)[r]{$\epsilon_{\mbox{vac}}=(250 \mbox{MeV})^{4}$}}
\multiput(658,856)(20.756,0.000){4}{\usebox{\plotpoint}}
\put(724,856){\usebox{\plotpoint}}
\put(222,308){\usebox{\plotpoint}}
\put(222.00,308.00){\usebox{\plotpoint}}
\multiput(237,308)(20.120,5.097){8}{\usebox{\plotpoint}}
\multiput(387,346)(19.970,5.654){8}{\usebox{\plotpoint}}
\multiput(553,393)(20.264,4.490){8}{\usebox{\plotpoint}}
\multiput(720,430)(20.524,3.091){8}{\usebox{\plotpoint}}
\multiput(886,455)(20.693,1.611){8}{\usebox{\plotpoint}}
\multiput(1053,468)(20.749,0.500){4}{\usebox{\plotpoint}}
\put(1136,470){\usebox{\plotpoint}}
\sbox{\plotpoint}{\rule[-0.400pt]{0.800pt}{0.800pt}}%
\put(636,811){\makebox(0,0)[r]{$\epsilon_{\mbox{vac}}=(300 \mbox{MeV})^{4}$}}
\put(658.0,811.0){\rule[-0.400pt]{15.899pt}{0.800pt}}
\put(222,311){\usebox{\plotpoint}}
\put(222,310.84){\rule{7.468pt}{0.800pt}}
\multiput(222.00,309.34)(15.500,3.000){2}{\rule{3.734pt}{0.800pt}}
\multiput(253.00,315.41)(1.647,0.502){85}{\rule{2.809pt}{0.121pt}}
\multiput(253.00,312.34)(144.170,46.000){2}{\rule{1.404pt}{0.800pt}}
\multiput(403.00,361.41)(1.476,0.502){107}{\rule{2.544pt}{0.121pt}}
\multiput(403.00,358.34)(161.720,57.000){2}{\rule{1.272pt}{0.800pt}}
\multiput(570.00,418.41)(1.764,0.502){79}{\rule{2.991pt}{0.121pt}}
\multiput(570.00,415.34)(143.793,43.000){2}{\rule{1.495pt}{0.800pt}}
\multiput(720.00,461.41)(2.103,0.502){73}{\rule{3.520pt}{0.121pt}}
\multiput(720.00,458.34)(158.694,40.000){2}{\rule{1.760pt}{0.800pt}}
\multiput(886.00,501.41)(2.746,0.503){55}{\rule{4.510pt}{0.121pt}}
\multiput(886.00,498.34)(157.640,31.000){2}{\rule{2.255pt}{0.800pt}}
\multiput(1053.00,532.40)(4.079,0.512){15}{\rule{6.236pt}{0.123pt}}
\multiput(1053.00,529.34)(70.056,11.000){2}{\rule{3.118pt}{0.800pt}}
\sbox{\plotpoint}{\rule[-0.500pt]{1.000pt}{1.000pt}}%
\put(636,766){\makebox(0,0)[r]{$\epsilon_{\mbox{vac}}=(350 \mbox{MeV})^{4}$}}
\multiput(658,766)(20.756,0.000){4}{\usebox{\plotpoint}}
\put(724,766){\usebox{\plotpoint}}
\put(222,313){\usebox{\plotpoint}}
\multiput(222,313)(20.409,3.779){4}{\usebox{\plotpoint}}
\multiput(303,328)(19.529,7.030){6}{\usebox{\plotpoint}}
\multiput(403,364)(19.446,7.256){3}{\usebox{\plotpoint}}
\multiput(470,389)(19.520,7.055){4}{\usebox{\plotpoint}}
\multiput(553,419)(19.749,6.386){9}{\usebox{\plotpoint}}
\multiput(720,473)(20.002,5.543){8}{\usebox{\plotpoint}}
\multiput(886,519)(20.212,4.720){8}{\usebox{\plotpoint}}
\multiput(1053,558)(20.380,3.929){4}{\usebox{\plotpoint}}
\put(1136,574){\usebox{\plotpoint}}
\sbox{\plotpoint}{\rule[-0.600pt]{1.200pt}{1.200pt}}%
\put(636,721){\makebox(0,0)[r]{$\epsilon_{\mbox{vac}}=(800 \mbox{MeV})^{4}$}}
\put(658.0,721.0){\rule[-0.600pt]{15.899pt}{1.200pt}}
\put(222,314){\usebox{\plotpoint}}
\put(222,312.01){\rule{3.614pt}{1.200pt}}
\multiput(222.00,311.51)(7.500,1.000){2}{\rule{1.807pt}{1.200pt}}
\multiput(237.00,317.24)(1.636,0.500){82}{\rule{4.213pt}{0.121pt}}
\multiput(237.00,312.51)(141.256,46.000){2}{\rule{2.107pt}{1.200pt}}
\multiput(387.00,363.24)(1.221,0.500){58}{\rule{3.229pt}{0.121pt}}
\multiput(387.00,358.51)(76.297,34.000){2}{\rule{1.615pt}{1.200pt}}
\multiput(470.00,397.24)(1.299,0.500){54}{\rule{3.413pt}{0.121pt}}
\multiput(470.00,392.51)(75.917,32.000){2}{\rule{1.706pt}{1.200pt}}
\multiput(553.00,429.24)(1.342,0.500){52}{\rule{3.513pt}{0.121pt}}
\multiput(553.00,424.51)(75.709,31.000){2}{\rule{1.756pt}{1.200pt}}
\multiput(636.00,460.24)(1.454,0.500){48}{\rule{3.776pt}{0.121pt}}
\multiput(636.00,455.51)(76.163,29.000){2}{\rule{1.888pt}{1.200pt}}
\multiput(720.00,489.24)(1.541,0.500){98}{\rule{3.989pt}{0.120pt}}
\multiput(720.00,484.51)(157.721,54.000){2}{\rule{1.994pt}{1.200pt}}
\multiput(886.00,543.24)(1.746,0.500){86}{\rule{4.475pt}{0.121pt}}
\multiput(886.00,538.51)(157.712,48.000){2}{\rule{2.238pt}{1.200pt}}
\multiput(1053.00,591.24)(2.002,0.501){32}{\rule{5.043pt}{0.121pt}}
\multiput(1053.00,586.51)(72.533,21.000){2}{\rule{2.521pt}{1.200pt}}
\end{picture}

%% file: fa8a.tex
\setlength{\unitlength}{0.240900pt}
\ifx\plotpoint\undefined\newsavebox{\plotpoint}\fi
\sbox{\plotpoint}{\rule[-0.200pt]{0.400pt}{0.400pt}}%
\begin{picture}(1200,1200)(0,0)
\font\gnuplot=cmr10 at 10pt
\gnuplot
\sbox{\plotpoint}{\rule[-0.200pt]{0.400pt}{0.400pt}}%
\put(220.0,113.0){\rule[-0.200pt]{220.664pt}{0.400pt}}
\put(220.0,113.0){\rule[-0.200pt]{0.400pt}{245.477pt}}
\put(220.0,113.0){\rule[-0.200pt]{4.818pt}{0.400pt}}
\put(198,113){\makebox(0,0)[r]{0.0}}
\put(1116.0,113.0){\rule[-0.200pt]{4.818pt}{0.400pt}}
\put(220.0,259.0){\rule[-0.200pt]{4.818pt}{0.400pt}}
\put(198,259){\makebox(0,0)[r]{10}}
\put(1116.0,259.0){\rule[-0.200pt]{4.818pt}{0.400pt}}
\put(220.0,404.0){\rule[-0.200pt]{4.818pt}{0.400pt}}
\put(198,404){\makebox(0,0)[r]{20}}
\put(1116.0,404.0){\rule[-0.200pt]{4.818pt}{0.400pt}}
\put(220.0,550.0){\rule[-0.200pt]{4.818pt}{0.400pt}}
\put(198,550){\makebox(0,0)[r]{30}}
\put(1116.0,550.0){\rule[-0.200pt]{4.818pt}{0.400pt}}
\put(220.0,695.0){\rule[-0.200pt]{4.818pt}{0.400pt}}
\put(198,695){\makebox(0,0)[r]{40}}
\put(1116.0,695.0){\rule[-0.200pt]{4.818pt}{0.400pt}}
\put(220.0,841.0){\rule[-0.200pt]{4.818pt}{0.400pt}}
\put(198,841){\makebox(0,0)[r]{50}}
\put(1116.0,841.0){\rule[-0.200pt]{4.818pt}{0.400pt}}
\put(220.0,986.0){\rule[-0.200pt]{4.818pt}{0.400pt}}
\put(198,986){\makebox(0,0)[r]{60}}
\put(1116.0,986.0){\rule[-0.200pt]{4.818pt}{0.400pt}}
\put(220.0,1132.0){\rule[-0.200pt]{4.818pt}{0.400pt}}
\put(198,1132){\makebox(0,0)[r]{70}}
\put(1116.0,1132.0){\rule[-0.200pt]{4.818pt}{0.400pt}}
\put(220.0,113.0){\rule[-0.200pt]{0.400pt}{4.818pt}}
\put(220,68){\makebox(0,0){0}}
\put(220.0,1112.0){\rule[-0.200pt]{0.400pt}{4.818pt}}
\put(351.0,113.0){\rule[-0.200pt]{0.400pt}{4.818pt}}
\put(351,68){\makebox(0,0){0.1}}
\put(351.0,1112.0){\rule[-0.200pt]{0.400pt}{4.818pt}}
\put(482.0,113.0){\rule[-0.200pt]{0.400pt}{4.818pt}}
\put(482,68){\makebox(0,0){0.2}}
\put(482.0,1112.0){\rule[-0.200pt]{0.400pt}{4.818pt}}
\put(613.0,113.0){\rule[-0.200pt]{0.400pt}{4.818pt}}
\put(613,68){\makebox(0,0){0.3}}
\put(613.0,1112.0){\rule[-0.200pt]{0.400pt}{4.818pt}}
\put(743.0,113.0){\rule[-0.200pt]{0.400pt}{4.818pt}}
\put(743,68){\makebox(0,0){0.4}}
\put(743.0,1112.0){\rule[-0.200pt]{0.400pt}{4.818pt}}
\put(874.0,113.0){\rule[-0.200pt]{0.400pt}{4.818pt}}
\put(874,68){\makebox(0,0){0.5}}
\put(874.0,1112.0){\rule[-0.200pt]{0.400pt}{4.818pt}}
\put(1005.0,113.0){\rule[-0.200pt]{0.400pt}{4.818pt}}
\put(1005,68){\makebox(0,0){0.6}}
\put(1005.0,1112.0){\rule[-0.200pt]{0.400pt}{4.818pt}}
\put(1136.0,113.0){\rule[-0.200pt]{0.400pt}{4.818pt}}
\put(1136,68){\makebox(0,0){0.7}}
\put(1136.0,1112.0){\rule[-0.200pt]{0.400pt}{4.818pt}}
\put(220.0,113.0){\rule[-0.200pt]{220.664pt}{0.400pt}}
\put(1136.0,113.0){\rule[-0.200pt]{0.400pt}{245.477pt}}
\put(220.0,1132.0){\rule[-0.200pt]{220.664pt}{0.400pt}}
\put(45,622){\makebox(0,0){\shortstack{$\sigma$\\ (MeV)}}}
\put(678,23){\makebox(0,0){$\rho_{B}$ (fm$^{-3})$}}
\put(678,1177){\makebox(0,0){Fig. 8(a)}}
\put(325,1059){\makebox(0,0)[l]{Cold nuclear matter}}
\put(220.0,113.0){\rule[-0.200pt]{0.400pt}{245.477pt}}
\put(639,986){\makebox(0,0)[r]{$\epsilon_{\mbox{vac}}=(200 \mbox{MeV})^{4}$}}
\put(661.0,986.0){\rule[-0.200pt]{15.899pt}{0.400pt}}
\put(221,116){\usebox{\plotpoint}}
\multiput(221.58,116.00)(0.499,1.026){125}{\rule{0.120pt}{0.919pt}}
\multiput(220.17,116.00)(64.000,129.093){2}{\rule{0.400pt}{0.459pt}}
\multiput(285.58,247.00)(0.499,0.873){129}{\rule{0.120pt}{0.797pt}}
\multiput(284.17,247.00)(66.000,113.346){2}{\rule{0.400pt}{0.398pt}}
\multiput(351.58,362.00)(0.493,0.814){23}{\rule{0.119pt}{0.746pt}}
\multiput(350.17,362.00)(13.000,19.451){2}{\rule{0.400pt}{0.373pt}}
\multiput(364.58,383.00)(0.493,0.814){23}{\rule{0.119pt}{0.746pt}}
\multiput(363.17,383.00)(13.000,19.451){2}{\rule{0.400pt}{0.373pt}}
\multiput(377.58,404.00)(0.493,0.814){23}{\rule{0.119pt}{0.746pt}}
\multiput(376.17,404.00)(13.000,19.451){2}{\rule{0.400pt}{0.373pt}}
\multiput(390.58,425.00)(0.493,0.774){23}{\rule{0.119pt}{0.715pt}}
\multiput(389.17,425.00)(13.000,18.515){2}{\rule{0.400pt}{0.358pt}}
\multiput(403.58,445.00)(0.493,0.774){23}{\rule{0.119pt}{0.715pt}}
\multiput(402.17,445.00)(13.000,18.515){2}{\rule{0.400pt}{0.358pt}}
\multiput(416.58,465.00)(0.493,0.774){23}{\rule{0.119pt}{0.715pt}}
\multiput(415.17,465.00)(13.000,18.515){2}{\rule{0.400pt}{0.358pt}}
\multiput(429.58,485.00)(0.493,0.774){23}{\rule{0.119pt}{0.715pt}}
\multiput(428.17,485.00)(13.000,18.515){2}{\rule{0.400pt}{0.358pt}}
\multiput(442.58,505.00)(0.494,0.680){25}{\rule{0.119pt}{0.643pt}}
\multiput(441.17,505.00)(14.000,17.666){2}{\rule{0.400pt}{0.321pt}}
\multiput(456.58,524.00)(0.493,0.734){23}{\rule{0.119pt}{0.685pt}}
\multiput(455.17,524.00)(13.000,17.579){2}{\rule{0.400pt}{0.342pt}}
\multiput(469.58,543.00)(0.493,0.774){23}{\rule{0.119pt}{0.715pt}}
\multiput(468.17,543.00)(13.000,18.515){2}{\rule{0.400pt}{0.358pt}}
\multiput(482.58,563.00)(0.493,0.734){23}{\rule{0.119pt}{0.685pt}}
\multiput(481.17,563.00)(13.000,17.579){2}{\rule{0.400pt}{0.342pt}}
\multiput(495.58,582.00)(0.493,0.734){23}{\rule{0.119pt}{0.685pt}}
\multiput(494.17,582.00)(13.000,17.579){2}{\rule{0.400pt}{0.342pt}}
\multiput(508.59,601.00)(0.482,0.762){9}{\rule{0.116pt}{0.700pt}}
\multiput(507.17,601.00)(6.000,7.547){2}{\rule{0.400pt}{0.350pt}}
\multiput(514.58,610.00)(0.496,0.727){37}{\rule{0.119pt}{0.680pt}}
\multiput(513.17,610.00)(20.000,27.589){2}{\rule{0.400pt}{0.340pt}}
\multiput(534.58,639.00)(0.493,0.774){23}{\rule{0.119pt}{0.715pt}}
\multiput(533.17,639.00)(13.000,18.515){2}{\rule{0.400pt}{0.358pt}}
\multiput(547.58,659.00)(0.499,0.781){129}{\rule{0.120pt}{0.724pt}}
\multiput(546.17,659.00)(66.000,101.497){2}{\rule{0.400pt}{0.362pt}}
\multiput(613.58,762.00)(0.497,0.941){61}{\rule{0.120pt}{0.850pt}}
\multiput(612.17,762.00)(32.000,58.236){2}{\rule{0.400pt}{0.425pt}}
\multiput(645.58,822.00)(0.497,1.203){63}{\rule{0.120pt}{1.058pt}}
\multiput(644.17,822.00)(33.000,76.805){2}{\rule{0.400pt}{0.529pt}}
\put(639,941){\makebox(0,0)[r]{$\epsilon_{\mbox{vac}}=(250 \mbox{MeV})^{4}$}}
\multiput(661,941)(20.756,0.000){4}{\usebox{\plotpoint}}
\put(727,941){\usebox{\plotpoint}}
\put(221,116){\usebox{\plotpoint}}
\multiput(221,116)(8.720,18.835){3}{\usebox{\plotpoint}}
\multiput(246,170)(9.762,18.316){11}{\usebox{\plotpoint}}
\multiput(351,367)(10.925,17.648){6}{\usebox{\plotpoint}}
\multiput(416,472)(11.842,17.046){6}{\usebox{\plotpoint}}
\multiput(482,567)(12.332,16.695){5}{\usebox{\plotpoint}}
\multiput(547,655)(13.014,16.169){5}{\usebox{\plotpoint}}
\multiput(613,737)(13.388,15.860){5}{\usebox{\plotpoint}}
\multiput(678,814)(13.593,15.685){5}{\usebox{\plotpoint}}
\multiput(743,889)(14.025,15.300){4}{\usebox{\plotpoint}}
\multiput(809,961)(13.697,15.594){5}{\usebox{\plotpoint}}
\multiput(874,1035)(13.508,15.759){5}{\usebox{\plotpoint}}
\put(946.40,1121.14){\usebox{\plotpoint}}
\put(954,1132){\usebox{\plotpoint}}
\sbox{\plotpoint}{\rule[-0.400pt]{0.800pt}{0.800pt}}%
\put(639,896){\makebox(0,0)[r]{$\epsilon_{\mbox{vac}}=(300 \mbox{MeV})^{4}$}}
\put(661.0,896.0){\rule[-0.400pt]{15.899pt}{0.800pt}}
\put(221,116){\usebox{\plotpoint}}
\multiput(222.41,116.00)(0.501,1.052){121}{\rule{0.121pt}{1.875pt}}
\multiput(219.34,116.00)(64.000,130.108){2}{\rule{0.800pt}{0.938pt}}
\multiput(286.41,250.00)(0.501,0.905){125}{\rule{0.121pt}{1.642pt}}
\multiput(283.34,250.00)(66.000,115.591){2}{\rule{0.800pt}{0.821pt}}
\multiput(352.41,369.00)(0.501,0.810){123}{\rule{0.121pt}{1.492pt}}
\multiput(349.34,369.00)(65.000,101.903){2}{\rule{0.800pt}{0.746pt}}
\multiput(417.41,474.00)(0.501,0.721){125}{\rule{0.121pt}{1.352pt}}
\multiput(414.34,474.00)(66.000,92.195){2}{\rule{0.800pt}{0.676pt}}
\multiput(483.41,569.00)(0.501,0.654){123}{\rule{0.121pt}{1.246pt}}
\multiput(480.34,569.00)(65.000,82.414){2}{\rule{0.800pt}{0.623pt}}
\multiput(548.41,654.00)(0.501,0.591){125}{\rule{0.121pt}{1.145pt}}
\multiput(545.34,654.00)(66.000,75.623){2}{\rule{0.800pt}{0.573pt}}
\multiput(614.41,732.00)(0.501,0.526){253}{\rule{0.121pt}{1.043pt}}
\multiput(611.34,732.00)(130.000,134.835){2}{\rule{0.800pt}{0.522pt}}
\multiput(743.00,870.41)(0.555,0.501){229}{\rule{1.088pt}{0.121pt}}
\multiput(743.00,867.34)(128.742,118.000){2}{\rule{0.544pt}{0.800pt}}
\multiput(874.00,988.41)(0.630,0.501){201}{\rule{1.208pt}{0.121pt}}
\multiput(874.00,985.34)(128.493,104.000){2}{\rule{0.604pt}{0.800pt}}
\multiput(1005.00,1092.41)(0.684,0.502){75}{\rule{1.293pt}{0.121pt}}
\multiput(1005.00,1089.34)(53.317,41.000){2}{\rule{0.646pt}{0.800pt}}
\sbox{\plotpoint}{\rule[-0.500pt]{1.000pt}{1.000pt}}%
\put(639,851){\makebox(0,0)[r]{$\epsilon_{\mbox{vac}}=(350 \mbox{MeV})^{4}$}}
\multiput(661,851)(20.756,0.000){4}{\usebox{\plotpoint}}
\put(727,851){\usebox{\plotpoint}}
\put(221,116){\usebox{\plotpoint}}
\multiput(221,116)(8.945,18.729){8}{\usebox{\plotpoint}}
\multiput(285,250)(10.002,18.186){6}{\usebox{\plotpoint}}
\multiput(351,370)(10.925,17.648){6}{\usebox{\plotpoint}}
\multiput(416,475)(11.842,17.046){6}{\usebox{\plotpoint}}
\multiput(482,570)(12.702,16.415){5}{\usebox{\plotpoint}}
\multiput(547,654)(13.609,15.671){5}{\usebox{\plotpoint}}
\multiput(613,730)(14.508,14.843){9}{\usebox{\plotpoint}}
\multiput(743,863)(15.835,13.418){8}{\usebox{\plotpoint}}
\multiput(874,974)(16.802,12.185){8}{\usebox{\plotpoint}}
\multiput(1005,1069)(17.561,11.063){5}{\usebox{\plotpoint}}
\put(1105,1132){\usebox{\plotpoint}}
\sbox{\plotpoint}{\rule[-0.600pt]{1.200pt}{1.200pt}}%
\put(639,806){\makebox(0,0)[r]{$\epsilon_{\mbox{vac}}=(800 \mbox{MeV})^{4}$}}
\put(661.0,806.0){\rule[-0.600pt]{15.899pt}{1.200pt}}
\put(233,142){\usebox{\plotpoint}}
\multiput(235.24,142.00)(0.500,0.969){226}{\rule{0.120pt}{2.629pt}}
\multiput(230.51,142.00)(118.000,223.544){2}{\rule{1.200pt}{1.314pt}}
\multiput(353.24,371.00)(0.501,0.822){16}{\rule{0.121pt}{2.331pt}}
\multiput(348.51,371.00)(13.000,17.162){2}{\rule{1.200pt}{1.165pt}}
\multiput(366.24,393.00)(0.501,0.822){16}{\rule{0.121pt}{2.331pt}}
\multiput(361.51,393.00)(13.000,17.162){2}{\rule{1.200pt}{1.165pt}}
\multiput(379.24,415.00)(0.501,0.781){16}{\rule{0.121pt}{2.238pt}}
\multiput(374.51,415.00)(13.000,16.354){2}{\rule{1.200pt}{1.119pt}}
\multiput(392.24,436.00)(0.501,0.739){16}{\rule{0.121pt}{2.146pt}}
\multiput(387.51,436.00)(13.000,15.546){2}{\rule{1.200pt}{1.073pt}}
\multiput(405.24,456.00)(0.501,0.739){16}{\rule{0.121pt}{2.146pt}}
\multiput(400.51,456.00)(13.000,15.546){2}{\rule{1.200pt}{1.073pt}}
\multiput(418.24,476.00)(0.501,0.739){16}{\rule{0.121pt}{2.146pt}}
\multiput(413.51,476.00)(13.000,15.546){2}{\rule{1.200pt}{1.073pt}}
\multiput(431.24,496.00)(0.501,0.698){16}{\rule{0.121pt}{2.054pt}}
\multiput(426.51,496.00)(13.000,14.737){2}{\rule{1.200pt}{1.027pt}}
\multiput(444.24,515.00)(0.501,0.647){18}{\rule{0.121pt}{1.929pt}}
\multiput(439.51,515.00)(14.000,14.997){2}{\rule{1.200pt}{0.964pt}}
\multiput(458.24,534.00)(0.501,0.657){16}{\rule{0.121pt}{1.962pt}}
\multiput(453.51,534.00)(13.000,13.929){2}{\rule{1.200pt}{0.981pt}}
\multiput(471.24,552.00)(0.501,0.657){16}{\rule{0.121pt}{1.962pt}}
\multiput(466.51,552.00)(13.000,13.929){2}{\rule{1.200pt}{0.981pt}}
\multiput(484.24,570.00)(0.501,0.657){16}{\rule{0.121pt}{1.962pt}}
\multiput(479.51,570.00)(13.000,13.929){2}{\rule{1.200pt}{0.981pt}}
\multiput(497.24,588.00)(0.501,0.616){16}{\rule{0.121pt}{1.869pt}}
\multiput(492.51,588.00)(13.000,13.120){2}{\rule{1.200pt}{0.935pt}}
\multiput(510.24,605.00)(0.501,0.575){16}{\rule{0.121pt}{1.777pt}}
\multiput(505.51,605.00)(13.000,12.312){2}{\rule{1.200pt}{0.888pt}}
\multiput(523.24,621.00)(0.501,0.616){16}{\rule{0.121pt}{1.869pt}}
\multiput(518.51,621.00)(13.000,13.120){2}{\rule{1.200pt}{0.935pt}}
\multiput(536.24,638.00)(0.501,0.575){16}{\rule{0.121pt}{1.777pt}}
\multiput(531.51,638.00)(13.000,12.312){2}{\rule{1.200pt}{0.888pt}}
\multiput(549.24,654.00)(0.501,0.534){16}{\rule{0.121pt}{1.685pt}}
\multiput(544.51,654.00)(13.000,11.503){2}{\rule{1.200pt}{0.842pt}}
\multiput(562.24,669.00)(0.501,0.575){16}{\rule{0.121pt}{1.777pt}}
\multiput(557.51,669.00)(13.000,12.312){2}{\rule{1.200pt}{0.888pt}}
\multiput(575.24,685.00)(0.501,0.534){16}{\rule{0.121pt}{1.685pt}}
\multiput(570.51,685.00)(13.000,11.503){2}{\rule{1.200pt}{0.842pt}}
\multiput(588.24,700.00)(0.501,0.493){16}{\rule{0.121pt}{1.592pt}}
\multiput(583.51,700.00)(13.000,10.695){2}{\rule{1.200pt}{0.796pt}}
\multiput(601.24,714.00)(0.501,0.495){18}{\rule{0.121pt}{1.586pt}}
\multiput(596.51,714.00)(14.000,11.709){2}{\rule{1.200pt}{0.793pt}}
\multiput(613.00,731.24)(0.505,0.500){246}{\rule{1.519pt}{0.120pt}}
\multiput(613.00,726.51)(126.848,128.000){2}{\rule{0.759pt}{1.200pt}}
\multiput(743.00,859.24)(0.621,0.500){200}{\rule{1.797pt}{0.120pt}}
\multiput(743.00,854.51)(127.270,105.000){2}{\rule{0.899pt}{1.200pt}}
\multiput(874.00,964.24)(0.741,0.500){166}{\rule{2.086pt}{0.120pt}}
\multiput(874.00,959.51)(126.670,88.000){2}{\rule{1.043pt}{1.200pt}}
\multiput(1005.00,1052.24)(0.883,0.500){138}{\rule{2.424pt}{0.120pt}}
\multiput(1005.00,1047.51)(125.968,74.000){2}{\rule{1.212pt}{1.200pt}}
\put(1136,1124){\usebox{\plotpoint}}
\end{picture}

%% file: fa8b.tex
\setlength{\unitlength}{0.240900pt}
\ifx\plotpoint\undefined\newsavebox{\plotpoint}\fi
\sbox{\plotpoint}{\rule[-0.200pt]{0.400pt}{0.400pt}}%
\begin{picture}(1200,1200)(0,0)
\font\gnuplot=cmr10 at 10pt
\gnuplot
\sbox{\plotpoint}{\rule[-0.200pt]{0.400pt}{0.400pt}}%
\put(220.0,113.0){\rule[-0.200pt]{220.664pt}{0.400pt}}
\put(220.0,113.0){\rule[-0.200pt]{0.400pt}{245.477pt}}
\put(220.0,113.0){\rule[-0.200pt]{4.818pt}{0.400pt}}
\put(198,113){\makebox(0,0)[r]{0.0}}
\put(1116.0,113.0){\rule[-0.200pt]{4.818pt}{0.400pt}}
\put(220.0,240.0){\rule[-0.200pt]{4.818pt}{0.400pt}}
\put(198,240){\makebox(0,0)[r]{10}}
\put(1116.0,240.0){\rule[-0.200pt]{4.818pt}{0.400pt}}
\put(220.0,368.0){\rule[-0.200pt]{4.818pt}{0.400pt}}
\put(198,368){\makebox(0,0)[r]{20}}
\put(1116.0,368.0){\rule[-0.200pt]{4.818pt}{0.400pt}}
\put(220.0,495.0){\rule[-0.200pt]{4.818pt}{0.400pt}}
\put(198,495){\makebox(0,0)[r]{30}}
\put(1116.0,495.0){\rule[-0.200pt]{4.818pt}{0.400pt}}
\put(220.0,623.0){\rule[-0.200pt]{4.818pt}{0.400pt}}
\put(198,623){\makebox(0,0)[r]{40}}
\put(1116.0,623.0){\rule[-0.200pt]{4.818pt}{0.400pt}}
\put(220.0,750.0){\rule[-0.200pt]{4.818pt}{0.400pt}}
\put(198,750){\makebox(0,0)[r]{50}}
\put(1116.0,750.0){\rule[-0.200pt]{4.818pt}{0.400pt}}
\put(220.0,877.0){\rule[-0.200pt]{4.818pt}{0.400pt}}
\put(198,877){\makebox(0,0)[r]{60}}
\put(1116.0,877.0){\rule[-0.200pt]{4.818pt}{0.400pt}}
\put(220.0,1005.0){\rule[-0.200pt]{4.818pt}{0.400pt}}
\put(198,1005){\makebox(0,0)[r]{70}}
\put(1116.0,1005.0){\rule[-0.200pt]{4.818pt}{0.400pt}}
\put(220.0,1132.0){\rule[-0.200pt]{4.818pt}{0.400pt}}
\put(198,1132){\makebox(0,0)[r]{80}}
\put(1116.0,1132.0){\rule[-0.200pt]{4.818pt}{0.400pt}}
\put(220.0,113.0){\rule[-0.200pt]{0.400pt}{4.818pt}}
\put(220,68){\makebox(0,0){0}}
\put(220.0,1112.0){\rule[-0.200pt]{0.400pt}{4.818pt}}
\put(387.0,113.0){\rule[-0.200pt]{0.400pt}{4.818pt}}
\put(387,68){\makebox(0,0){0.1}}
\put(387.0,1112.0){\rule[-0.200pt]{0.400pt}{4.818pt}}
\put(553.0,113.0){\rule[-0.200pt]{0.400pt}{4.818pt}}
\put(553,68){\makebox(0,0){0.2}}
\put(553.0,1112.0){\rule[-0.200pt]{0.400pt}{4.818pt}}
\put(720.0,113.0){\rule[-0.200pt]{0.400pt}{4.818pt}}
\put(720,68){\makebox(0,0){0.3}}
\put(720.0,1112.0){\rule[-0.200pt]{0.400pt}{4.818pt}}
\put(886.0,113.0){\rule[-0.200pt]{0.400pt}{4.818pt}}
\put(886,68){\makebox(0,0){0.4}}
\put(886.0,1112.0){\rule[-0.200pt]{0.400pt}{4.818pt}}
\put(1053.0,113.0){\rule[-0.200pt]{0.400pt}{4.818pt}}
\put(1053,68){\makebox(0,0){0.5}}
\put(1053.0,1112.0){\rule[-0.200pt]{0.400pt}{4.818pt}}
\put(220.0,113.0){\rule[-0.200pt]{220.664pt}{0.400pt}}
\put(1136.0,113.0){\rule[-0.200pt]{0.400pt}{245.477pt}}
\put(220.0,1132.0){\rule[-0.200pt]{220.664pt}{0.400pt}}
\put(45,622){\makebox(0,0){\shortstack{$\sigma$\\ (MeV)}}}
\put(678,23){\makebox(0,0){$\rho_{B}$ (fm$^{-3})$}}
\put(678,1177){\makebox(0,0){Fig. 8(b)}}
\put(387,1005){\makebox(0,0)[l]{Temperature 200 MeV}}
\put(220.0,113.0){\rule[-0.200pt]{0.400pt}{245.477pt}}
\put(636,941){\makebox(0,0)[r]{$\epsilon_{\mbox{vac}}=(200 \mbox{MeV})^{4}$}}
\put(658.0,941.0){\rule[-0.200pt]{15.899pt}{0.400pt}}
\put(222,228){\usebox{\plotpoint}}
\put(222,227.67){\rule{3.614pt}{0.400pt}}
\multiput(222.00,227.17)(7.500,1.000){2}{\rule{1.807pt}{0.400pt}}
\multiput(237.00,229.58)(1.029,0.499){143}{\rule{0.922pt}{0.120pt}}
\multiput(237.00,228.17)(148.087,73.000){2}{\rule{0.461pt}{0.400pt}}
\multiput(387.00,302.58)(0.849,0.498){95}{\rule{0.778pt}{0.120pt}}
\multiput(387.00,301.17)(81.386,49.000){2}{\rule{0.389pt}{0.400pt}}
\multiput(470.00,351.58)(0.925,0.498){87}{\rule{0.838pt}{0.120pt}}
\multiput(470.00,350.17)(81.261,45.000){2}{\rule{0.419pt}{0.400pt}}
\multiput(553.00,396.58)(1.068,0.498){75}{\rule{0.951pt}{0.120pt}}
\multiput(553.00,395.17)(81.026,39.000){2}{\rule{0.476pt}{0.400pt}}
\multiput(636.00,435.58)(1.242,0.498){65}{\rule{1.088pt}{0.120pt}}
\multiput(636.00,434.17)(81.741,34.000){2}{\rule{0.544pt}{0.400pt}}
\multiput(720.00,469.58)(1.543,0.498){105}{\rule{1.330pt}{0.120pt}}
\multiput(720.00,468.17)(163.240,54.000){2}{\rule{0.665pt}{0.400pt}}
\multiput(886.00,523.58)(2.273,0.498){71}{\rule{1.905pt}{0.120pt}}
\multiput(886.00,522.17)(163.045,37.000){2}{\rule{0.953pt}{0.400pt}}
\multiput(1053.00,560.59)(4.805,0.489){15}{\rule{3.789pt}{0.118pt}}
\multiput(1053.00,559.17)(75.136,9.000){2}{\rule{1.894pt}{0.400pt}}
\put(636,896){\makebox(0,0)[r]{$\epsilon_{\mbox{vac}}=(250 \mbox{MeV})^{4}$}}
\multiput(658,896)(20.756,0.000){4}{\usebox{\plotpoint}}
\put(724,896){\usebox{\plotpoint}}
\put(222,228){\usebox{\plotpoint}}
\put(222.00,228.00){\usebox{\plotpoint}}
\multiput(237,229)(18.364,9.672){8}{\usebox{\plotpoint}}
\multiput(387,308)(17.493,11.170){10}{\usebox{\plotpoint}}
\multiput(553,414)(18.225,9.931){9}{\usebox{\plotpoint}}
\multiput(720,505)(18.915,8.546){9}{\usebox{\plotpoint}}
\multiput(886,580)(19.420,7.326){8}{\usebox{\plotpoint}}
\multiput(1053,643)(19.806,6.204){5}{\usebox{\plotpoint}}
\put(1136,669){\usebox{\plotpoint}}
\sbox{\plotpoint}{\rule[-0.400pt]{0.800pt}{0.800pt}}%
\put(636,851){\makebox(0,0)[r]{$\epsilon_{\mbox{vac}}=(300 \mbox{MeV})^{4}$}}
\put(658.0,851.0){\rule[-0.400pt]{15.899pt}{0.800pt}}
\put(222,227){\usebox{\plotpoint}}
\multiput(222.00,228.39)(3.253,0.536){5}{\rule{4.333pt}{0.129pt}}
\multiput(222.00,225.34)(22.006,6.000){2}{\rule{2.167pt}{0.800pt}}
\multiput(253.00,234.41)(0.854,0.501){169}{\rule{1.564pt}{0.121pt}}
\multiput(253.00,231.34)(146.755,88.000){2}{\rule{0.782pt}{0.800pt}}
\multiput(403.00,322.41)(0.760,0.501){213}{\rule{1.415pt}{0.121pt}}
\multiput(403.00,319.34)(164.064,110.000){2}{\rule{0.707pt}{0.800pt}}
\multiput(570.00,432.41)(0.874,0.501){165}{\rule{1.595pt}{0.121pt}}
\multiput(570.00,429.34)(146.689,86.000){2}{\rule{0.798pt}{0.800pt}}
\multiput(720.00,518.41)(1.003,0.501){159}{\rule{1.800pt}{0.121pt}}
\multiput(720.00,515.34)(162.264,83.000){2}{\rule{0.900pt}{0.800pt}}
\multiput(886.00,601.41)(1.182,0.501){135}{\rule{2.082pt}{0.121pt}}
\multiput(886.00,598.34)(162.679,71.000){2}{\rule{1.041pt}{0.800pt}}
\multiput(1053.00,672.41)(1.404,0.503){53}{\rule{2.413pt}{0.121pt}}
\multiput(1053.00,669.34)(77.991,30.000){2}{\rule{1.207pt}{0.800pt}}
\sbox{\plotpoint}{\rule[-0.500pt]{1.000pt}{1.000pt}}%
\put(636,806){\makebox(0,0)[r]{$\epsilon_{\mbox{vac}}=(350 \mbox{MeV})^{4}$}}
\multiput(658,806)(20.756,0.000){4}{\usebox{\plotpoint}}
\put(724,806){\usebox{\plotpoint}}
\put(222,227){\usebox{\plotpoint}}
\multiput(222,227)(19.541,6.996){5}{\usebox{\plotpoint}}
\multiput(303,256)(17.323,11.433){5}{\usebox{\plotpoint}}
\multiput(403,322)(17.111,11.748){4}{\usebox{\plotpoint}}
\multiput(470,368)(17.206,11.609){5}{\usebox{\plotpoint}}
\multiput(553,424)(17.854,10.584){10}{\usebox{\plotpoint}}
\multiput(720,523)(18.429,9.548){9}{\usebox{\plotpoint}}
\multiput(886,609)(18.976,8.409){8}{\usebox{\plotpoint}}
\multiput(1053,683)(19.366,7.466){5}{\usebox{\plotpoint}}
\put(1136,715){\usebox{\plotpoint}}
\sbox{\plotpoint}{\rule[-0.600pt]{1.200pt}{1.200pt}}%
\put(636,761){\makebox(0,0)[r]{$\epsilon_{\mbox{vac}}=(800 \mbox{MeV})^{4}$}}
\put(658.0,761.0){\rule[-0.600pt]{15.899pt}{1.200pt}}
\put(222,227){\usebox{\plotpoint}}
\put(222,225.51){\rule{3.614pt}{1.200pt}}
\multiput(222.00,224.51)(7.500,2.000){2}{\rule{1.807pt}{1.200pt}}
\multiput(237.00,231.24)(0.902,0.500){156}{\rule{2.469pt}{0.120pt}}
\multiput(237.00,226.51)(144.876,83.000){2}{\rule{1.234pt}{1.200pt}}
\multiput(387.00,314.24)(0.711,0.500){106}{\rule{2.017pt}{0.120pt}}
\multiput(387.00,309.51)(78.813,58.000){2}{\rule{1.009pt}{1.200pt}}
\multiput(470.00,372.24)(0.723,0.500){104}{\rule{2.047pt}{0.120pt}}
\multiput(470.00,367.51)(78.751,57.000){2}{\rule{1.024pt}{1.200pt}}
\multiput(553.00,429.24)(0.778,0.500){96}{\rule{2.179pt}{0.120pt}}
\multiput(553.00,424.51)(78.477,53.000){2}{\rule{1.090pt}{1.200pt}}
\multiput(636.00,482.24)(0.853,0.500){88}{\rule{2.357pt}{0.121pt}}
\multiput(636.00,477.51)(79.108,49.000){2}{\rule{1.179pt}{1.200pt}}
\multiput(720.00,531.24)(0.931,0.500){168}{\rule{2.538pt}{0.120pt}}
\multiput(720.00,526.51)(160.732,89.000){2}{\rule{1.269pt}{1.200pt}}
\multiput(886.00,620.24)(1.070,0.500){146}{\rule{2.869pt}{0.120pt}}
\multiput(886.00,615.51)(161.045,78.000){2}{\rule{1.435pt}{1.200pt}}
\multiput(1053.00,698.24)(1.185,0.500){60}{\rule{3.146pt}{0.121pt}}
\multiput(1053.00,693.51)(76.471,35.000){2}{\rule{1.573pt}{1.200pt}}
\end{picture}

%% file: fa9.tex
\setlength{\unitlength}{0.240900pt}
\ifx\plotpoint\undefined\newsavebox{\plotpoint}\fi
\sbox{\plotpoint}{\rule[-0.200pt]{0.400pt}{0.400pt}}%
\begin{picture}(1200,1200)(0,0)
\font\gnuplot=cmr10 at 10pt
\gnuplot
\sbox{\plotpoint}{\rule[-0.200pt]{0.400pt}{0.400pt}}%
\put(220.0,453.0){\rule[-0.200pt]{220.664pt}{0.400pt}}
\put(220.0,113.0){\rule[-0.200pt]{0.400pt}{245.477pt}}
\put(220.0,113.0){\rule[-0.200pt]{4.818pt}{0.400pt}}
\put(198,113){\makebox(0,0)[r]{-40}}
\put(1116.0,113.0){\rule[-0.200pt]{4.818pt}{0.400pt}}
\put(220.0,283.0){\rule[-0.200pt]{4.818pt}{0.400pt}}
\put(198,283){\makebox(0,0)[r]{-20}}
\put(1116.0,283.0){\rule[-0.200pt]{4.818pt}{0.400pt}}
\put(220.0,453.0){\rule[-0.200pt]{4.818pt}{0.400pt}}
\put(198,453){\makebox(0,0)[r]{0.0}}
\put(1116.0,453.0){\rule[-0.200pt]{4.818pt}{0.400pt}}
\put(220.0,623.0){\rule[-0.200pt]{4.818pt}{0.400pt}}
\put(198,623){\makebox(0,0)[r]{20}}
\put(1116.0,623.0){\rule[-0.200pt]{4.818pt}{0.400pt}}
\put(220.0,792.0){\rule[-0.200pt]{4.818pt}{0.400pt}}
\put(198,792){\makebox(0,0)[r]{40}}
\put(1116.0,792.0){\rule[-0.200pt]{4.818pt}{0.400pt}}
\put(220.0,962.0){\rule[-0.200pt]{4.818pt}{0.400pt}}
\put(198,962){\makebox(0,0)[r]{60}}
\put(1116.0,962.0){\rule[-0.200pt]{4.818pt}{0.400pt}}
\put(220.0,1132.0){\rule[-0.200pt]{4.818pt}{0.400pt}}
\put(198,1132){\makebox(0,0)[r]{80}}
\put(1116.0,1132.0){\rule[-0.200pt]{4.818pt}{0.400pt}}
\put(220.0,113.0){\rule[-0.200pt]{0.400pt}{4.818pt}}
\put(220,68){\makebox(0,0){0}}
\put(220.0,1112.0){\rule[-0.200pt]{0.400pt}{4.818pt}}
\put(351.0,113.0){\rule[-0.200pt]{0.400pt}{4.818pt}}
\put(351,68){\makebox(0,0){0.1}}
\put(351.0,1112.0){\rule[-0.200pt]{0.400pt}{4.818pt}}
\put(482.0,113.0){\rule[-0.200pt]{0.400pt}{4.818pt}}
\put(482,68){\makebox(0,0){0.2}}
\put(482.0,1112.0){\rule[-0.200pt]{0.400pt}{4.818pt}}
\put(613.0,113.0){\rule[-0.200pt]{0.400pt}{4.818pt}}
\put(613,68){\makebox(0,0){0.3}}
\put(613.0,1112.0){\rule[-0.200pt]{0.400pt}{4.818pt}}
\put(743.0,113.0){\rule[-0.200pt]{0.400pt}{4.818pt}}
\put(743,68){\makebox(0,0){0.4}}
\put(743.0,1112.0){\rule[-0.200pt]{0.400pt}{4.818pt}}
\put(874.0,113.0){\rule[-0.200pt]{0.400pt}{4.818pt}}
\put(874,68){\makebox(0,0){0.5}}
\put(874.0,1112.0){\rule[-0.200pt]{0.400pt}{4.818pt}}
\put(1005.0,113.0){\rule[-0.200pt]{0.400pt}{4.818pt}}
\put(1005,68){\makebox(0,0){0.6}}
\put(1005.0,1112.0){\rule[-0.200pt]{0.400pt}{4.818pt}}
\put(1136.0,113.0){\rule[-0.200pt]{0.400pt}{4.818pt}}
\put(1136,68){\makebox(0,0){0.7}}
\put(1136.0,1112.0){\rule[-0.200pt]{0.400pt}{4.818pt}}
\put(220.0,113.0){\rule[-0.200pt]{220.664pt}{0.400pt}}
\put(1136.0,113.0){\rule[-0.200pt]{0.400pt}{245.477pt}}
\put(220.0,1132.0){\rule[-0.200pt]{220.664pt}{0.400pt}}
\put(45,622){\makebox(0,0){\shortstack{$E_{tot}-M_{N}$\\$(\mbox{MeV})^{4}$}}}
\put(678,23){\makebox(0,0){$\rho_{B}$ (fm$^{-3})$}}
\put(678,1177){\makebox(0,0){Fig. 9}}
\put(285,1005){\makebox(0,0)[l]{Cold nuclear matter}}
\put(220.0,113.0){\rule[-0.200pt]{0.400pt}{245.477pt}}
\put(613,920){\makebox(0,0)[r]{$\epsilon_{\mbox{vac}}=(200 \mbox{MeV})^{4}$}}
\put(635.0,920.0){\rule[-0.200pt]{15.899pt}{0.400pt}}
\put(221,463){\usebox{\plotpoint}}
\multiput(221.58,460.17)(0.499,-0.727){125}{\rule{0.120pt}{0.681pt}}
\multiput(220.17,461.59)(64.000,-91.586){2}{\rule{0.400pt}{0.341pt}}
\multiput(285.00,368.92)(0.523,-0.499){123}{\rule{0.519pt}{0.120pt}}
\multiput(285.00,369.17)(64.923,-63.000){2}{\rule{0.260pt}{0.400pt}}
\multiput(351.00,305.93)(0.950,-0.485){11}{\rule{0.843pt}{0.117pt}}
\multiput(351.00,306.17)(11.251,-7.000){2}{\rule{0.421pt}{0.400pt}}
\multiput(364.00,298.93)(1.123,-0.482){9}{\rule{0.967pt}{0.116pt}}
\multiput(364.00,299.17)(10.994,-6.000){2}{\rule{0.483pt}{0.400pt}}
\multiput(377.00,292.93)(1.378,-0.477){7}{\rule{1.140pt}{0.115pt}}
\multiput(377.00,293.17)(10.634,-5.000){2}{\rule{0.570pt}{0.400pt}}
\multiput(390.00,287.95)(2.695,-0.447){3}{\rule{1.833pt}{0.108pt}}
\multiput(390.00,288.17)(9.195,-3.000){2}{\rule{0.917pt}{0.400pt}}
\put(403,284.67){\rule{3.132pt}{0.400pt}}
\multiput(403.00,285.17)(6.500,-1.000){2}{\rule{1.566pt}{0.400pt}}
\put(429,285.17){\rule{2.700pt}{0.400pt}}
\multiput(429.00,284.17)(7.396,2.000){2}{\rule{1.350pt}{0.400pt}}
\multiput(442.00,287.61)(2.918,0.447){3}{\rule{1.967pt}{0.108pt}}
\multiput(442.00,286.17)(9.918,3.000){2}{\rule{0.983pt}{0.400pt}}
\multiput(456.00,290.59)(1.378,0.477){7}{\rule{1.140pt}{0.115pt}}
\multiput(456.00,289.17)(10.634,5.000){2}{\rule{0.570pt}{0.400pt}}
\multiput(469.00,295.59)(0.950,0.485){11}{\rule{0.843pt}{0.117pt}}
\multiput(469.00,294.17)(11.251,7.000){2}{\rule{0.421pt}{0.400pt}}
\multiput(482.00,302.59)(0.950,0.485){11}{\rule{0.843pt}{0.117pt}}
\multiput(482.00,301.17)(11.251,7.000){2}{\rule{0.421pt}{0.400pt}}
\multiput(495.00,309.58)(0.652,0.491){17}{\rule{0.620pt}{0.118pt}}
\multiput(495.00,308.17)(11.713,10.000){2}{\rule{0.310pt}{0.400pt}}
\multiput(508.00,319.59)(0.491,0.482){9}{\rule{0.500pt}{0.116pt}}
\multiput(508.00,318.17)(4.962,6.000){2}{\rule{0.250pt}{0.400pt}}
\multiput(514.00,325.58)(0.498,0.496){37}{\rule{0.500pt}{0.119pt}}
\multiput(514.00,324.17)(18.962,20.000){2}{\rule{0.250pt}{0.400pt}}
\multiput(534.58,345.00)(0.493,0.616){23}{\rule{0.119pt}{0.592pt}}
\multiput(533.17,345.00)(13.000,14.771){2}{\rule{0.400pt}{0.296pt}}
\multiput(547.58,361.00)(0.499,0.918){129}{\rule{0.120pt}{0.833pt}}
\multiput(546.17,361.00)(66.000,119.270){2}{\rule{0.400pt}{0.417pt}}
\multiput(613.58,482.00)(0.497,1.652){61}{\rule{0.120pt}{1.413pt}}
\multiput(612.17,482.00)(32.000,102.068){2}{\rule{0.400pt}{0.706pt}}
\multiput(645.58,587.00)(0.497,2.950){63}{\rule{0.120pt}{2.439pt}}
\multiput(644.17,587.00)(33.000,187.937){2}{\rule{0.400pt}{1.220pt}}
\put(416.0,285.0){\rule[-0.200pt]{3.132pt}{0.400pt}}
\put(613,875){\makebox(0,0)[r]{$\epsilon_{\mbox{vac}}=(250 \mbox{MeV})^{4}$}}
\multiput(635,875)(20.756,0.000){4}{\usebox{\plotpoint}}
\put(701,875){\usebox{\plotpoint}}
\put(221,463){\usebox{\plotpoint}}
\multiput(221,463)(14.101,-15.230){2}{\usebox{\plotpoint}}
\multiput(246,436)(14.746,-14.606){7}{\usebox{\plotpoint}}
\multiput(351,332)(19.567,-6.924){4}{\usebox{\plotpoint}}
\multiput(416,309)(20.640,2.189){3}{\usebox{\plotpoint}}
\multiput(482,316)(18.275,9.840){3}{\usebox{\plotpoint}}
\multiput(547,351)(15.358,13.962){5}{\usebox{\plotpoint}}
\multiput(613,411)(12.332,16.695){5}{\usebox{\plotpoint}}
\multiput(678,499)(10.146,18.107){6}{\usebox{\plotpoint}}
\multiput(743,615)(8.267,19.038){8}{\usebox{\plotpoint}}
\multiput(809,767)(6.503,19.710){10}{\usebox{\plotpoint}}
\multiput(874,964)(4.807,20.191){9}{\usebox{\plotpoint}}
\put(914,1132){\usebox{\plotpoint}}
\sbox{\plotpoint}{\rule[-0.400pt]{0.800pt}{0.800pt}}%
\put(613,830){\makebox(0,0)[r]{$\epsilon_{\mbox{vac}}=(300 \mbox{MeV})^{4}$}}
\put(635.0,830.0){\rule[-0.400pt]{15.899pt}{0.800pt}}
\put(221,464){\usebox{\plotpoint}}
\multiput(222.41,459.49)(0.501,-0.554){121}{\rule{0.121pt}{1.087pt}}
\multiput(219.34,461.74)(64.000,-68.743){2}{\rule{0.800pt}{0.544pt}}
\multiput(285.00,391.09)(0.623,-0.502){99}{\rule{1.196pt}{0.121pt}}
\multiput(285.00,391.34)(63.517,-53.000){2}{\rule{0.598pt}{0.800pt}}
\multiput(351.00,338.09)(1.442,-0.505){39}{\rule{2.461pt}{0.122pt}}
\multiput(351.00,338.34)(59.892,-23.000){2}{\rule{1.230pt}{0.800pt}}
\put(416,316.84){\rule{15.899pt}{0.800pt}}
\multiput(416.00,315.34)(33.000,3.000){2}{\rule{7.950pt}{0.800pt}}
\multiput(482.00,321.41)(1.135,0.504){51}{\rule{1.993pt}{0.121pt}}
\multiput(482.00,318.34)(60.863,29.000){2}{\rule{0.997pt}{0.800pt}}
\multiput(547.00,350.41)(0.674,0.502){91}{\rule{1.278pt}{0.121pt}}
\multiput(547.00,347.34)(63.348,49.000){2}{\rule{0.639pt}{0.800pt}}
\multiput(614.41,398.00)(0.501,0.604){253}{\rule{0.121pt}{1.166pt}}
\multiput(611.34,398.00)(130.000,154.580){2}{\rule{0.800pt}{0.583pt}}
\multiput(744.41,555.00)(0.501,0.872){255}{\rule{0.121pt}{1.592pt}}
\multiput(741.34,555.00)(131.000,224.695){2}{\rule{0.800pt}{0.796pt}}
\multiput(875.41,783.00)(0.501,1.144){255}{\rule{0.121pt}{2.026pt}}
\multiput(872.34,783.00)(131.000,294.795){2}{\rule{0.800pt}{1.013pt}}
\multiput(1006.41,1082.00)(0.507,1.515){27}{\rule{0.122pt}{2.553pt}}
\multiput(1003.34,1082.00)(17.000,44.701){2}{\rule{0.800pt}{1.276pt}}
\sbox{\plotpoint}{\rule[-0.500pt]{1.000pt}{1.000pt}}%
\put(613,785){\makebox(0,0)[r]{$\epsilon_{\mbox{vac}}=(350 \mbox{MeV})^{4}$}}
\multiput(635,785)(20.756,0.000){4}{\usebox{\plotpoint}}
\put(701,785){\usebox{\plotpoint}}
\put(221,464){\usebox{\plotpoint}}
\multiput(221,464)(14.225,-15.114){5}{\usebox{\plotpoint}}
\multiput(285,396)(16.183,-12.996){4}{\usebox{\plotpoint}}
\multiput(351,343)(19.567,-6.924){3}{\usebox{\plotpoint}}
\multiput(416,320)(20.746,0.629){4}{\usebox{\plotpoint}}
\multiput(482,322)(19.271,7.708){3}{\usebox{\plotpoint}}
\multiput(547,348)(17.149,11.692){4}{\usebox{\plotpoint}}
\multiput(613,393)(13.962,15.358){9}{\usebox{\plotpoint}}
\multiput(743,536)(11.333,17.388){12}{\usebox{\plotpoint}}
\multiput(874,737)(9.603,18.400){13}{\usebox{\plotpoint}}
\multiput(1005,988)(8.430,18.967){8}{\usebox{\plotpoint}}
\put(1069,1132){\usebox{\plotpoint}}
\sbox{\plotpoint}{\rule[-0.600pt]{1.200pt}{1.200pt}}%
\put(613,740){\makebox(0,0)[r]{$\epsilon_{\mbox{vac}}=(800 \mbox{MeV})^{4}$}}
\put(635.0,740.0){\rule[-0.600pt]{15.899pt}{1.200pt}}
\put(233,456){\usebox{\plotpoint}}
\multiput(233.00,453.26)(0.538,-0.500){208}{\rule{1.599pt}{0.120pt}}
\multiput(233.00,453.51)(114.681,-109.000){2}{\rule{0.800pt}{1.200pt}}
\multiput(351.00,344.26)(0.835,-0.505){4}{\rule{2.529pt}{0.122pt}}
\multiput(351.00,344.51)(7.752,-7.000){2}{\rule{1.264pt}{1.200pt}}
\multiput(364.00,337.25)(0.962,-0.509){2}{\rule{2.900pt}{0.123pt}}
\multiput(364.00,337.51)(6.981,-6.000){2}{\rule{1.450pt}{1.200pt}}
\put(377,329.01){\rule{3.132pt}{1.200pt}}
\multiput(377.00,331.51)(6.500,-5.000){2}{\rule{1.566pt}{1.200pt}}
\put(390,325.01){\rule{3.132pt}{1.200pt}}
\multiput(390.00,326.51)(6.500,-3.000){2}{\rule{1.566pt}{1.200pt}}
\put(403,322.01){\rule{3.132pt}{1.200pt}}
\multiput(403.00,323.51)(6.500,-3.000){2}{\rule{1.566pt}{1.200pt}}
\put(416,319.51){\rule{3.132pt}{1.200pt}}
\multiput(416.00,320.51)(6.500,-2.000){2}{\rule{1.566pt}{1.200pt}}
\put(429,318.01){\rule{3.132pt}{1.200pt}}
\multiput(429.00,318.51)(6.500,-1.000){2}{\rule{1.566pt}{1.200pt}}
\put(442,318.01){\rule{3.373pt}{1.200pt}}
\multiput(442.00,317.51)(7.000,1.000){2}{\rule{1.686pt}{1.200pt}}
\put(456,319.01){\rule{3.132pt}{1.200pt}}
\multiput(456.00,318.51)(6.500,1.000){2}{\rule{1.566pt}{1.200pt}}
\put(469,320.51){\rule{3.132pt}{1.200pt}}
\multiput(469.00,319.51)(6.500,2.000){2}{\rule{1.566pt}{1.200pt}}
\put(482,323.01){\rule{3.132pt}{1.200pt}}
\multiput(482.00,321.51)(6.500,3.000){2}{\rule{1.566pt}{1.200pt}}
\put(495,326.51){\rule{3.132pt}{1.200pt}}
\multiput(495.00,324.51)(6.500,4.000){2}{\rule{1.566pt}{1.200pt}}
\put(508,330.51){\rule{3.132pt}{1.200pt}}
\multiput(508.00,328.51)(6.500,4.000){2}{\rule{1.566pt}{1.200pt}}
\multiput(521.00,337.24)(0.962,0.509){2}{\rule{2.900pt}{0.123pt}}
\multiput(521.00,332.51)(6.981,6.000){2}{\rule{1.450pt}{1.200pt}}
\multiput(534.00,343.24)(0.962,0.509){2}{\rule{2.900pt}{0.123pt}}
\multiput(534.00,338.51)(6.981,6.000){2}{\rule{1.450pt}{1.200pt}}
\multiput(547.00,349.24)(0.835,0.505){4}{\rule{2.529pt}{0.122pt}}
\multiput(547.00,344.51)(7.752,7.000){2}{\rule{1.264pt}{1.200pt}}
\multiput(560.00,356.24)(0.835,0.505){4}{\rule{2.529pt}{0.122pt}}
\multiput(560.00,351.51)(7.752,7.000){2}{\rule{1.264pt}{1.200pt}}
\multiput(573.00,363.24)(0.732,0.503){6}{\rule{2.250pt}{0.121pt}}
\multiput(573.00,358.51)(8.330,8.000){2}{\rule{1.125pt}{1.200pt}}
\multiput(586.00,371.24)(0.587,0.502){10}{\rule{1.860pt}{0.121pt}}
\multiput(586.00,366.51)(9.139,10.000){2}{\rule{0.930pt}{1.200pt}}
\multiput(599.00,381.24)(0.715,0.502){8}{\rule{2.167pt}{0.121pt}}
\multiput(599.00,376.51)(9.503,9.000){2}{\rule{1.083pt}{1.200pt}}
\multiput(613.00,390.24)(0.497,0.500){250}{\rule{1.500pt}{0.120pt}}
\multiput(613.00,385.51)(126.887,130.000){2}{\rule{0.750pt}{1.200pt}}
\multiput(745.24,518.00)(0.500,0.677){252}{\rule{0.120pt}{1.931pt}}
\multiput(740.51,518.00)(131.000,173.993){2}{\rule{1.200pt}{0.965pt}}
\multiput(876.24,696.00)(0.500,0.827){252}{\rule{0.120pt}{2.288pt}}
\multiput(871.51,696.00)(131.000,212.252){2}{\rule{1.200pt}{1.144pt}}
\multiput(1007.24,913.00)(0.500,0.935){224}{\rule{0.120pt}{2.546pt}}
\multiput(1002.51,913.00)(117.000,213.715){2}{\rule{1.200pt}{1.273pt}}
\end{picture}